\documentclass[11pt]{article}
\usepackage[utf8]{inputenc}
\usepackage[left=1in,right=1in,top=1in,bottom=1in]{geometry}
\usepackage{lmodern}
\usepackage{fancyhdr}
\usepackage{setspace}   
\usepackage{hhline}     
\usepackage{float}      
\usepackage{array}      
\usepackage{eurosym}    
\usepackage{arydshln}   
\usepackage{csquotes}   
\usepackage{nccmath}    
\usepackage{xcolor}     
\usepackage{multirow}   
\usepackage{bm}         
\usepackage{physics}    
\usepackage{systeme}    
\usepackage{mathtools}  
\usepackage{amsmath, amsthm, amssymb, amsfonts} 
\usepackage{graphicx}   
\usepackage{pgfplots}   
\usepackage{tikz} 
\usetikzlibrary{intersections,calc} 
\usepackage[labelfont=sc,textfont=it]{caption} 
\usepackage{soul} 
\usepackage{sidecap} 
\usepackage[space]{grffile} 
\usepackage{subcaption} 
\usepackage{makebox} 
\usepackage{abraces} 
\usepackage{comment} 
\usepackage{cancel} 
\usepackage{accents} 
\usepackage{stackrel} 
\usepackage[export]{adjustbox}
\usepackage{centernot} 
\usepackage{scalerel} 
\usepackage{makeidx}
\usepackage[flushleft]{threeparttable}
\usepackage{listings}
\usepackage[round]{natbib}
\usepackage{hyperref,etoolbox}
\usepackage{xfrac}
\usepackage{wrapfig}
\usepackage{paralist}
\usepackage{titlesec}
\usepackage{lastpage}

\AtBeginDocument{
\addtolength{\abovedisplayskip}{-5pt}
\addtolength{\abovedisplayshortskip}{-5pt}
\addtolength{\belowdisplayskip}{-5pt}
\addtolength{\belowdisplayshortskip}{-5pt}
}

\hypersetup{
    colorlinks=true,
    linkcolor=black,
    filecolor=magenta,
    citecolor=black, 
    urlcolor=black,
    }
\makeatletter
\def\@footnotecolor{red}
\define@key{Hyp}{footnotecolor}{%
 \HyColor@HyperrefColor{#1}\@footnotecolor%
}
\patchcmd{\@footnotemark}{\hyper@linkstart{link}}{\hyper@linkstart{footnote}}{}{}
\makeatother
\hypersetup{footnotecolor=black}
\usepackage{pdflscape}

\usepackage[ruled,vlined]{algorithm2e}
\SetKwInput{KwInput}{Input}                
\SetKwInput{KwOutput}{Output}

\newlength{\bibitemsep}
\newlength{\bibparskip}
\let\oldthebibliography\thebibliography
\renewcommand\thebibliography[1]{%
  \oldthebibliography{#1}%
  \setlength{\parskip}{\bibitemsep}%
  \setlength{\itemsep}{\bibparskip}%
}

\newcommand{\parT}[1]{\left(#1\right)}

\newcommand{\parG}[1]{\left\{#1\right\}}

\newcommand\bw[1]{\textcolor{white}{#1}}

\newcommand{\E}{\mathbb{E}}
\newcommand{\I}{\mathbb{I}}

\newcommand*{\giv}{\hspace{1pt}|\hspace{1pt}}

\newcommand{\Cov}{\operatorname{Cov}}
\newcommand{\Var}{\operatorname{Var}}

\newcommand{\bs}[1]{\boldsymbol{#1}}

\newcolumntype{L}[1]{>{\raggedright\let\newline\\\arraybackslash\hspace{0pt}}m{#1}}
\newcolumntype{C}[1]{>{\centering\let\newline\\\arraybackslash\hspace{0pt}}m{#1}}
\newcolumntype{R}[1]{>{\raggedleft\let\newline\\\arraybackslash\hspace{0pt}}m{#1}}

\theoremstyle{definition}

\newtheorem*{assumption*}{\assumptionnumber}
\providecommand{\assumptionnumber}{}


\theoremstyle{definition}

\theoremstyle{definition}

\theoremstyle{definition}

\theoremstyle{definition}

\theoremstyle{theorem}

\theoremstyle{definition}

\theoremstyle{definition}

\definecolor{blue}{RGB}{0,114,178}
\definecolor{red}{RGB}{204,51,17}
\definecolor{yellow}{RGB}{240,228,66}
\definecolor{green}{RGB}{0,158,115}

\tikzset{
solid node/.style={circle,draw,inner sep=1.5,fill=black},
hollow node/.style={circle,draw,inner sep=1.5}
}

\title{
\bf Measuring the Euro Area Output Gap$^\ast$\\[-10pt]
}
\author{\normalsize \textsc{Matteo Barigozzi}\\[-2pt] \small University of Bologna \\[-3pt] \footnotesize matteo.barigozzi@unibo.it
\and
\normalsize  \textsc{Claudio Lissona}\\[-2pt] \small  University of Bologna \\[-3pt] \footnotesize claudio.lissona2@unibo.it
\and
\normalsize \textsc{Matteo Luciani}\\[-2pt] \small Federal Reserve Board \\[-3pt] \footnotesize matteo.luciani@frb.gov}

\date{\vspace{11pt} \small{Last update: \today} \\}

\begin{document}
\maketitle

\vspace{-20pt}
\begin{abstract}
\noindent We measure the Euro Area (EA) output gap and potential output using a non-stationary dynamic factor model estimated on a large dataset of macroeconomic and financial variables. Our results indicate that, between 2012 and 2024, the EA economy was consistently tighter than suggested by institutional estimates, implying that its weak growth reflects a potential output problem rather than a business-cycle one. Moreover, we find that the decline in trend inflation---rather than economic slack---kept core inflation below 2\% before the pandemic, while demand forces explain at least 30\% of the post-pandemic rise in core inflation.

\end{abstract}

\renewcommand{\thefootnote}{$\ast$} 
\thispagestyle{empty}

\footnotetext{We would like to thank  for helpful comment Travis Berge, Danilo Cascaldi Garcia, Antonio Conti, Thiago Ferreira, Domenico Giannone, Manuel Gonzalez-Astudillo, Michele Lenza, Giovanni Pellegrino, and Riccardo Trezzi. This paper has benefited also from discussions with seminar participants at the Federal Reserve Board and with participants of several conferences. M. Barigozzi and C. Lissona gratefully acknowledge financial support from MIUR (PRIN2020, Grant 2020N9YFFE). Of course, any error is our responsibility.\smallskip

\noindent \textsc{Disclaimer:} the views expressed in this paper are those of the authors and do not necessarily reflect the views and policies of the Board of Governors or the Federal Reserve System.} 

\renewcommand{\thefootnote}{\arabic{footnote}
}

\section{Introduction}
The decomposition of GDP in potential output---the level of output consistent with current technologies and ``normal'' use of capital and labor---and the output gap---the percentage deviation of GDP from its potential---is a fundamental task for policymakers. Potential output tells us how fast an economy can grow in the long run; the output gap helps assess the cyclical position of the economy and, thus, potential inflationary pressures \citep[e.g.,][]{jarocinski2018inflation, banbura_does_2023}. Both measures are central to the common Euro Area (EA) monetary policy and the fiscal policy of individual countries---they are among the main pillars of the EA fiscal surveillance framework, ultimately affecting the fiscal capacity of each member country  \citep{EC_Stability}. However, since both quantities are unobserved, policymakers need a model to extract them from the data.

This paper proposes a new measure of potential output and the output gap for the EA based on a non-stationary dynamic factor model estimated on a large dataset of macroeconomic and financial variables. Compared to the prevailing literature, which focuses on theoretical structural models with few variables of interest, we adopt a distinct approach because we let the data speak by leveraging a large information set conditional on a few key macroeconomic priors---for example, the long-run slowdown in output growth \citep{cette2016pre}. 

We conduct our analysis on a new large-dimensional dataset comprising 118 EA economic indicators from 2001:Q1 to 2025:Q1. Four main results emerge from our analysis: first, our output gap estimate is in line with those published by the European Commission (EC) and the International Monetary Fund (IMF) until the 2011--2012 Sovereign Debt Recession (henceforth, SDR), after which our output gap measure suggests that the EA economy was tighter than estimated by the EC and the IMF. Moreover, we estimate that potential output growth decelerated after the 2008--2009 Global Financial Crisis (henceforth, GFC), and as of 2025:Q1, potential output growth has yet to return to the pre-GFC pace. In other words, our results suggest that the EA has a potential output issue, not a business cycle issue. Hence, if the goal is to achieve better economic conditions in the EA, European countries should both implement structural reforms and promote productivity-enhancing investments, and support aggregate demand more forcefully during downturns to mitigate recession-induced output losses.  


Second, we find that, on average, the Okun's law relationship and the Phillips Curve are satisfied in our model, even though we do not impose either of them; hence, cyclical movements in real activity, unemployment, and inflation are interconnected \citep[see, e.g.,][]{BianchiNicoloSong}. 

Third, we find that core inflation remained below 2\% after the GFC, not because there was slack in the economy, but rather because trend inflation decreased by one percentage point---in line with the idea that inflation expectations de-anchored on the downside after the GFC \citep{CiccarelliOsbat, CorselloNeriTagliabracci}. Moreover, we show that the output gap contributed to at least 30\% of the post-pandemic increase in core inflation, thus supporting existing literature that suggests demand forces played a substantial role in the rise of post-pandemic inflation \citep{AscariTrezzi,GiannonePrimicieri,Canovaetal}. Finally, our output gap measure yields better inflation forecasts than those derived from commonly used methods. These results confirm that our data-driven measure is economically meaningful and valuable for policy analysis.


Fourth, we find that growth financed through household debt is not sustainable in the long run; hence, policies aimed at boosting debt-financed household consumption or residential investment would deliver only short-term gains.

To measure potential output and the output gap, we first estimate the non-stationary dynamic factor model by Quasi Maximum Likelihood using the EM algorithm jointly with the Kalman smoother \citep{doz2012quasi,barigozzi_quasi_2022}. Then, we extract a common trend from the estimated common factors and compute the cyclical component by subtracting the common trend from the common factors. Having estimated the common trend and the common cyclical component, we measure potential output as the part of GDP explained by the common trends and the output gap as the part of GDP explained by the common cyclical component. Our model belongs to the class of unobserved component models, which is among the most indicated ones for extracting the transitory component of GDP \citep{canova2022faq}.

This model is an enhanced version of the model \citet{barigozzi_measuring_2021} used to measure the US output gap, as it does not require a long sample to identify the trend and deals with the Covid pandemic. Specifically, we use a three-step estimation procedure to account for the latter. First, we estimate the model using only pre-Covid data. Second, we estimate the effects induced by the Covid shock (level-shift and increased volatility). Third, we re-estimate the model on the full dataset after purging the data from Covid-induced dynamics. 

Our model is both large-dimensional and non-stationary. A large-dimensional model enables us to capture well-established co-movements in macroeconomic variables. It is now widely recognized that cross-sectional aggregation of a large number of series allows to consistently disentangle the co-movements in the data from idiosyncratic dynamics \citep{stock_chapter_2016}, and that a rich information set is necessary to obtain meaningful estimates of potential output and the output gap \citep{buncic_discovering_2022}. To the best of our knowledge, this is the first paper that estimates the EA output gap with such a rich information set. 

Allowing for non-stationarity enables us to go beyond the common practice of pre-transforming data into stationary, mean-centered variables. This approach preserves features that are crucial for identifying the common trend, which would otherwise be lost if the data were differenced to achieve stationarity \citep{ng2018comments}. This is important not only for estimating the output gap but also because accurately accounting for low-frequency movements helps eliminate any potential confounding effect when estimating the relationship between real activity and inflation over the business cycle \citep{BianchiNicoloSong}.

\paragraph{Related literature.} How to estimate potential output and the output gap has been a hotly debated topic for several decades \citep[see, e.g.,][]{canova2022faq}. The literature has proposed two main approaches: a theoretical approach and a statistical approach. 

The theoretical approach uses theoretical models, such as production-function-based models used by the EC \citep{havik2014production} and the IMF \citep{demasi1997} or New-Keynesian DSGE models \citep{JPT,burlon2020reliable,furlanetto2021output}. 

The statistical approach uses (univariate or multivariate) statistical models, sometimes paired with some macroeconomic relationships of interest, e.g., the Phillips Curve. For example, many papers rely on univariate models \citep[e.g.,][]{morley2003beveridge,kamber2018intuitive,hamilton2018you,phillips2021business,phillips2021boosting,hartl2022fractional}, while few others employ multivariate non-stationary methods, which are either low-dimensional models \citep{jarocinski2018inflation,ManuelGA,toth_multivariate_2021, hasenzagl2022model}, or medium-size, but stationary, models \citep{aastveit2014estimating,morley2020estimating,morley_estimating_2023,OgapBilkent}. Most of these works focus on the US, while only a few focus on the EA, the most recent being \citet{morley_estimating_2023}. 

\paragraph{Structure of the paper.} The rest of the paper is organized as follows. In Section \ref{sec::data}, we present the data used in our analysis, and in Section \ref{sec::method}, we present the model and the estimation strategy. Then, in Section \ref{sec::poutgap}, we present our estimate of potential output and the output gap, and in Section \ref{sec::OL&PC}, we dive deep into the economic content of these estimates, focusing on the Okun's law and the Phillips correlation. Next, we focus on inflation. Section \ref{sec::InflationDynamics} interprets inflation dynamics after the GFC through the lens of our model, and Section \ref{sec::InflationForecasting} assesses the ability of our output gap estimate to forecast inflation. Finally, Section \ref{sec::Credit} looks into what signals our model takes from financial indicators to estimate the output gap, and Section \ref{sec::conclude} concludes.

A Supplemental Appendix contains additional details on the model, showing various robustness analyses, and reporting additional results. Specifically, Appendix \ref{datades} provides full details on the dataset, while Appendixes \ref{app:ass}, \ref{sec::estdetail}, and \ref{app::confbands} provide additional details about the model. Next, Appendixes \ref{sec::covid} and \ref{app::notvpars} provide robustness analysis. Appendix \ref{sec::altTC} compares our output gap estimate with that obtained with alternative statistical methodologies. Lastly, Appendix \ref{sec::altTCest} compares our output gap estimate with that obtained with alternative estimation strategies for the common trend, and Appendix \ref{sec::realtime} assesses the reliability of our output gap estimate.

%
%
\section{A large Euro Area dataset} \label{sec::data}
We construct a large macroeconomic dataset of $n=118$ EA series, observed from 2001:Q1 to 2025:Q1 ($T=97$). The dataset contains a wide range of macroeconomic indicators, including national account statistics, industrial production and turnover indicators, labor market and compensation indicators, price indexes, oil prices, natural gas prices, house prices, exchange rates, interest rates, a stock market index, monetary aggregates, non-financial assets and liabilities, and confidence indexes. 

In terms of broad categories of data, we include in the dataset the usual suspects normally considered for high-dimensional macroeconomic analysis \citep[see, for example,][]{mccracken_fred-md_2016,mccracken_fred-qd_2020}. 

In terms of which and how many series to include for each category, we use a mix of economic and statistical reasoning. On the one hand, to identify the common factors driving the co-movement in the data, it is crucial to pool information from many indicators; hence, a larger information set should be preferred. On the other hand, a key assumption of the model is the presence of mild cross-sectional correlation among the idiosyncratic components: violating this assumption leads to a deterioration of the model's performance \citep{boivin2006more,lucioADFM}. Thus, when building a dataset for factor analysis, we face a trade-off between the need for a larger information set and the risk of introducing too much idiosyncratic correlation. For this reason, we selected the variables to include in the dataset to maximize economic signal while limiting the noise, with exceptions motivated by economic reasoning. For instance, we include GDP and its components since their informational content justifies a relatively high level of idiosyncratic correlations. Similarly, consumption and employment are decomposed according to durability and sectoral composition, respectively, while assets and liabilities are decomposed by ownership. In contrast, we keep the consumer price index for energy while dropping the producer price index for energy, as they carry the same signal (their correlation is greater than 0.95). Likewise, we drop the consumer price index for industrial goods because it has a correlation greater than 0.95 with the goods consumer price index.

As for the treatment of the series, we take logarithms for all variables except for confidence indicators and those already expressed in percentage points. We keep all variables in levels except for price indicators, for which we take first differences; i.e., we work with inflation rates. This is a common approach used in the literature to avoid spurious dynamics resulting from the potential $I(2)$ behavior in price indexes \citep{stock_chapter_2016,mccracken_fred-md_2016,mccracken_fred-qd_2020}. Appendix \ref{datades} provides the complete list of variables included in the dataset, along with their sources and treatment.

%
%
\section{Methodology}\label{sec::method}
In Section \ref{sec::mod}, we outline the model and its main features---we discuss all the details, formal assumptions, and further comments in Appendix \ref{app:ass}. In Section \ref{sec::fattoni}, we sketch how we estimate the model while referring the reader to Appendix \ref{sec::estdetail} for a step-by-step guide on how estimation is carried out in practice and Appendix \ref{app::confbands} for the bootstrap procedure used to measure uncertainty around our estimates.

\subsection{The model}\label{sec::mod}
We denote the observed $i$-th time series at a given quarter $t$ as $y_{it}$, with $1\le i\le 118$ and $\text{2001:Q1} \le t\le \text{2025:Q1}$. In our non-stationary dynamic factor model, each variable is the sum of (i) a secular component $\mathrm{D}_{it}$, which is treated either as deterministic or stochastic, (ii) $q$ common factors $\mathbf{f}_t=(f_{1t}\cdots f_{qt})'$, which capture the macroeconomic long- and short-run co-movements and have a dynamics governed by a VAR, and (iii) an idiosyncratic component $\xi_{it}$, which captures local dynamics or measurement errors and is possibly correlated across $i$ and $t$.  

We partition the $n$ series according to two features. First, according to the nature of the secular component, that is, whether $\mathrm{D}_{it}$ is a stochastic or a deterministic process. Second, according to the nature of the idiosyncratic component, that is, whether $\xi_{it}$ has a stochastic trend or it is stationary.

In particular, we model $\mathrm{D}_{it}$ as a local linear trend for GDP to account for the well-documented slowdown in productivity \citep{cette2016pre} and for households' financial liabilities ({\small HHLB}) and households' long-term loans ({\small HHLB.LLN}), which make up more than 80\% of total household's liabilities, whose average growth rate has slowed down consistently since the GFC. {For these variables, we say that $i\in \mathcal{L}_1$}. Moreover, we model $\mathrm D_{it}$ as a local level model for the unemployment rate ({\small UNETOT}) to account for relevant labor market features \citep{cette2016pre}---this includes the reallocation of employees across sectors, which contributed to the slowdown in the EA productivity growth---and for all consumer price inflation indexes ({\small HICPOV, HICPNEF, HICPG, HICPSV, HICPNG, HICPFD}) as well as oil and natural gas prices ({\small POIL, PNGAS}) to account for the slowdown in inflation occurred after the GFC---accounting for low-frequency dynamics in inflation is crucial to avoid potential confounding effects which may alter the relation between inflation and real activity over the business cycle \citep{BianchiNicoloSong}. For these variables, we say that $i\in \mathcal{L}_0$.\footnote{Appendix \ref{app::notvpars} shows  results for the estimate of the output gap when removing the time variation in the secular trends.} For all other series, $\mathrm{D}_{it}$ is either a linear trend with a constant slope, in which case we say that $i\in\mathcal I_b$, or $\mathrm D_{it}$ is just a constant equal to $\mathrm D_{i0}$. To determine $\mathcal I_b$, we test the significance of the sample mean of $\Delta{y}_{it}$ (see Appendix \ref{datades}). 

As for the idiosyncratic components, if $\xi_{it}\sim I(1)$, then we say that $i\in\mathcal I_1$ and we model $\xi_{it}$ as a random walk, while if $\xi_{it}\sim I(0)$, we say that $i\notin\mathcal I_1$ and we leave its dynamics unspecified to avoid over-parametrization of the model. To determine $\mathcal I_1$, we employ the test proposed by \cite{bai2004panic} for the null hypothesis of an idiosyncratic unit root (see Appendix \ref{datades}).

Furthermore, we capture the effect of the Covid shock, which generated a large shift both in the levels \citep{ng2021modeling,stockcomovement}, and in the volatility \citep{lenza_how_2022,carriero_addressing_2022} of most macroeconomic EA series, through an additional common factor $g_t$ \citep{stockcomovement}, and a scalar $s_t$ scaling the conditional volatility of the latent factors \citep{lenza_how_2022}. While we model the former to have an impact on all series only in 2020 and 2021, we allow the latter to have an effect that persists even after the recovery from the pandemic. These choices reflect the fact that mobility restrictions and lockdowns in the EA have been on and off until early 2022, while in the US, they were enforced only at the beginning of the pandemic.

Formally, the model reads as follows: 
\begin{align}
y_{it}\ &=\ \mathrm{D}_{it} + \bm\lambda_{i}^\prime\mathbf {f}_{t} + \gamma_i\hspace{1pt} g_t \mathbb I_{\text{\tiny 2020:Q1$\le\! t\!\le$2021:Q4}} +  \xi_{it}, \hspace{10pt}&& 1\le i\le 118,\;\;\text{\footnotesize 2001:Q1}\le t \le \text{\footnotesize 2025:Q1},
\label{eq::obseq} \\[3pt]
\mathrm{D}_{it}\ &=\ \mathrm{D}_{it-1} + b_{i,t-1}\mathbb I_{i\in\mathcal I_b} + \epsilon_{it}\hspace{1pt}, && \epsilon_{it}\sim (0,\sigma_{\epsilon_i}^2\mathbb I_{i\in \mathcal{L}_0}),\label{eq::seccomp}\\[3pt] 
b_{it}\ &=\ b_{it-1} +\eta_{it}\hspace{1pt}, && \eta_{it}\sim(0,\sigma_{\eta_i}^2\mathbb I_{i\in \mathcal{L}_1}), \label{eq::betat} \\[3pt] 
\mathbf{f}_t\ &=\sum_{j=1}^{p}\mathbf A_j\mathbf{f}_{t-j} +  \{s_t \mathbb I_{\,\text{\tiny $t\!\ge\!$ 2020:Q1}} + \mathbb I_{\,\text{\tiny $t\!<\!$ 2020:Q1} })\}\mathbf{u}_t\hspace{1pt}, &&\mathbf{u}_t\stackrel{{i.i.d.}}{\sim}(\mathbf{0},\bs{\Sigma}_u), \label{eq::faceq}\\[3pt] 
\xi_{it}\ &=\xi_{it-1} \mathbb I_{i\in\mathcal I_1}+ e_{it},  &&  e_{it}\stackrel{}{\sim} ({0},\sigma_{e_i}^2), \label{eq::idioeq}
\end{align}
where $\mathbb I_A=1$, if $A$ is true, and $\mathbb I_A=0$, otherwise, $\bm\lambda_i$, $\mathbf f_t$, $\mathbf u_t$ are $q$-dimensional vectors, and $\mathbf A_j$ and $\bm\Sigma_u$ are $q\times q$. We set $p=2$ based on the BIC criterion for a VAR on the estimated factors, and $q=4$ based on standard criteria implemented on differenced data \citep{bai2002determining,hallin_determining_2007,ABC}. Furthermore, following the criteria by \citet{barigozzi_large-dimensional_2021} and \citet{ACFZ} applied on the zero-frequency spectral density matrix of the differenced data, we impose the presence of one common trend, ${\tau}_t$, driving the non-stationarity in the factors $\mathbf{f}_t$. This finding  is consistent with many theoretical models in which a common trend is the sole driver of long-run dynamics \citep[e.g.,][]{del2007fit}. 

Given the above results, if follows that $\mathbf{f}_t$ is a cointegrated vector with cointegration rank $q-1=3$. Consequently, the VAR characteristic polynomial in \eqref{eq::faceq} has one root in $z=1$, while the remaining roots lie inside the unit circle. Hence, $\mathbf{f}_t$ also admits the following decomposition \citep{escribano1994cointegration}:
\begin{align}
\mathbf{f}_t\ &=\ \bm\psi \tau_t+\boldsymbol{\omega}_t, \qquad \boldsymbol{\omega}_t  \sim (\mathbf{0}, \bm\Sigma_\omega), \label{sbeq::obsTR}
\end{align}
where $\bm\psi$ and $\bm\omega_t$ are $q$-dimensional vectors and $\bm\Sigma_\omega$ is $q\times q$.

In this paper, we assume that the common trend evolves as: 
\begin{align}
\tau_t\ &=\ \tau_{t-1}+\nu_t, \qquad \nu_t \sim ( 0,  \sigma_\nu^2).\label{sbeq::st1TR}
\end{align}
It is important to emphasize that, under our estimation approach, we do not need to impose a specific parametric model for the dynamic evolution of $\nu_t$; in fact, $\nu_t$ can still be an autocorrelated process. In this respect, our specification differs from the pure random walk assumption \citep[e.g.,][]{stockwatson88JASA} and is instead compatible with the ARIMA specification \citep[e.g.,][]{lippi_diffusion_1994,barigozzi_measuring_2021,morley2023simple}. Finally, we remain agnostic about the law of motion of the residual stationary component, $\boldsymbol{\omega}_t$.\footnote{Modeling the cycle as an AR(2) poses identification problems because the same state-space representation can equivalently be obtained with an ARMA(2,1) specification \citep{kim2022trend}.}

The model we just described is a modified version of the model \citet{barigozzi_measuring_2021} (BL) used to estimate the output gap in the US. We modified BL's model to overcome two important limitations. First, their model relies on estimating cointegrating relationships between the factors to retrieve the common trends, which require longer time series to get a reliable estimate. As such, BL's model can be estimated only on US macroeconomic data for which more than 50 years of quarterly data are available. Second, BL estimate their model on pre-Covid data; thus, to incorporate more recent observations, some modification is needed to handle the different co-movements brought about by the Covid pandemic. In this paper, we solve both limitations by introducing \eqref{sbeq::obsTR}-\eqref{sbeq::st1TR}, which we can estimate even on short samples, and by incorporating recently proposed methods to handle the Covid period in the estimation strategy. 

Combining Equations \eqref{eq::obseq} and \eqref{sbeq::obsTR}, we obtain the decomposition of each observed variable:
\begin{equation}
y_{it}\ =\ \mathrm{D}_{it} + \bs{\lambda}_i'\bs{\psi} {\tau}_t + \bs{\lambda}_i' \bs{\omega}_t + \gamma_{i}g_t \I_{\text{\tiny 2020:Q1$\le\! t\!\le$2021:Q4}}  + \xi_{it}.\label{eq::TrendCycleCovidIdioDec}
\end{equation}
Focusing on GDP, we  define potential output, $\text{PO}_t$, and the output gap, $\text{OG}_t$, as:
\begin{align}
\text{PO}_t\ &=\ \mathrm{D}_{\text{\tiny GDP},t} + \bs{\lambda}'_{\text{\tiny GDP} }\bs{\psi}{\tau}_t,\label{POT}\\
\text{OG}_t\ &=\ \bs{\lambda}'_{\text{\tiny GDP}}\bs{\omega}_t.\label{OGT}
\end{align}
Hence, in our framework, potential output is the sum of the time-varying secular trend of GDP ($\mathrm{D}_{\text{\tiny GDP},t}$), which captures the long-run decline in EA output growth, and the part of GDP driven by the common trend component (${\tau}_t$); the output gap is the part of GDP driven by the stationary cyclical component ($\bs{\omega}_t$). 

From the definition of potential output \eqref{POT} and output gap \eqref{OGT}, we left out the idiosyncratic component, $\xi_{\text{\tiny GDP},t}$, and the Covid component, $\gamma_{\text{\tiny GDP}}g_t$. While the idiosyncratic component is likely to be just a measurement error \citep{aruoba2016improving}, hence, it is clear why we are leaving it out; the exclusion of the Covid shock deserves an explanation.

The Covid component represents the co-movements from 2020:Q1 to 2021:Q4 that neither potential output nor the output gap captures. In principle, this component could be allocated to the output gap, which would be equivalent to assuming that the productive capacity of the EA \enquote{froze} due to the lockdowns. While this view is commonly accepted by European institutions \citep{thum2022ii}, it is still unclear whether, and by what amount, the EA productive capacity has been affected by the Covid shock. Thus, we remain agnostic on the allocation of the Covid component between potential output and the output gap, and we will present it as a standalone component.

\subsection{Estimating the model}\label{sec::fattoni}
To estimate potential output and the output gap, we first estimate the DFM \eqref{eq::obseq}-\eqref{eq::idioeq} using a three-step estimation procedure, and then we estimate the model for the common trend \eqref{sbeq::obsTR}-\eqref{sbeq::st1TR} on the estimated factors.

\paragraph{Estimating the dynamic factor model.}
To estimate the model in \eqref{eq::obseq}-\eqref{eq::idioeq}, we need to extract the latent states $\mathbf f_t$, $g_t$, $\mathrm D_{\text{\tiny i},t}$ (if $i\in\mathcal{L}_1$ or $i\in \mathcal{L}_0$), and $\xi_{it}$ (if $i\in\mathcal I_1$), and estimate the parameters $\bm\lambda_{i}$, $\gamma_i$, $\mathbf A_j$, $s_t$, $\bm\Sigma_u$, $\sigma_{e_i}^2$, $a_i$, and $b_i$ (if $i\in\mathcal I_b$), while we calibrate $\sigma^2_{\epsilon_i}$  (if $i\in\mathcal L_0$) and $\sigma^2_{\eta_i}$ (if $i\in\mathcal L_1$) following \cite{del2017safety}.\footnote{We calibrate the variances of the stochastic secular components so that, for $i\in \mathcal{L}_1$, the standard deviation of the secular trend is approximately $1\%$ over 100 years, while for $i\in \mathcal{L}_0$, itis approximately $1\%$ over 50 years. This calibration, \textit{de facto}, defines the secular trends by allowing the average growth rate of the variables in $\mathcal{L}_1$ and the average level of the variables in $\mathcal{L}_0$ to drift slowly over time. In principle, these variances could be estimated; however, given the short sample, we consider it more reliable to define the trends upfront as slow-moving processes, an approach also adopted by \citet{AhnLuciani2024}, which avoids the overfitting ``pile-up'' problem highlighted by \citet{kim2022trend}. Appendix \ref{app::notvpars} shows that the results are robust to reasonable changes to this calibration.} To do so, we use a three-step estimation procedure that we summarize below. 
\medskip

\leftskip 1em
\parindent -1em

\textsc{Step 1: Estimate the model up to 2019:Q4 (pre-Covid step).} We obtain a preliminary estimate of the parameters using PCA for $I(1)$ data \citep{bai2004panic,barigozzi_large-dimensional_2021,onatski_spurious_2021}. Then, we run the EM algorithm, jointly with the Kalman smoother, as described in \citet{barigozzi_quasi_2022} in the high-dimensional case.\medskip

\textsc{Step 2: Estimate the Covid factor and volatility (Covid step).}

\leftskip 1em
\parindent 0em

\textsc{Covid factor} \citep{stockcomovement}. Using the parameter estimated over the pre-Covid period, we extract the latent states using data up to the end of the sample by running the Kalman smoother separately for the pre- and post-pandemic periods to prevent the pandemic observations from changing the pre-pandemic inference, following the approach of \citet{AhnLuciani2024}. This yields an estimate of the factors during Covid as if they had continued along their pre-pandemic dynamics, while all the Covid-specific co-movements are absorbed by the idiosyncratic component $\xi_{it}$. Since Covid represented a common shock affecting most (if not all) series in the dataset, we then estimate the Covid factor $\widehat g_t$ and its loadings $\widehat \gamma_i$ by PCA on the variance-covariance matrix of the idiosyncratic component for the period 2020:Q1-2021:Q4. \smallskip 

\textsc{Covid volatilities} \citep{lenza_how_2022}. Lastly, we estimate the Covid volatilities $\widehat s_t$ for the period 2020:Q1-2025:Q1 by maximum likelihood and by using the factors extracted using pre-Covid parameters. We find that $\widehat s_t$ jumps from 1 to about 3.5 at the onset of the Covid pandemic, and then remain larger than 2 until the end of 2022, thus justifying our choice of imposing a time-varying volatility until the end of the sample. In Appendix \ref{sec::covid}, we show how our measures would change if we do not explicitly model the effect of Covid, or if we use the exponential decay parametrization proposed by \citet{lenza_how_2022}. \smallskip 

\leftskip 1em
\parindent 1em

The procedure just described is general and could be applied again should another large, non-economic event occur. Its purpose is twofold: first, to prevent an exceptional, non-economic event from contaminating the model's estimates going forward; and second, to provide a way to assess the impact of such a non-economic event on the economy. If the researcher's sole objective is to insulate the model estimates from the effects of the event, a simpler alternative is to append post Covid observations to pre-Covid ones.\footnote{It is common practice in dynamic factor models---whether stationary or non-stationary---to adjust for outliers on a series-by-series basis. However, applying this procedure to Covid data would be ill-advised, since much of the additional volatility during this period is pandemic induced rather than the result of measurement error \citep{ng2021modeling}. A univariate outlier adjustment method cannot distinguish between the two and would therefore risk removing economically relevant information. Indeed, when we apply such an adjustment, the estimated Covid factor is essentially zero, and the output gap appears flat, as if nothing had happened.}\medskip

\leftskip 1em
\parindent -1em

\textsc{Step 3: Full sample estimation.}  We estimate all the parameters and latent states up to the end of the sample, by using data net of the Covid component, i.e., with $y_{it} - \widehat \gamma_i \widehat g_t$.  Specifically, by using the factors estimated over the whole sample in Step 2 rescaled by $\widehat s_t$ in the last part of the sample, we estimate the parameters, $\widehat{\bm\lambda}_{i}$, $\widehat{\mathbf A}_j$, $\widehat{\bm\Sigma}_u$, $\widehat\sigma_{e_i}^2$, $\widehat a_i$, and $\widehat b_i$, by maximizing the expected likelihood. Finally, with the estimated parameters in hand, we obtain a final estimate of the states, $\widehat{\mathbf f}_t$, $\widehat {\mathrm  D}_{i,t}$, and $\widehat{\xi}_{it}$, through the Kalman smoother again truncated in 2020:Q1 and reinitialized before iterating backward, as explained in Step 2.

\leftskip 0em
\parindent 1.5em

\paragraph{Estimating the common trend.} 
Having estimated the model parameters and unobserved states, we can now estimate the common trend. To this end, we estimate the state-space model in \eqref{sbeq::obsTR}-\eqref{sbeq::st1TR} using the EM algorithm by replacing the true factors with the estimated ones. Since we are extracting one trend from four factors, we can identify the trend by properly initializing the loadings $\bm\psi$. Specifically, we set all entries of $\bm\psi$ to zero except for the one corresponding to the factor with the largest share of variance at frequencies below eight years, which is set equal to one. As for the initialization of $\sigma^2_\nu$, our experimentation has shown that it does not matter.

At convergence of the EM algorithm, we obtain a final estimate of the parameters, $\widehat{\bm\psi}$, $\widehat{\bm\Sigma}_\omega$, and $\widehat{\sigma}^2_\nu$, and using these estimates, we have a final estimate of the trend $\widehat \tau_t$ and of the cyclical component $\widehat{\bm\omega}_t=\widehat{\mathbf f}_t-\widehat{\bm\psi}\widehat \tau_t$, obtained through the Kalman smoother. Given the estimates of the common trend and the cyclical common component, we compute our final estimates of potential output and the output gap according to \eqref{POT} and \eqref{OGT}, respectively.\bigskip

As shown in \citet{barigozzi_quasi_2022}, the estimation procedure we just outlined delivers consistent estimators of all parameters and of the factors, provided that $n$ and $T$ grow to infinity and the EM algorithm is initialized with the estimator of the loadings and factors introduced by \citet{barigozzi_large-dimensional_2021}. Furthermore, to prove this result, we neither have to impose the Gaussianity assumption nor have to require uncorrelatedness of the idiosyncratic components ($\xi_{it}$ if $i\notin \mathcal I_1$ or $e_{it}$ if $i\in\mathcal I_1$) across $i$ or $t$. Rather, we just have to impose standard moment conditions, such as existence and summability of 4th-order cumulants.

We conclude by comparing our approach with two alternative estimation strategies. First, our model implies that, in principle, the dynamics of the factors could be modeled directly using a VECM, thereby estimating the cointegration space in the first step. However, this approach appears to be quite unstable possibly due to the short sample (see Appendix \ref{sbsec::VECM} for a comparison).  Second, in a recent paper, \citet{morley2023simple} suggest to smooth a preliminary estimate of the trend by appropriately rescaling the parameters of an ARMA fitted on $\nu_t$. While their univariate approach is prone to overfitting, our multivariate approach isolates low-frequency dynamics without the need for ex-post smoothing (see  Appendix \ref{app:morley} for a comparison).

%
%
\section{Potential output and output gap of the Euro Area}\label{sec::poutgap}
\begin{figure}[ht] \caption{Potential output} \label{fig::pout}
\centering \footnotesize \sc \smallskip
\setlength{\tabcolsep}{.01\textwidth}
\begin{tabular}{cc}
$100\times\log$ levels & Year-on-year growth rates\\
\includegraphics[trim={2cm 9.1cm 2.2cm 9.5cm},clip,width = 0.475\textwidth]{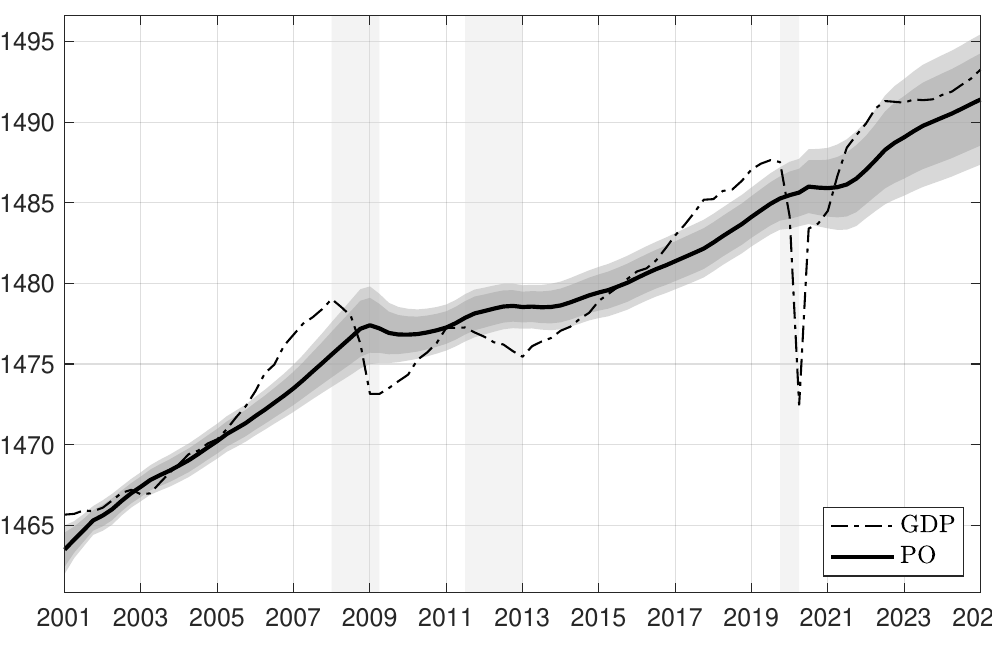}&
\includegraphics[trim={2cm 9.1cm 2.2cm 9.5cm},clip,width = 0.475\textwidth]{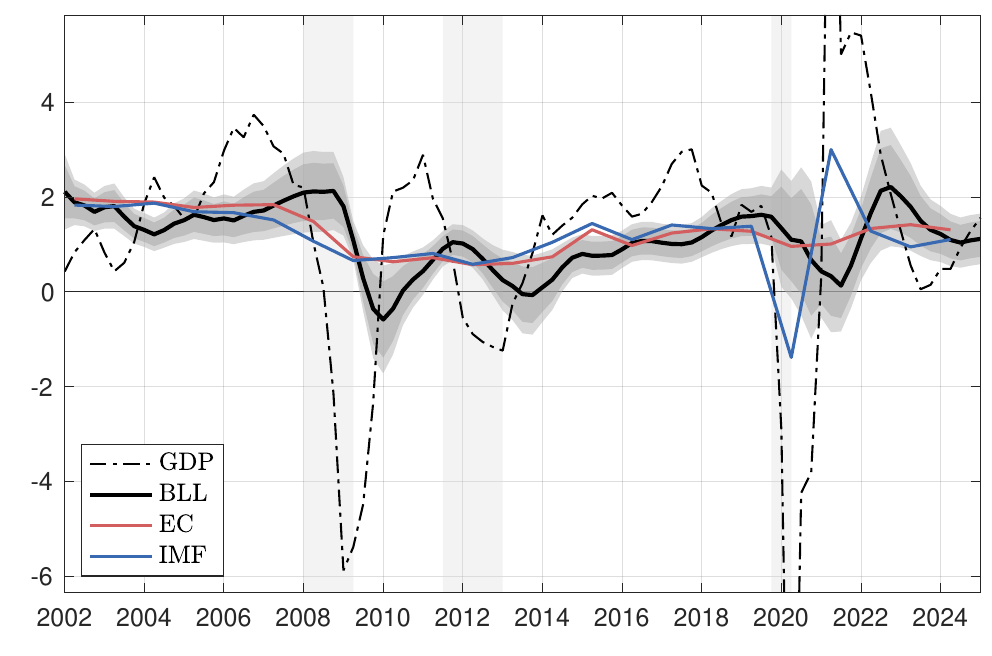} \\
\end{tabular} 
\begin{tabular}{p{.98\textwidth}} \scriptsize Notes: \rm In all charts, the black solid line is our estimate of potential output, the grey shaded areas are the 68$\%$ and 84$\%$ confidence bands, and the dashed black line is GDP---we truncated the y-axis in the right chart for readability. In the right chart, the blue and red lines are the potential output estimates published by the European Commission and the IMF, respectively. The IMF estimate of YoY potential output growth reported in the right chart is the result of our own calculation. Indeed, the IMF publishes only an estimate of the output gap from which we backed out potential output. Thus, the blue line in the right chart does not account for any adjustment for Covid that the IMF might have done.
\end{tabular}
\end{figure}

Figure \ref{fig::pout} presents our potential output estimate (black line), both in $100\times\log$ levels (left plot) and in year-on-year (YoY) growth rates (right plot). The right plot  also includes potential output growth estimates from the EC (red line) and IMF (blue line). 

Four main results emerge from Figure \ref{fig::pout}. First, the GFC and the SDR had a permanent negative effect on the level of potential output (a hysteresis effect), consistent with \citet{schmoller2021deep}.   Second, potential output growth decelerated after the GFC and slowed further following the SDR---average potential output growth was 2.1\% in 2008, while the subsequent peak was 1.6\% on the eve of the Covid pandemic.  Third, although both the EC and IMF estimate a slowdown in potential growth after the GFC, neither estimate a slowdown after the SDR. Fourth, it is still too soon to determine whether the Covid recession had a long-term effect on potential output growth. The prevailing institutional view is that it did not \citep{thum2022ii}; however, while potential growth rebounded quickly after the Covid shock---peaking at 2\% in 2022:Q4---it subsequently declined to 0.95\% in 2025:Q1, below its pre-Covid pace. Overall, these results indicate a gradual slowdown of potential growth over the past two decades, punctuated by persistent losses following major crises.

\begin{figure}[ht] \caption{Output gap}\label{fig::ogapEC}
\centering \footnotesize \sc \smallskip
\setlength{\tabcolsep}{.01\textwidth}
\begin{tabular}{cc}
Levels & Year-on-year growth rates\\
\includegraphics[trim={2cm 9.1cm 2.2cm 9.5cm},clip,width = 0.475\textwidth]{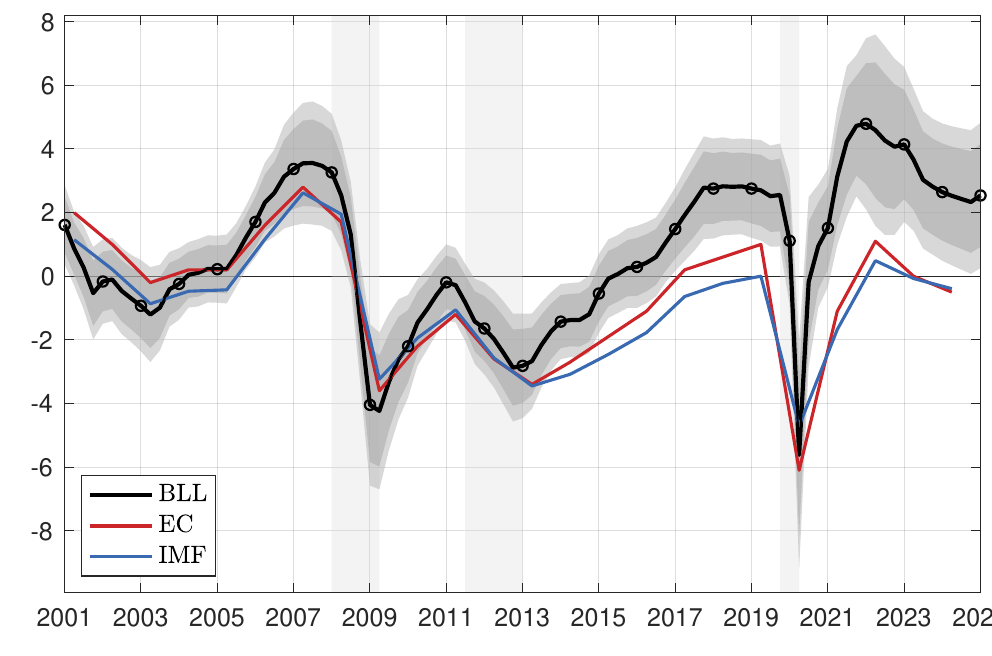}&
\includegraphics[trim={2cm 9.1cm 2.2cm 9.5cm},clip,width = 0.475\textwidth]{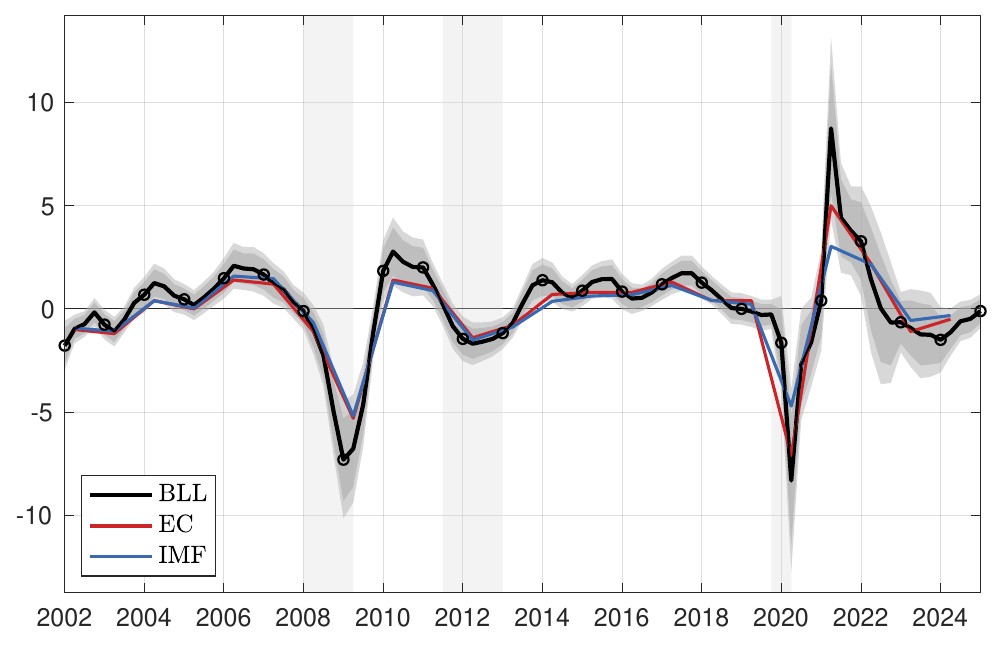} 
\end{tabular}

\begin{tabular}{p{.98\textwidth}} \scriptsize Notes: \rm The black line is our estimate of the output gap (OG) in levels (left plot) and YoY growth rates (right plot)---the level of the output gap is the percentage deviation from potential, the YoY growth rates is $\text{OG}_t-\text{OG}_{t-4}$. Each black marker denotes one year (four quarters), starting from $2001$:Q1. The grey shaded areas are the 68$\%$ and 84$\%$ confidence bands. The red and blue lines are the output gap estimates published by the European Commission and the IMF, respectively.
\end{tabular}
\end{figure}

Figure \ref{fig::ogapEC} shows our estimated output gap alongside the the EC and IMF estimates. The three measures display similar dynamics, with turning points occurring at the same dates. They also align closely up to the SDR, all indicating substantial overheating before the GFC and a persistently negative output gap during and between the two recessions. However, our estimate rises after the SDR, remaining around 2\% from 2017 until the Covid pandemic, implying a considerably tighter economy than suggested by the EC or IMF estimates. Similarly, our measure points to a much tighter post-Covid economy than those from the EC and the IMFs.

\begin{figure}[h!] \caption{Decomposition of GDP growth} \label{fig::decompgdp}
\centering \footnotesize \smallskip
\setlength{\tabcolsep}{.01\textwidth}
\begin{tabular}{cc}
\textsc{Year-on-year growth rate} & \textsc{Quarter-on-quarter annualized growth rates}\\
\includegraphics[trim={1cm 9cm 1.4cm 9cm},clip,width = 0.475\textwidth]{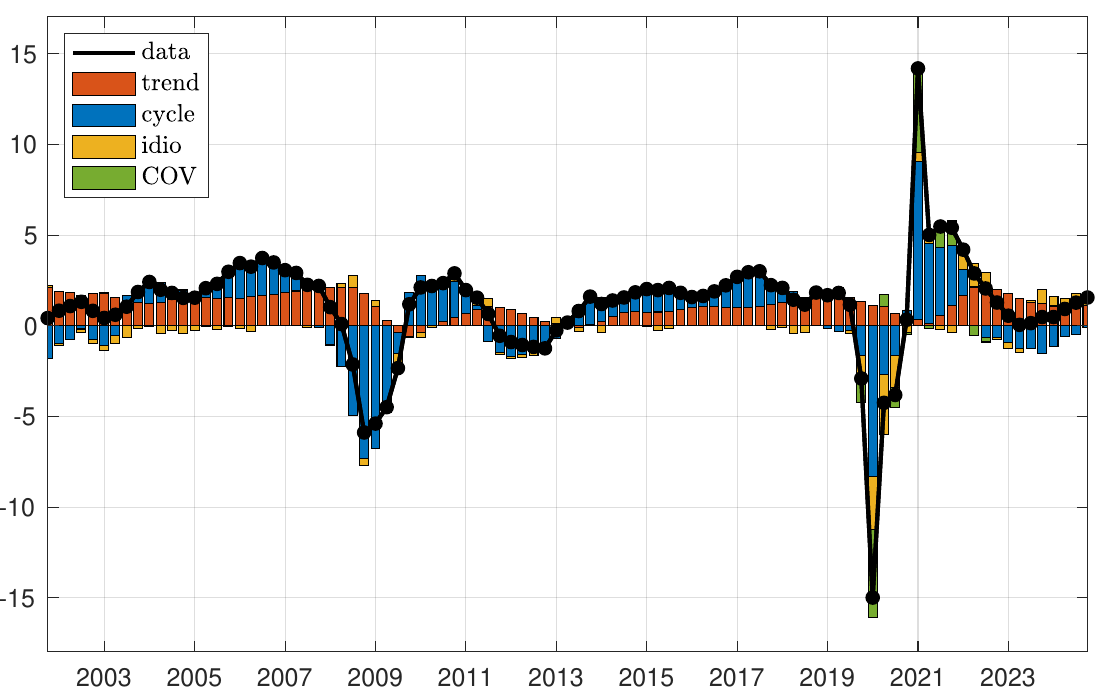}&
\begin{tikzpicture}
  \node[anchor=south west,inner sep=0] (image) at (0,0) {\includegraphics[trim={1cm 9cm 1.4cm 9cm},clip,width = 0.475\textwidth]{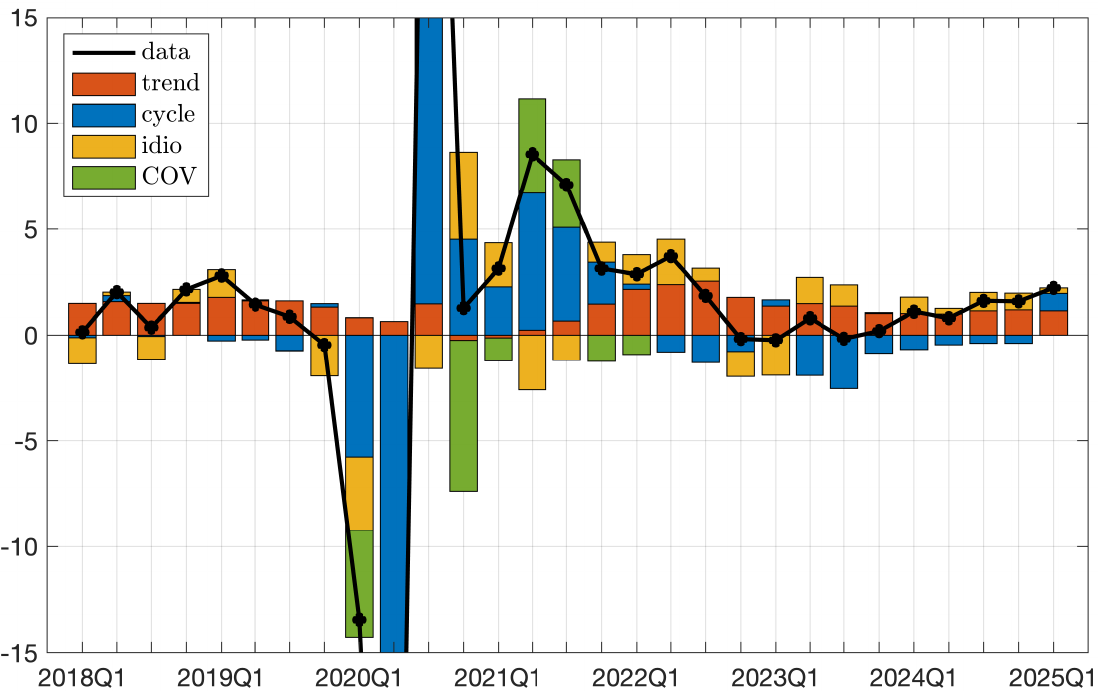}};
    \begin{scope}[x={(image.south east)},y={(image.north west)}]                
    \node[anchor=south west,inner sep=0] (image) at (0.41,0.08) 
            {\setlength{\tabcolsep}{.0\textwidth} \tiny
      \begin{tabular}{|C{.06\textwidth}C{.05\textwidth}C{.05\textwidth}C{.05\textwidth}C{.05\textwidth}|}\hline
            & trend & cycle & idio & COV \\\hline
            2020:Q2 & 0.8 &-27.2 &-5.5  &-15.0  \\
            2020:Q3 & 1.0 &~22.0 &-2.4  &~23.3  \\\hline
      \end{tabular}};  
  \end{scope}
\end{tikzpicture}

\end{tabular}

\begin{tabular}{p{.98\textwidth}} \scriptsize Notes: \rm The black line with dot markers is GDP growth. The bars represent the contribution of each component to GDP growth. The left plot shows YoY growth rates, while the right plot shows QoQ growth at an annual rate. Growth rates are computed using the log approximation. 
\end{tabular}
\end{figure}

To conclude, Figure \ref{fig::decompgdp} shows GDP growth decomposed into the contribution of potential output, the output gap, the Covid factor, and the idiosyncratic component. During the Covid pandemic, the output gap subtracted 27.2 percentage points (p.p.) from quarter-on-quarter (QoQ) annualized GDP growth in 2020:Q2 (inset box, right chart). Given the -47\% QoQ annualized GDP decline, such a contribution may appear implausible. However, as explained in Section \ref{sec::fattoni}, our output gap captures only business-as-usual co-movements, while the Covid factor isolates the extraordinary co-movements induced by the pandemic. As shown in the inset box in the right plot in Figure \ref{fig::decompgdp}, the Covid factor accounts for an additional -15 p.p. of the 2020:Q2 contraction. Since then it contributed about +23 p.p. in 2020:Q3, before alternating between negative and positive values over the next six quarters, reflecting the tightening and easing of mobility restrictions.

In summary, Figures \ref{fig::pout}--\ref{fig::decompgdp} show that the EA has a potential output issue, not a business cycle issue.  Since the seminal work of \citet{BQ89}, the prevailing view has been that supply shocks have permanent effects on output, whereas demand shocks have only transitory effects. Empirical studies mostly support this perspective. For instance, \citet{ForniAmerican} show that the impulse response functions to a supply (demand) shock are almost identical to those of a permanent (transitory shock), and \citet{BenatiLubik} find that it is essentially impossible to detect aggregate demand shocks that permanently affect GDP. Nevertheless, some recent studies challenge this view. For example, \citet{FurlanettoetaAEJmacro} provide evidence of hysteresis effects, showing that demand-driven recessions can lead to lasting output losses.

Based on this evidence, we assume that growth in the common trend $\tau_t$---and thus potential output growth---is primarily driven by supply forces, except during recessions, while the cyclical common component $\bm \omega_t$---and thus the output gap---is mainly driven by demand forces. This distinction implies that, to foster long-run growth, European countries should implement structural reforms and promote productivity-enhancing investments, whereas to mitigate recession-induced output losses, they should support aggregate demand more forcefully during downturns. In contrast, policies aiming at stimulating household consumption and residential investments in normal times will have only short-term effects at best because, as we will show in Section \ref{sec::Credit}, growth financed through household debt is not sustainable in the long run.

%
%
\section{What about the Okun’s law and the Phillips curve?}\label{sec::OL&PC}

Our estimate of the output gap has a different meaning than that of the EC and IMF, which derive the output gap and potential output according to the so-called \enquote{production function approach} \citep{kiley2013output}.  In production-function-based models, the output and unemployment gaps are related through the Okun's law, and the output gap is related to inflation through the Phillips curve. Thus, in these models, the unemployment gap decreases whenever the output gap increases, and vice-versa, and low inflation suggests a negative output gap, while high inflation indicates a positive gap. 

In our model (like any statistical model), we do not impose any Okun's law or Phillips curve. Thus, we must be careful when we compare our estimate with that of the EC or the IMF because their output gap measures are designed to indicate potential inflation pressure, whereas ours is not. Nonetheless, in Sections \ref{sec::OL} and \ref{sec::PC}, we show that, on average, our model satisfies the Okun's law relationship and exhibits Phillips correlation.

\subsection{The Okun's law}\label{sec::OL}
\begin{figure}[ht!]\caption{Okun's Law}\label{fig::OkunLaw}
\centering \footnotesize\sc \smallskip
\setlength{\tabcolsep}{.01\textwidth}
\begin{tabular}{cc}  
Unemployment rate gap vs. output gap & Okun's law coefficient (expanding window)\\
\includegraphics[trim={1.8cm 9.1cm 2.2cm 9.5cm},clip,width = 0.475\textwidth]{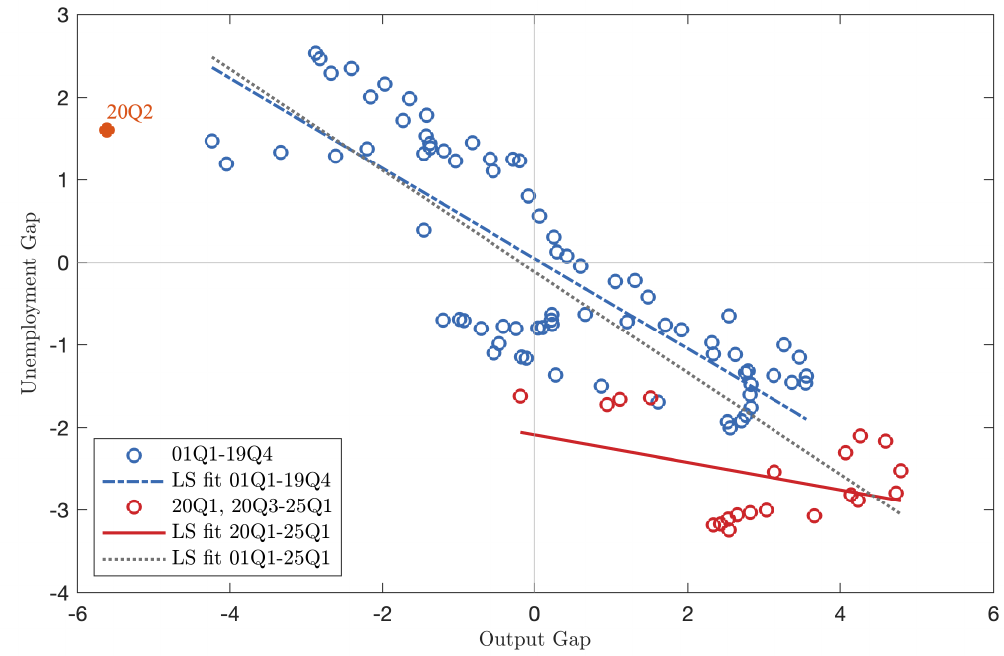} & 
\includegraphics[trim={2cm 9.1cm 2cm 9.5cm},clip,width = 0.475\textwidth]{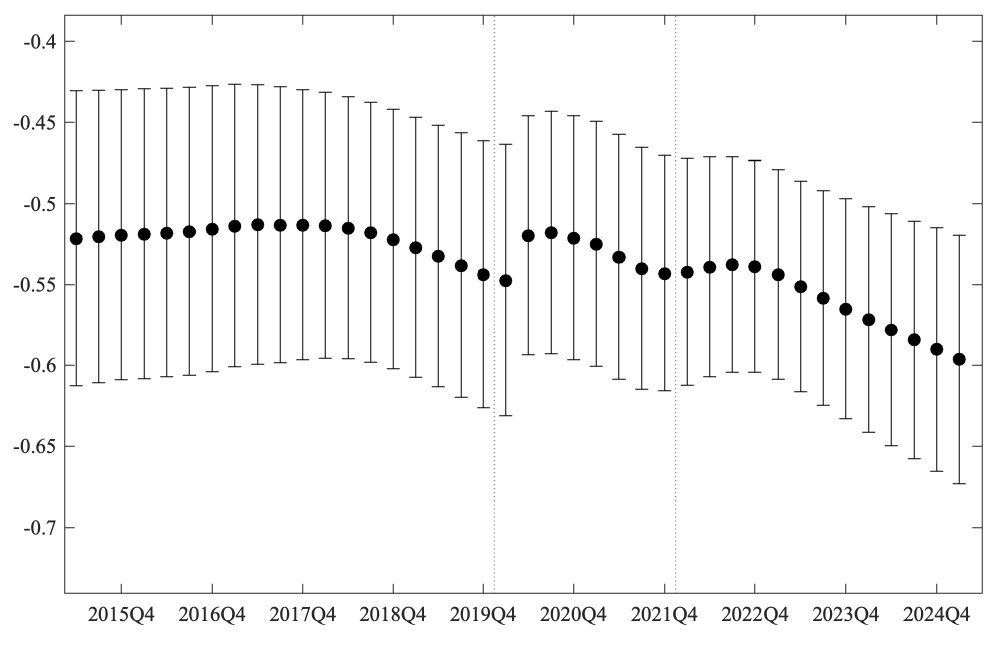}\\
\end{tabular}  

\begin{tabular}{p{.98\textwidth}} \scriptsize Notes: \rm 
The left chart shows the Okun’s law relationship with the output gap on the horizontal axis and the unemployment rate gap on the vertical axis. Each circle corresponds to an output gap - unemployment gap pair at a given time $t$. The grey dotted, blue dashed-dotted, and red solid lines are the least squares fit lines for the full, pre-Covid, and post-Covid samples, respectively, obtained by omitting the observation for 2020:Q2 (orange dot).

~~~The right chart shows the least squares estimate, based on an expanding window starting from 2015:Q1, of the Okun’s law slope $\beta$ given by the regression $(\text{UR}_t - D_{\text{UR},t}) = \alpha + \beta \text{OG}_t + \varepsilon_{\text{UR},t}$, where $D_{\text{UR},t}$ is the time-varying mean of the unemployment rate defined in \eqref{eq::seccomp}. Each dot is an estimate of $\beta$ while the whiskers are $\pm$ one HAC standard errors.
\end{tabular} 
\end{figure}

The left plot in Figure \ref{fig::OkunLaw} shows that in our model, the unemployment rate gap and the output gap are negatively correlated; that is, our model captures the Okun’s law relationship in the data. Over the whole sample, on average, for every percentage point increase in the output gap, the unemployment gap decreases 0.6 p.p. (grey dotted line). Moreover, this correlation has decreased after Covid from -0.56 (blue dash-dotted line) to -0.23 (red solid line), suggesting a (temporary) disconnect between the labor and goods and services market post-Covid in the EA as also noted by \citet{berson2025explaining}. The right plot in Figure \ref{fig::OkunLaw} shows the expanding window estimate of the Okun’s law coefficient $\beta$ in the regression $(\text{UR}_t - D_{\text{UR},t}) = \alpha + \beta \text{OG}_t + \varepsilon_{\text{UR},t}$, where $D_{\text{UR},t}$ is the time-varying mean of the unemployment rate defined in \eqref{eq::seccomp}. On average, the Okun's law coefficient varies between -0.55 to -0.60.\footnote{We also estimated the relationship between the output gap and hours worked gap and find a positive correlation in line with the results of \citet{morley_estimating_2023}.}

\begin{figure}[ht!]\caption{Generalized Impulse Response Functions to a shock to the unemployment rate}\label{fig::urshock_g}
\centering \scriptsize \sc \smallskip
\setlength{\tabcolsep}{0\textwidth}
\begin{tabular}{ccc}
Common component: UR & Common component: GDP & Potential output and output gap \\
\includegraphics[trim={2cm 9.1cm 2.2cm 9.5cm},clip,width = 0.33\textwidth]{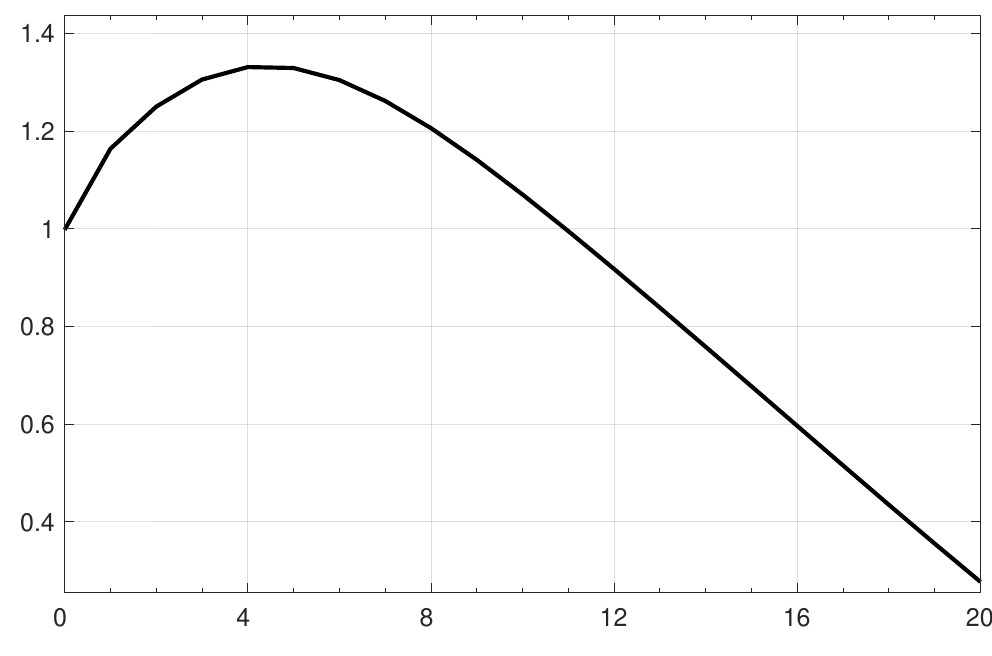} &
\includegraphics[trim={2cm 9.1cm 2.2cm 9.5cm},clip,width = 0.33\textwidth]{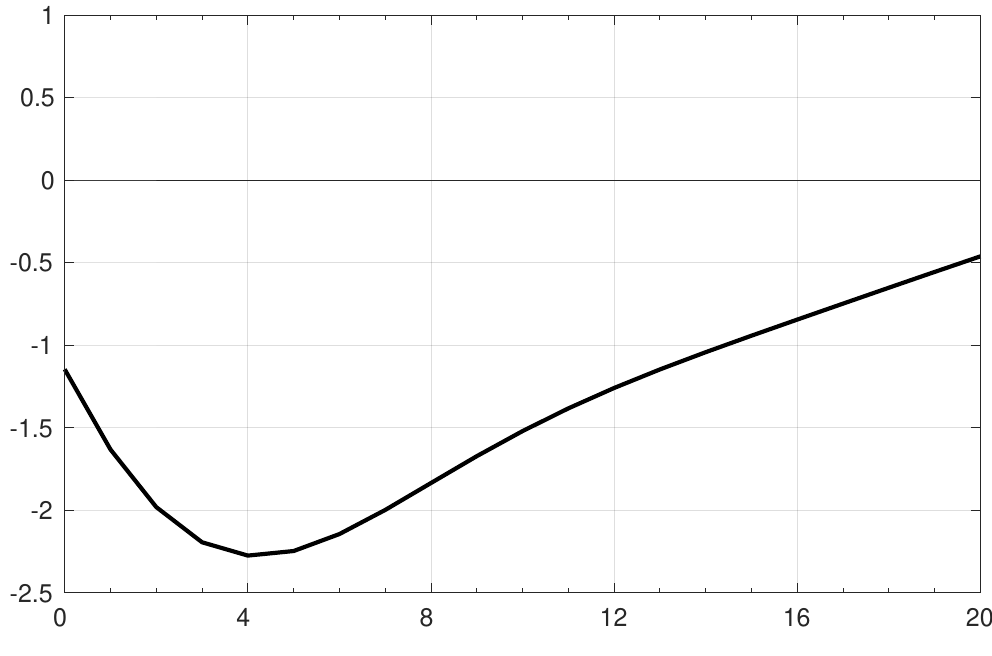} &
\includegraphics[trim={2cm 9.1cm 2.2cm 9.5cm},clip,width = 0.33\textwidth]{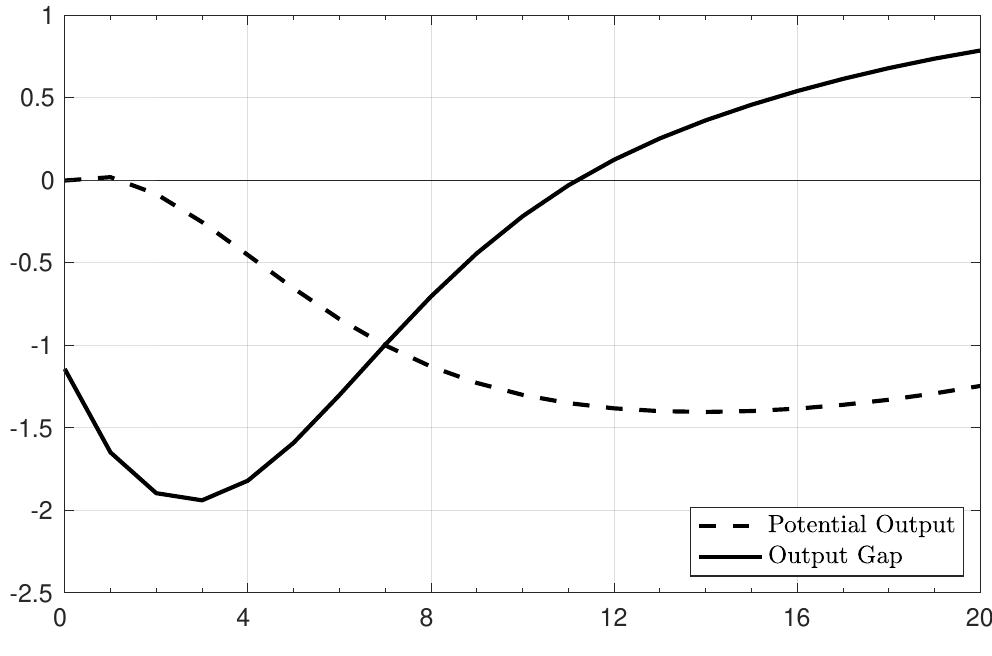} \\
\end{tabular}
\begin{tabular}{p{.98\textwidth}} \scriptsize Notes: \rm 
The black solid/dashed lines are the GIRFs to a 1 p.p. shock to the common component of the unemployment rate. The major ticks in the x-axis represent quarters after the shock. 
\end{tabular} 
\end{figure}

To further corroborate the intuition that there is a tight relationship between our output gap estimate and labor market indicators, Figure \ref{fig::urshock_g} shows the Generalized Impulse Response Functions (GIRFs) of the common component of the unemployment rate, GDP, potential output, and the output gap to a 1 p.p. shock to the common component of the unemployment rate.\footnote{The lag-$h$ GIRF of all variables is obtained by computing the differences between the $h$-step ahead forecast of their common component conditional on a shock to a given variable at time $T+1$ minus the $h$-step ahead unconditional forecast of the common component, i.e., when no shock is imposed. Both forecasts are computed conditional on all information available at time $T$ (the last observation in our sample) by means of the Kalman filter \citep{banbura2015conditional,crump2021large}.} Results confirm that our model, on average, associates an unemployment rate increase with an output gap decrease. After the shock, the common component of the unemployment rate remains 1 p.p. (or more) above the baseline for about a year and a half before decreasing and slowly returning to zero. In response, GDP decreases and keeps decreasing, reaching a through a year after the shock; then, it slowly returns to baseline. The model attributes most of the GDP response to movements in the output gap, while potential output slightly decreases only after a few quarters. The shock is fully absorbed in about 4 years.

\subsection{The Phillips curve}\label{sec::PC}
\begin{figure}[ht!]\caption{Phillips Curve}\label{fig::PhillipsCurve}
\centering \sc \smallskip

\setlength{\tabcolsep}{0\textwidth}
\begin{tabular}{cp{.02\textwidth}c}
\footnotesize Core Inflation gap vs Output gap && \footnotesize Phillips Curve slope (expanding window)\\
\includegraphics[trim={1.8cm 9.1cm 2.2cm 9.5cm},clip,width = 0.49\textwidth]{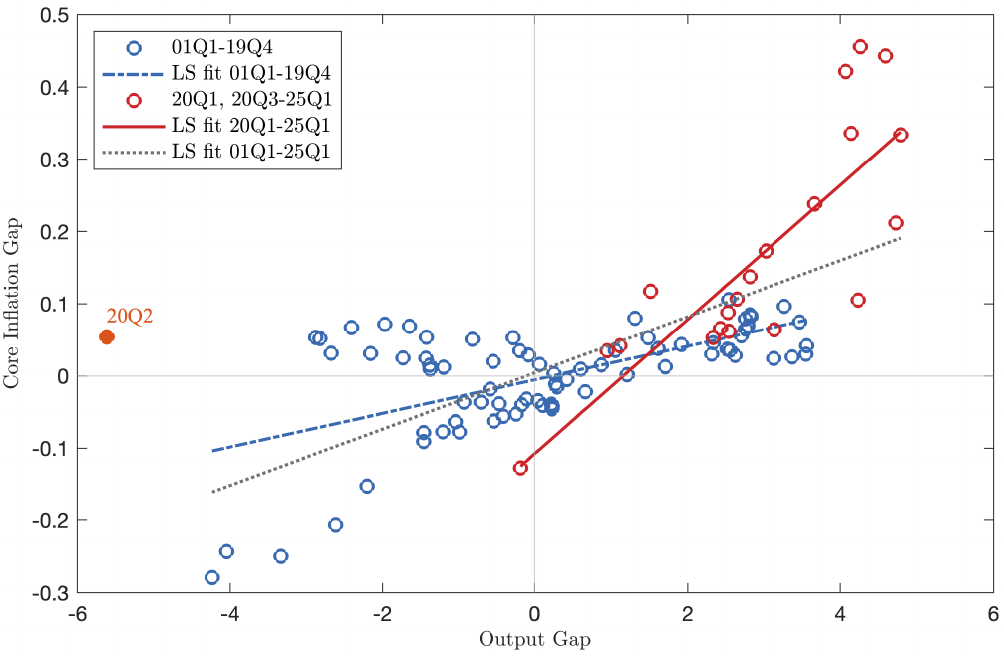} & &
\includegraphics[trim={2cm 9.1cm 2cm 9.5cm},clip,width = 0.49\textwidth]{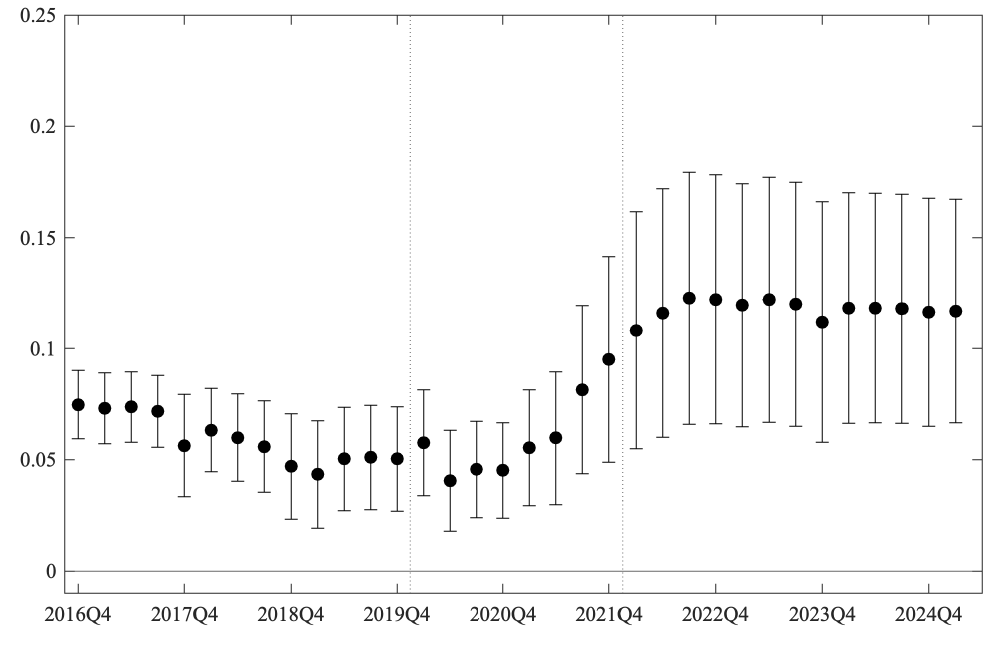}\\
\end{tabular}

\begin{tabular}{p{\textwidth}} \scriptsize Notes: \rm 
The left chart shows the Phillips curve relationship with the output gap on the horizontal axis and the core inflation gap on the vertical axis. Each circle corresponds to an output gap - core inflation gap pair at time $t$. The grey dotted, blue dashed-dotted, and red solid lines are the least squares fit lines for the full, pre-Covid, and post-Covid samples, respectively, obtained by omitting the observation for 2020:Q2 (orange dot).

~~~The right chart shows the least squares estimate, based on an expanding window starting from 2015:Q1, of the slope of the Phillips curve given by the following expectation-augmented specification: $\pi_t = c + \alpha \text{OG}_t + \beta \pi_{t-1} + \gamma \pi^e_t + \varepsilon_{\pi,t}$, where $\pi_t $ is core inflation and $\pi^e_t$ are the long-run (5-year ahead) inflation expectations in the Survey of Professional Forecasters. Each dot is an estimate of $\alpha$ while the whiskers are $\pm$ one HAC standard errors.
\end{tabular} 
\end{figure}

The left plot in Figure \ref{fig::PhillipsCurve} shows that in our model, the core inflation rate gap (i.e., the cyclical common component of core inflation) and the output gap are positively correlated; that is, there is Phillips correlation in the data, and our model captures it. Over the whole sample, on average, for every percentage point increase in the output gap, the core inflation gap increases 4 basis points (dotted grey line). Moreover, this correlation has increased significantly after Covid from 0.022 (dashed-dotted blue line) to 0.079 (solid red line). The right plot in Figure \ref{fig::PhillipsCurve} shows the expanding window estimate of the slope of the Phillips curve $\alpha$ in the following expectation-augmented specification \citep[e.g.,][]{conti2021resurrecting}: $\pi_t = c + \alpha \text{OG}_t + \beta \pi_{t-1} + \gamma \pi^e_t + \varepsilon_{\pi,t}$, where $\pi_t $ is core inflation and $\pi^e_t$ are the long-run (5-year ahead) inflation expectations in the Survey of Professional Forecasters. The results suggest that the relationship between inflation and the output gap has strengthened after Covid, a point also made by \citet{Lane2024}.

\begin{figure}[ht!]\caption{Generalized Impulse Response Functions to a shock to core inflation}\label{fig::coreshock_g}
\centering  \scriptsize \sc \smallskip
\setlength{\tabcolsep}{0\textwidth}
\begin{tabular}{ccc}
Common component: core HICP & Common component: GDP &  Potential output and output gap \\ 
\includegraphics[trim={2cm 9.1cm 2.2cm 9.5cm},clip,width = 0.33\textwidth]{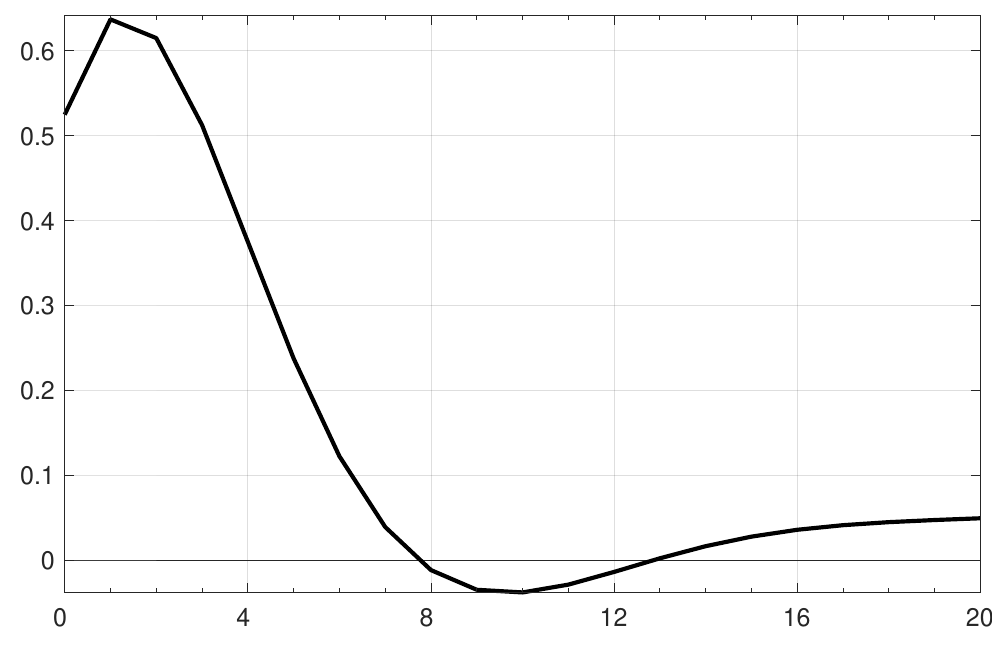} &
\includegraphics[trim={2cm 9.1cm 2.2cm 9.5cm},clip,width = 0.33\textwidth]{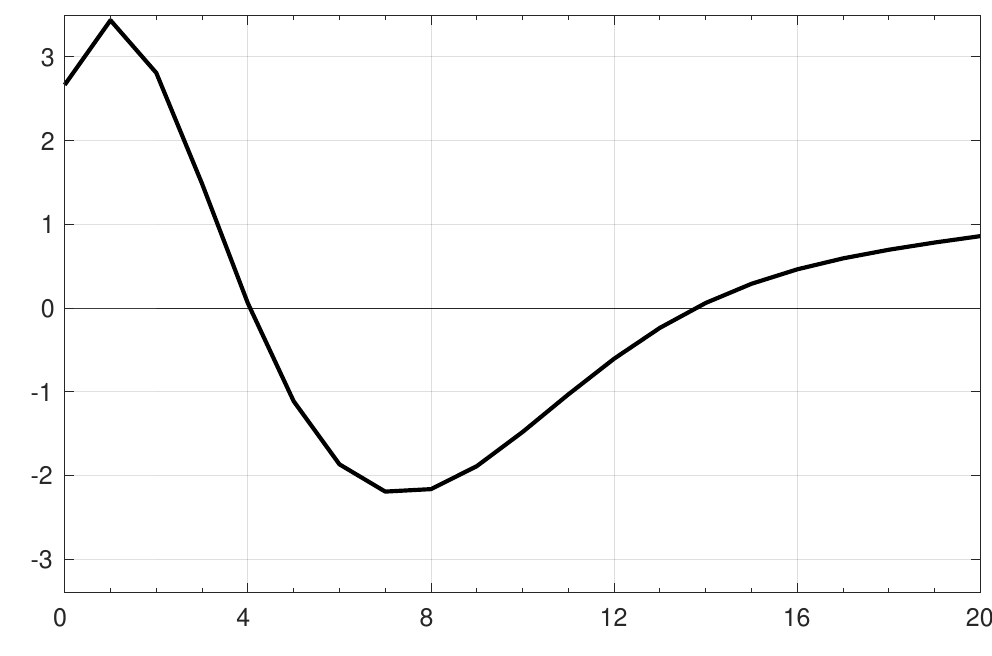} &
\includegraphics[trim={2cm 9.1cm 2.2cm 9.5cm},clip,width = 0.33\textwidth]{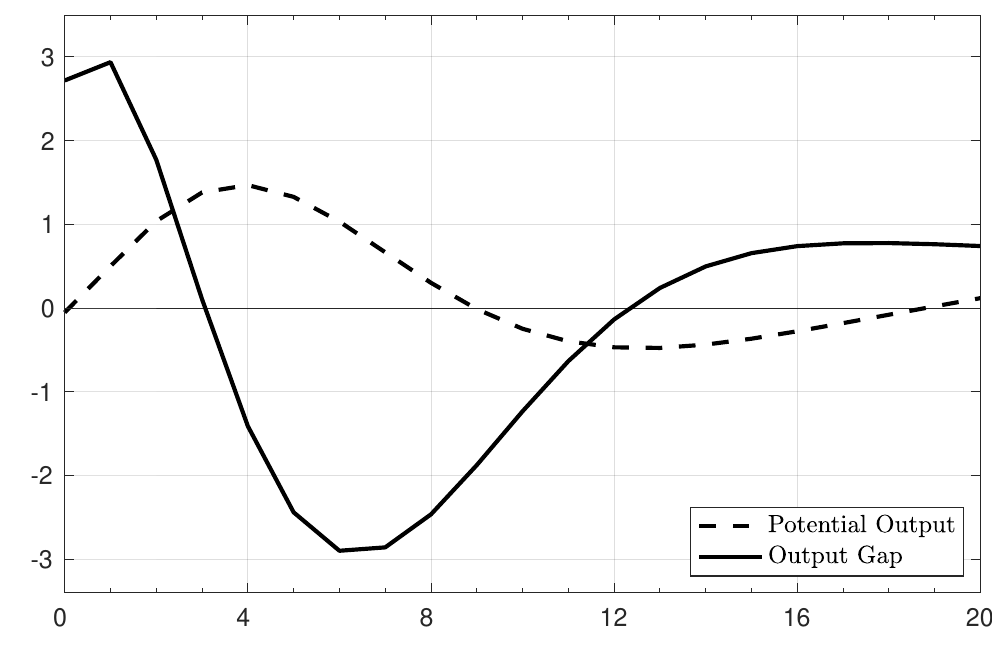}
\end{tabular}
\begin{tabular}{p{.98\textwidth}} \scriptsize Notes: \rm 
The black solid/dashed lines are the GIRF to a 0.5 p.p. shock to the common component of core inflation. The major ticks in the x-axis represent quarters after the shock.
\end{tabular} 
\end{figure}

To further corroborate the intuition that there is a relationship between our output gap estimate and inflation indicators, Figure \ref{fig::coreshock_g} shows the GIRFs  of the common component of core inflation, GDP, potential output, and the output gap to a 0.5 p.p. shock to the common component of core inflation. Results confirm that our model, on average, associates an increase in inflation with an output gap increase. The GIRF of the common component of core inflation peaks one quarter after the shocks before decreasing and slowly returning to zero. In response, GDP initially increases, but after about a year, it starts decreasing, reaching a trough about 2 years after the shock---the shock is fully absorbed in 5 years. The response of potential output is negative and persistent. The output gap initially increases, then decreases, and increases again before returning to zero.

%
%

\section{Inflation dynamics through the lens of our model}\label{sec::InflationDynamics}
In Section \ref{sec::PC}, we show that, although we did not create it specifically to signal inflationary pressures, our output gap measure does provide insights related to inflation dynamics. In light of these results, further inspection of Figure \ref{fig::ogapEC} raises two important questions: How can we reconcile an output gap of 2\% between 2017 and 2019 when inflation was just 1\%, well below the 2\% ECB target? How does our model interpret post-pandemic inflation dynamics?

\begin{figure}[ht!]\caption{Decomposition of core inflation }\label{fig::CoreDecomp}
\centering \footnotesize \smallskip
\setlength{\tabcolsep}{.01\textwidth}
\begin{tabular}{cc}
\textsc{Year-on-year inflation} & \textsc{Quarter-on-quarter annualized inflation}\\
\begin{tikzpicture}
  \node[anchor=south west,inner sep=0] (image) at (0,0) {\includegraphics[trim={1.3cm 9cm 1cm 9cm},clip,width = 0.485\textwidth]{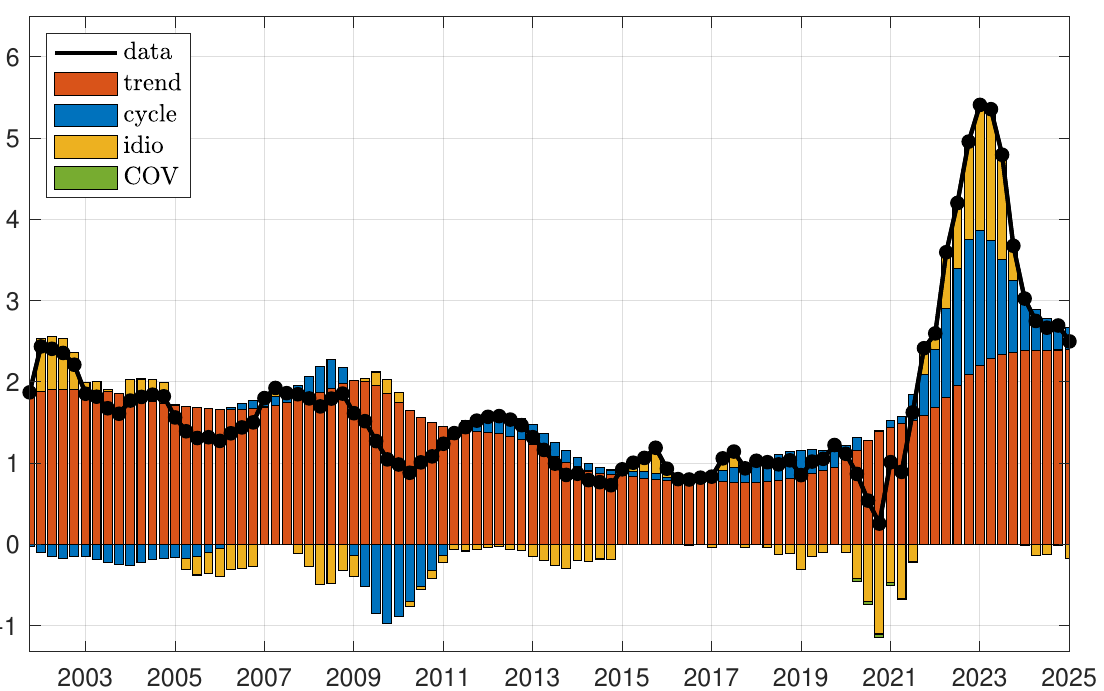}};
      \begin{scope}[x={(image.south east)},y={(image.north west)}]                
      \node[anchor=south west,inner sep=0] (image) at (0.19,0.74) 
              {\setlength{\tabcolsep}{.0\textwidth} \tiny
        \begin{tabular}{|C{.075\textwidth}C{.05\textwidth}C{.05\textwidth}C{.05\textwidth}C{.05\textwidth}|}\hline
                      & data  & trend & cycle & idio \\\hline
              2019:Q4 & 1.2   & 0.9   & 0.3   & ~0.0  \\
              2023:Q1 & 5.4   & 2.3   & 1.6   & ~1.5  \\
              2025:Q1 & 2.7   & 2.5   & 0.4   & -0.2  \\\hline
        \end{tabular}};  
    \end{scope}
\end{tikzpicture} &
\includegraphics[trim={1.3cm 9cm 1cm 9cm},clip,width = 0.485\textwidth]{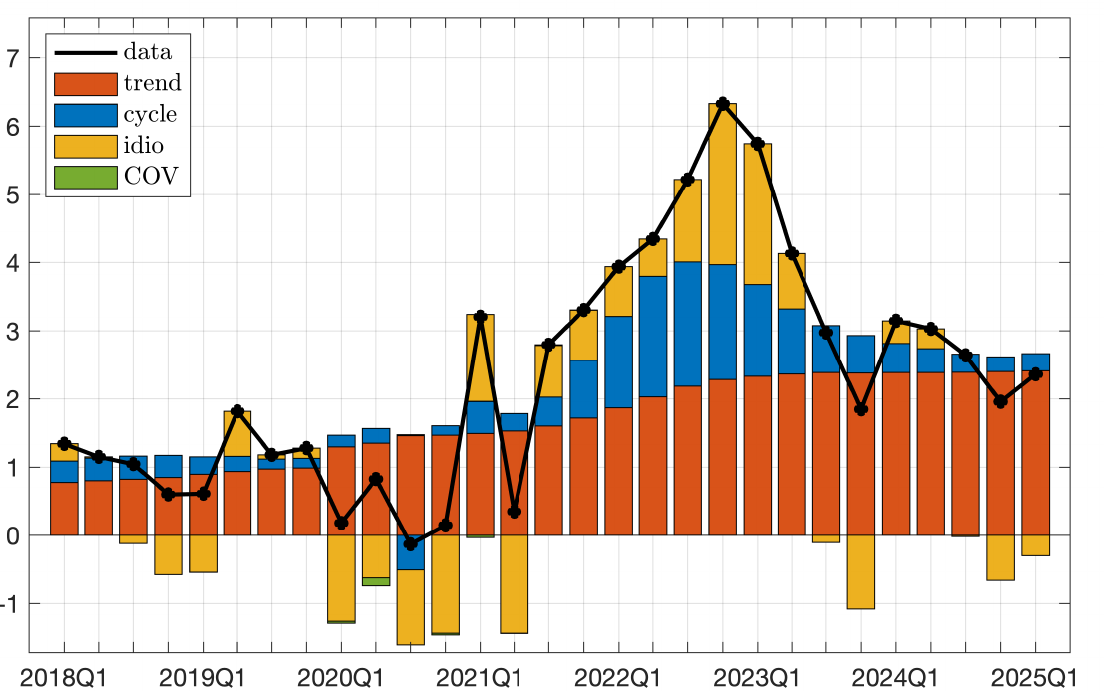} 
\end{tabular}

\begin{tabular}{p{.98\textwidth}} \scriptsize Notes: \rm The black line with dot markers is core inflation. The bars represent the contribution of each component to core inflation. The left plot shows YoY inflation, while the right plot shows QoQ inflation at an annual rate.
\end{tabular}
\end{figure}

To answer these questions, in Figure \ref{fig::CoreDecomp}, we plot the decomposition of core inflation (\small{HICPNEF}) implied by our model:
\begin{align}
y_{\text{\tiny HICPNEF},t}\ &=\ \bar{\pi}^c_t + \tilde{\pi}^c_t + \gamma_{\text{\tiny HICPNEF}}g_t \I_{\text{\tiny 2020:Q1$\le\! t\!\le$2021:Q4}}  + \xi_{\text{\tiny HICPNEF},t},\label{eq::CoreDecomp}\\
\bar{\pi}^c_t\ &=\ \mathrm{D}_{\text{\tiny HICPNEF},t} + \bs{\lambda}'_{\text{\tiny HICPNEF} }\bs{\psi}{\tau}_t,\label{TrendInf}\\
\tilde{\pi}^c_t\ &=\ \bs{\lambda}'_{\text{\tiny HICPNEF}}\bs{\omega}_t,\label{InfGap}
\end{align}
where $\bar{\pi}^c_t$ denotes trend core inflation---defined as the sum of the secular component of core inflation and the portion of core inflation driven by the common trend---and $\tilde{\pi}^c_t$ denotes the inflation gap---defined as the portion of core inflation driven by the common cyclical component.

Consistent with \citet{LucreziaInflation}, we estimate that, following the GFC, trend inflation decreased from 2\% to about 1\% in 2016, where it remained stable until the Covid pandemic. This finding is in line with \citet{CiccarelliOsbat} and \citet{CorselloNeriTagliabracci}, who show that inflation expectations de-anchored on the downside after the SDR. We therefore conclude that core inflation remained persistently below 2\% after the GFC primarily because trend core inflation decreased, not because there was slack in the economy, which explains why we estimate an output gap above 2\% when core inflation was 1\%.

Next, we turn our attention to post-pandemic inflation dynamics. As shown in the inset box of the left chart in Figure \ref{fig::CoreDecomp}, YoY core inflation increased about 4.2 p.p. between 2019:Q4 and 2023:Q1. Trend inflation and the core inflation gap jointly account for about 65\% of this 4.2 p.p. increase, while idiosyncratic factors drive the remaining portion. Since its 5.4\% peak in 2023:Q1, YoY core inflation decreased to 2.7\% in 2025:Q1. The core inflation gap accounts 40 basis points of this decline, and the idiosyncratic components for 250 basis points, while trend inflation increased 20 basis points. 

As we explained at the end of Section \ref{sec::poutgap}, the cyclical common component---therefore, the core inflation gap---primarily reflects demand forces. Under this assumption, the results of the decomposition in Figure \ref{fig::CoreDecomp} show that demand forces accounted for at least 30\% of the post-pandemic increase in core inflation, thus supporting existing literature that indicates that demand dynamics played a significant role in the inflation surge following the pandemic \citep{AscariTrezzi,GiannonePrimicieri,Canovaetal}.

Figure \ref{fig::CoreDecomp} clearly shows that other significant factors shaping post-pandemic inflation dynamics are idiosyncratic. For example, lingering Covid-specific effects (e.g., supply chain bottlenecks) that are not captured by the Covid factor. Another possibility is that the concurrent sharp rise in oil and natural gas prices following the onset of the Russia-Ukraine war induced second-round effects beyond those experienced in the pre-Covid sample. Finally, labor market tightness might have induced a non-linear response of prices that our linear model fails to capture. These factors might have a more or less persistent effect. Thus, our estimate of the role of demand forces should be considered a lower bound.

\section{Has our output gap measure predictive power for inflation?}\label{sec::InflationForecasting}

Any output gap measure \textit{must} be good at predicting inflation to be considered a credible indicator of current/future inflationary pressure. We therefore evaluate the ability of our output gap estimate to forecast year-on-year inflation one year ahead and compare its performance with that of other output gap estimates. In practice, we replicate the analysis conducted in \citet{banbura_does_2023}, and we employ the following model:
\begin{equation} \label{eq::ADL}
\pi_{t+4} = \alpha \pi_{t} + \beta OG_{t} + v_{t+4},
\end{equation}
where $\pi_{t} = 100\log({P_{t}}/{P_{t-1}})$ is the quarter-on-quarter inflation rate in quarter $t$, $P_t$ is the harmonized consumer price index (either headline or core), $\pi_{t+4} = \sum_{i=1}^4 \pi_{t+i}$ is year-on-year inflation in quarter $t+4$, and $OG_{t}$ is the output gap.\footnote{\citet{banbura_does_2023} labeled model \eqref{eq::ADL} the ``benchmark model,'' and they show that, despite being very simple, it delivers decent forecasts compared to more complex alternative models.} Alongside our output gap estimate, we consider estimates from different univariate and multivariate statistical models: the HP filter, the filter by \citet{hamilton2018you}, the boosted HP filter by \citet{phillips2021boosting}, the \cite{christianofitzgerald}filter, the Butterworth filter as recommended by \citet{canova2022faq}, and the large Bayesian VAR approach by \citet{morley_estimating_2023}---Appendix \ref{sec::altTC} describes these alternative models. 

We look at the forecasting performance of the different output gap measures over three distinct samples: the full sample (2015:Q4--2025:Q1), a pre-Covid sample, 2015:Q4--2019:Q4, and a post-Covid sample, 2022:Q1--2025:Q1. The forecasting exercise is an expanding window exercise, where the first window is a 60-quarter window. All models are estimated up to time $t$ and used to forecast inflation at $t+4$.

Table \ref{tab::s1519} compares the forecasting performance of the different output gap measures in terms of relative Root Mean Squared Error (RMSE), where values below one indicate superior forecasting performance when using our output gap estimate. As shown in rows (1)-(8), over the full sample our measure performs about as well as other statistical output gap estimates. However, this average masks important differences across subperiods. Excluding the Covid pandemic, in the pre-Covid sample---when inflation was low and stable---our output gap measure outperforms all alternatives in forecasting headline inflation and most alternatives in forecasting core inflation, often significantly so according to the \citet{DieboldMariano} test of equal predictive accuracy. In the post-Covid period, when inflation surged and then declined, our model continues to outperform all other measures, sometimes by a substantial margin, though not always significantly. It is worth noting, however, that the post-Covid sample includes only 13 observations, which limits the power of the  \citet{DieboldMariano} test. 

\begin{table}[ht!]\caption{4-quarter ahead year-on-year inflation forecasting} \label{tab::s1519} 
\centering 
\small \textit{Relative Root Mean Squared Errors}\\\smallskip

\setlength{\tabcolsep}{0pt}

\begin{tabular}{C{.05\textwidth}L{.35\textwidth}|C{.1\textwidth}C{.1\textwidth}|C{.1\textwidth}C{.1\textwidth}|C{.1\textwidth}C{.1\textwidth}} \hline\hline
&     &   \multicolumn{2}{c|}{2015:Q4-2025:Q1}      &   \multicolumn{2}{c}{2015:Q4-2019:Q4} \vline   &   \multicolumn{2}{c}{2022:Q1-2025:Q1} \\ \hline  
&   Output Gap Measure  &    Headline  &   Core     &    Headline  &   Core    &    Headline  &   Core\bw{$^*$} \\ \hline
\scriptsize (1) &   \small HP Filter ($\lambda = 1600$)           &  1.03  &  0.97  & 0.90  &  1.00\bw{$^*$}  &  1.02  &  0.93  \\
\scriptsize (2) &   \small HP Filter ($\lambda = 51200$)          &  1.02  &  0.96  & 0.91  &  1.01\bw{$^*$}  &  0.99  &  0.94  \\
\scriptsize (3) &   \small Hamilton Filter                        &  1.00  &  0.97  & 0.98  &  0.97$^*$  &  1.01  &  0.94  \\
\scriptsize (4) &   \small Boosted HP Filter ($\lambda = 1600$)   &  1.01  &  0.97  & 0.89  &  0.96\bw{$^*$}  &  0.93  &  0.93  \\
\scriptsize (5) &   \small Boosted HP Filter ($\lambda = 51200$)  &  1.01  &  0.97  & 0.89  &  0.99\bw{$^*$}  &  0.96  &  0.93  \\
\scriptsize (6) &   \small Christiano-Fitzgerald Filter           &  1.04  &  0.97  & 0.89  &  0.82$^*$  &  1.01  &  0.92  \\
\scriptsize (7) &   \small Butterworth Filter                     &  1.02  &  0.97  & 0.94  &  0.95$^*$  &  1.00  &  0.93  \\
\scriptsize (8) &   \small Multivariate Beveridge-Nelson          &   -    &   -    & 0.91  &  0.98$^*$  &   -    &   -     
\\  \hline
\end{tabular}

\begin{tabular}{p{\textwidth}} \scriptsize {\sc Notes}: \rm The table shows the relative RMSE of forecasting year-on-year inflation using \eqref{eq::ADL}, where $\text{OG}_t$ is either our output gap estimate, or an alternative output gap estimate. Our benchmark output gap estimate is always the numerator of the RMSE; therefore, values below 1 indicate a better forecasting performance when using our benchmark output gap estimate. Asterisks denote statistical significance at the 10\% confidence level according to the \citet{DieboldMariano} test of equal predictive accuracy. Rows (1)--(8) compare our benchmark estimate with alternative models. In row (8) forecasts are obtained with the output gap measure by \cite{morley_estimating_2023} which is available only until 2021:Q3; accordingly, we report results only for the pre-Covid forecasting exercise.
\end{tabular}
\end{table}

In conclusion, the results in this section demonstrate that our output gap measure is not only a measure of the cyclical position of the economy but also a reliable inflation gauge.

%
%
\section{The role of credit indicators}\label{sec::Credit}
As discussed in Section \ref{sec::OL&PC}, in production-function-based models, low inflation signals a negative output gap, whereas high inflation indicates a positive one. \citet{borio_rethinking_2017} challenge this view, arguing that credit expansions---especially since the late 1990s---have often resulted in unsustainable growth episodes without any corresponding rise in inflation. Hence, they argue that financial indicators are essential for a meaningful assessment of the business cycle. Supporting this perspective, \citet{berger2022unified} show that much of the U.S. economy's overheating before the GFC originated from credit and housing market imbalances. Similarly, \citet{claessens_how_2012}, \citet{runstler_business_2018}, and \citet{winter2022joint} find that business and financial cycles are correlated and tend to co-move over the medium run.

To explore the role of financial conditions in our framework, we examine which signals the model extracts from credit indicators when estimating the output gap. We focus on household liabilities, which have become a key driver of the business cycle in many advanced economies since the early 2000s \citep{mian2017household}. While in the 1990s non-financial corporations dominated the financial cycle, households drove both the pre-GFC leverage boom, which boosted demand, and the subsequent deleveraging, which curtailed it \citep{mian2018finance,reichlin_financial_2020,reichlin_financial_2020wp}. For the EA, \citet{gambetti2017loan} find that loan supply shocks significantly affect business cycle fluctuations. 

\begin{figure}[ht!]\caption{Scenario analysis} \label{fig::scenarioHH} \centering
\centering \footnotesize \sc \smallskip

\setlength{\tabcolsep}{.005\textwidth}
\begin{tabular}{cc}
\hspace{12pt}Household liabilities &\hspace{10pt} Household liabilities \\[-1pt]
\scriptsize \hspace{12pt}(Data and counterfactual) &  \hspace{12pt}\scriptsize (Conditional and unconditional forecast)\\
\includegraphics[trim={2.05cm 9.5cm 2.2cm 9.5cm},clip,width = 0.475\textwidth]{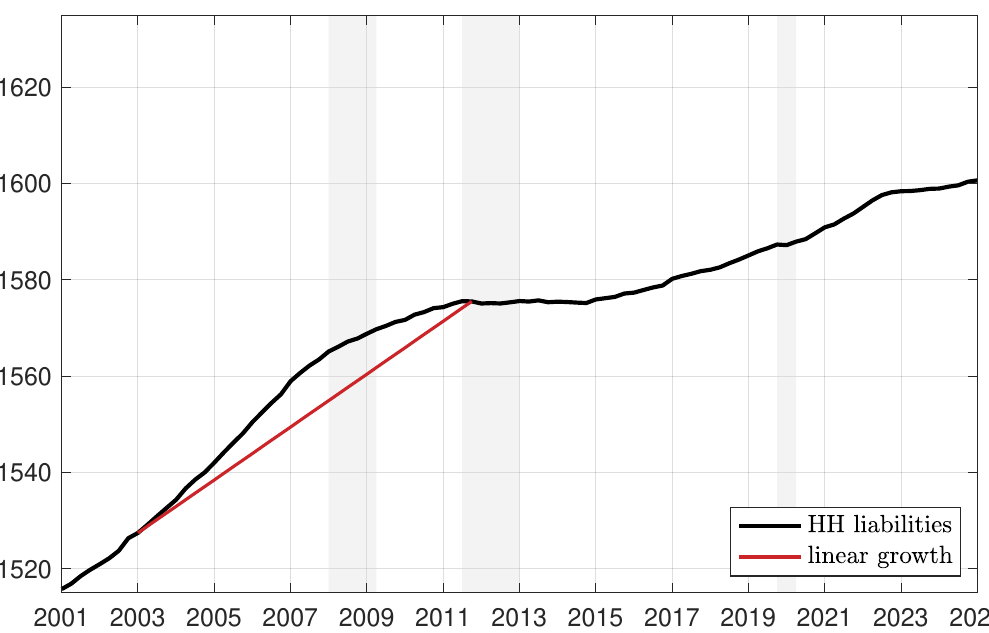} & 
\includegraphics[trim={2.05cm 9.5cm 2.2cm 9.5cm},clip,width = 0.475\textwidth]{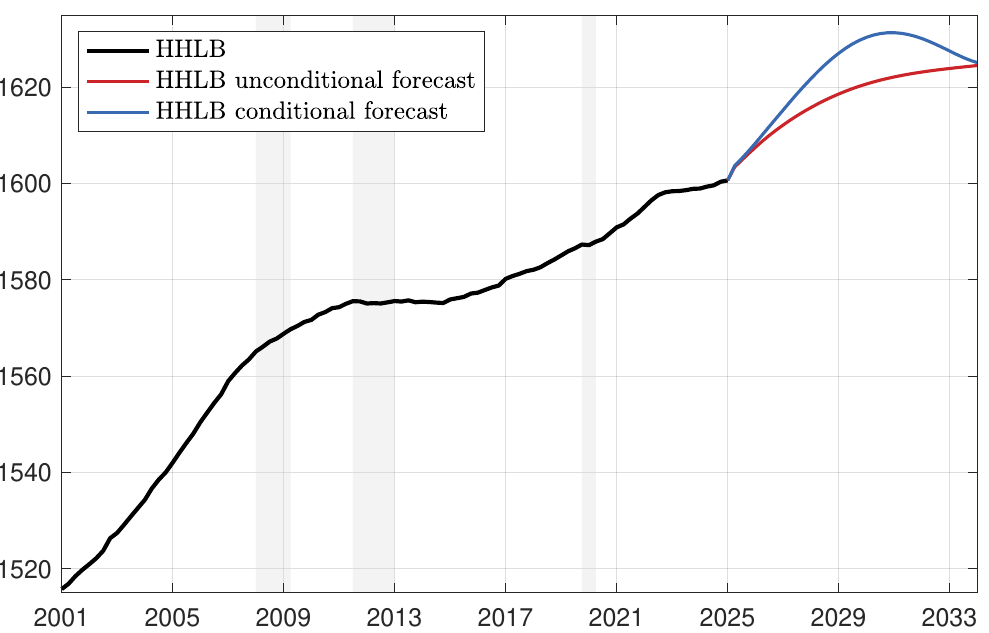}\\[5pt]
\hspace{12pt}GDP &\hspace{10pt} Potential output and output gap\\[-1pt]
\scriptsize \hspace{12pt}(Scenario dynamic effects) & \scriptsize \hspace{12pt}(Scenario dynamic effects)\\
\includegraphics[trim={2.05cm 9.5cm 2.2cm 9.5cm},clip,width = 0.475\textwidth]{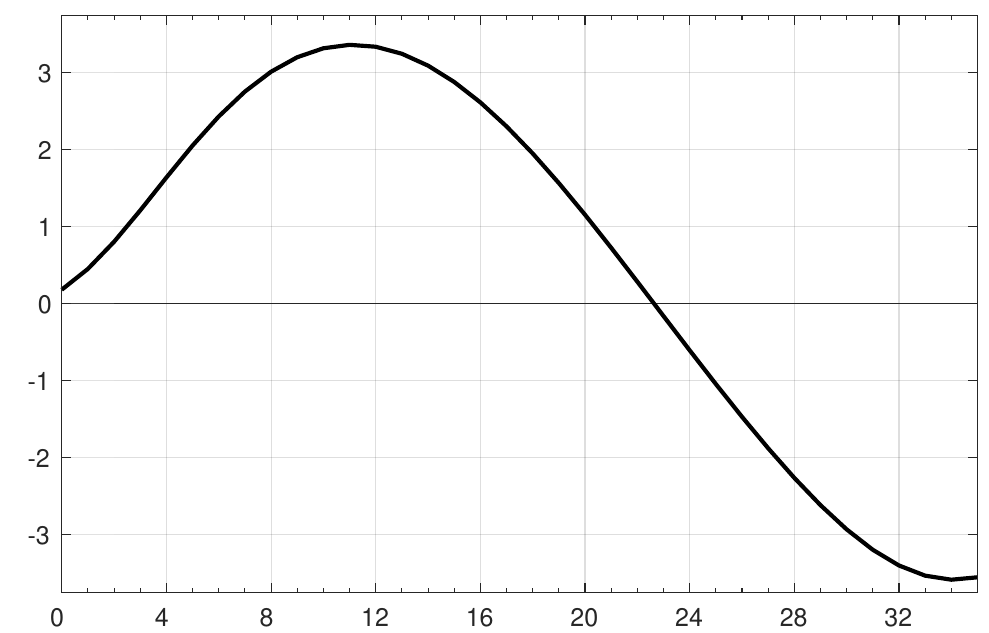} & 
\includegraphics[trim={2.05cm 9.5cm 2.2cm 9.5cm},clip,width = 0.475\textwidth]{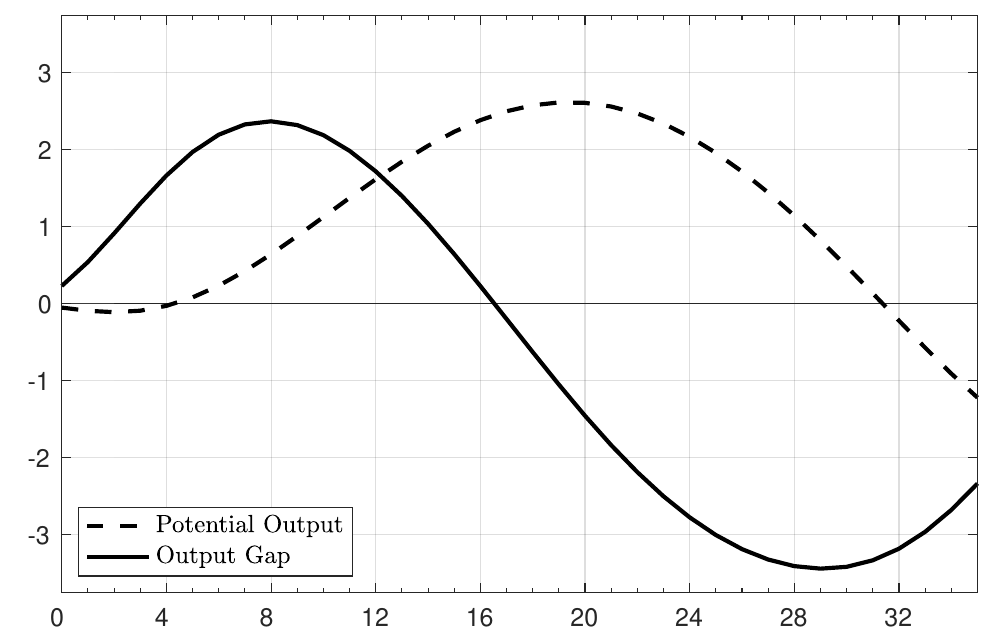}\\
\end{tabular}

\begin{tabular}{p{.975\textwidth}} \scriptsize Notes: \rm In the upper-left chart, the black line is the data (in 100$\times$log-levels), and the red line is a linear path starting from the value of household liabilities in 2003:Q1 and ending in 2011:Q4. In the upper-right chart, the black line are the data,the blue line is the scenario we simulate, and the red line is the forecast of household liabilities when no alternative scenario is imposed. In the lower charts, the black solid/dashed lines are the dynamic effects of the simulated scenario.
\end{tabular}
\end{figure}

Figure \ref{fig::scenarioHH} shows the impact of a scenario in which household liabilities increase faster than projected in the baseline for about 3\sfrac{1}{2} years. In this scenario, household liabilities reach a level about 6\sfrac{1}{4} p.p. higher than in the baseline before returning to baseline after about 8 years.\footnote{We calibrate this scenario by comparing the actual path of household liabilities between 2003:Q1 and 2011:Q4 with a counterfactual linear trend (red line, top-left panel of Figure \ref{fig::scenarioHH}). The smoothed difference (using a fifth-degree polynomial) is then added to the unconditional forecast (red line, top-right panel), generating the conditional scenario (blue line) used to simulate the dynamic responses of all variables in the model.}

The lower charts in Figure \ref{fig::scenarioHH} show the dynamic effects of this scenario on the log level of GDP and on the output gap and potential output.\footnote{This is equivalent to computing the effect of a sequence of shocks. Hence, it is estimated the same way we estimated the GIRFs in Section \ref{sec::OL&PC}.} Specifically, GDP increases for a little over three years, reaching a peak at 3 p.p. above the baseline. Then, it declines and, after six years, turns negative. The response of the output gap mimics that of GDP, but it is a little faster. Potential output slowly increases for the first five years and then returns to the baseline. These results show that that debt-driven expansions deliver only temporary gains, as growth financed through household debt is not sustainable in the long run.

\section{Conclusions} \label{sec::conclude}
This paper proposes a new measure of potential output and the output gap for the EA based on letting a large number of macroeconomic and financial indicators speak. To do so, we estimate a large-dimensional non-stationary dynamic factor model, which allows us to capture co-movements across series while incorporating relevant macroeconomic priors, such as the long-run decline in output growth. 

Our output gap estimate is in line with those published by the EC and the IMF in most of the sample. However, it diverges notably after the SDR, indicating that the EA economy was considerably tighter than estimated by the EC and the IMF. This result suggests that the EA has a potential output issue, not a business cycle issue. Hence, to achieve stronger and more resilient growth, European countries should prioritize structural reforms and promote productivity-enhancing investments, while supporting aggregate demand more decisively during downturns to prevent recession-induced output losses. In contrast, policies aiming at boosting debt-financed household consumption or residential investment would deliver only short-lived gains, as our findings indicate that growth financed through household debt is not sustainable in the long run.

Moreover, although we did not create our output gap measure specifically to signal inflationary pressures, it does provide insights related to inflation dynamics because, on average, the Phillips Curve is satisfied. In particular, we find that core inflation remained below 2\% after the GFC, not because there was slack in the economy, but rather because trend inflation decreased by one percentage point---in line with the idea that inflation expectations de-anchored on the downside after the GFC \citep{CiccarelliOsbat, CorselloNeriTagliabracci}. Finally, we show that the output gap contributed to at least 30\% of the post-pandemic increase in core inflation, thus supporting existing literature that suggests demand forces played a substantial role in the rise of post-pandemic inflation \citep{AscariTrezzi,GiannonePrimicieri,Canovaetal}. 


\clearpage

\titleformat{\section}
   {\titlerule[0.75pt]\normalfont\large\bfseries}{\thesection}{1em}{}[{\titlerule[0.75pt]}]
\titleformat{\subsection}
   {\normalfont\normalsize\bfseries}{\thesubsection}{1em}{}[{\titlerule[0.2pt]}]
\titleformat{\subsubsection}
   {\normalfont\normalsize\it}{\thesubsubsection}{1em}{}[{\titlerule[0.1pt]}]
\titlespacing*{\section}{0pt}{15pt}{6pt}
\titlespacing*{\subsection}{0pt}{12pt}{6pt}
\titlespacing*{\subsubsection}{0pt}{9pt}{6pt}

\pagestyle{fancy}
\fancyhf{}
\chead{\sc Supplementary material for the paper: Measuring the Euro Area Output Gap\\[-10pt]}
\cfoot{\sc Page \thepage\ of \pageref{LastPage}}
 \renewcommand{\footrulewidth}{0pt}
 \renewcommand{\headrulewidth}{0pt}

\gdef\thesection{\Alph{section}}
\gdef\thesubsection{\Alph{section}.\arabic{subsection}}
\gdef\thefigure{\Alph{section}\arabic{figure}}
\gdef\theequation{\Alph{section}\arabic{equation}}
\gdef\thetable{\Alph{section}\arabic{table}}

\setcounter{table}{0}
\setcounter{figure}{0}
\setcounter{equation}{0}
\setcounter{page}{1}

\begin{center}
\rule{\textwidth}{1pt}\\
\textit{Supplementary material for the paper:} \\

\Large{\bf Measuring the Euro Area Output Gap} \\[-10pt]
\rule{\textwidth}{1pt}\\[12pt]

\begin{tabular}{C{.3\textwidth}C{.3\textwidth}C{.3\textwidth}}
\normalsize Matteo Barigozzi & \normalsize  Claudio Lissona  & \normalsize Matteo Luciani \\[-6pt]
\small University of Bologna & \small  University of Bologna & \small Federal Reserve Board \\[-6pt]
\footnotesize matteo.barigozzi@unibo.it & \footnotesize claudio.lissona2@unibo.it & \footnotesize matteo.luciani@frb.gov \\
\end{tabular}

\end{center}

\renewcommand{\thefootnote}{ } 

\footnotetext{\noindent M. Barigozzi and C. Lissona gratefully acknowledges financial support from MIUR (PRIN2020, Grant 2020N9YFFE).\smallskip

\noindent \textsc{Disclaimer:} the views expressed in this paper are those of the authors and do not necessarily reflect the views and policies of the Board of Governors or the Federal Reserve System.} 

\gdef\thefootnote{(\roman{footnote})}
\appendix
\section*{Summary}
In Appendix \ref{datades}, we provide full details on the dataset. 
In Appendix \ref{app:ass}, we formally state all model assumptions and provide motivation and comments for each of them. 
Appendix \ref{sec::estdetail} describes the estimation procedure step by step, and Appendix \ref{app::confbands} details the computation of confidence bands used to quantify the uncertainty around our output gap estimate. 
In Appendix \ref{sec::covid}, we show how our measures would change if we did not model the effect of Covid explicitly, and we also compare our measure of the output gap with the ones obtained by using a different estimate of the Covid factor or when modeling the Covid induced volatility as suggested by \citet{lenza_how_2022}. 
Appendix \ref{app::notvpars}presents robustness results with respect to different calibrations of the variances of the secular trends. 
Appendices \ref{sec::altTC} and \ref{sec::altTCest} compare our output gap estimate with that obtained using alternative methodologies, and Appendix \ref{sec::realtime} assesses the real-time reliability of our measure.
\newpage

%
%

\section{Data Description}\label{datades}
Table \ref{tab::dataEA} provides a brief description for each of the 118 series in our dataset. Moreover, for each variable, Table \ref{tab::dataEA} indicates the source, the unit of measure, the seasonal adjustment treatment, the transformation (if any), model for the deterministic component for the idiosyncratic component. Table \ref{gloss} presents a glossary to proper understand the data description presented in Table \ref{tab::dataEA}.

All the series were retrieved in February 2025, with the sample starting in January 2000 and ending in October 2024, the last observation available for all the series. After dropping missing values and transforming the variables, the actual starting point for the analysis is 2001:Q1. Monthly series, which constitute around one-third of the dataset, are aggregated at the quarterly level by simple averages; hence, the sample used for the analysis is 2001:Q1-2025:Q1 ($T=97$). 

Most of the series in the dataset are available already seasonally adjusted from the source, while others, e.g., financial variables and producer price indexes, are only available not seasonality adjusted. In these cases, we deseasonalize the series using a simple dummy variable approach.

\begin{table}[H]
\caption{Glossary}
\label{gloss}
\resizebox{6.5in}{!}{%
\begin{tabular}{ l | l }
\hline
\hline
\textbf{Source} & \textbf{Unit} \\ 
\hline
\hline
EUR = Eurostat & CLV15 = Chain-linked volumes (2015=100) \\
OECD = Organization for Economic Co-operation and Development & $\text{CP}$ = Current Prices (Million\euro) \\
ECB = European Central Bank & 1000p = Thousands of persons  \\
FRED = Federal Reserve Economic Data &  1000U = Thousands of Units\\
& I\textit{xx} = Index, 20\textit{xx} = 100\\
\end{tabular}%
}
\end{table}
\begin{table}[H]
\resizebox{6.5in}{!}{%
\begin{tabular}{  l | l | l | l | l }
\hhline{=====}
\textbf{SA} & \textbf{F}  & \textbf{Trans} & \textbf{Trend} & \textbf{Idio} \\ \hhline{=====}
NSA = No Seasonal Adjustment & Q = Quarterly & 0 = No Transformation & 0 = No Trend & 0 = I(0)  \\
SA = Seasonal Adjustment & M = Monthly & 1 = First Differences & 1 = Deterministic Trend & 1 = I(1)  \\
SCA = Seasonal and Calendar Adjustment & & 2 = Log-Transformations  & 2 = Time-Varying Trend &  \\
MSA = Manual adjustment  & & 3 = First-Differences in Logs & 3 = Time-varying Mean \\\hline
\end{tabular}%
}
\end{table}
\newpage
\newgeometry{left=1.5cm,right=1.5cm,top=2cm,bottom=2.5cm}
\begin{table}[h!]
\caption{Data description: Euro Area}
\begin{footnotesize}
\begin{threeparttable}
\resizebox{7.4in}{!}{%
\begin{tabular}{  c | c | l | c | c | c | c | c | c | c}
\hline
\hline
\textbf{ID} & \textbf{Ticker} & \textbf{Series} & \textbf{Unit} & \textbf{SA} & \textbf{F} & \textbf{Source} & \textbf{Trans} & \textbf{Trend}  & \textbf{Idio}\\ 
\hline
\hline
\multicolumn{9}{c}{(1) \textbf{National Accounts}}\\
\hline
1  & GDP    & Real Gross Domestic Product                                             & CLV(2015) & SCA & Q & EUR & 2 & 2 & 0 \\ 
2  & EXPGS  & Real Export Goods and services                                          & CLV(2015) & SCA & Q & EUR & 2 & 1 & 0 \\ 
3  & IMPGS  & Real Import Goods and services                                          & CLV(2015) & SCA & Q & EUR & 2 & 1 & 0 \\ 
4  & GFCE   & Real Government Final consumption expenditure                           & CLV(2015) & SCA & Q & EUR & 2 & 1 & 0 \\ 
5  & HFCE   & Real Households consumption expenditure                                 & CLV(2015) & SCA & Q & EUR & 2 & 1 & 1 \\ 
6  & CONSD  & Real Households consumption expenditure: Durable Goods                  & CLV(2015) & SCA & Q & EUR & 2 & 1 & 1 \\ 
7  & CONSND & Real Households consumption expenditure: Non-Durable Goods and Services & CLV(2015) & SCA & Q & EUR & 2 & 1 & 1 \\ 
8  & GCF    & Real Gross capital formation                                            & CLV(2015) & SCA & Q & EUR & 2 & 0 & 0 \\ 
9  & GCFC   & Real Gross fixed capital formation                                      & CLV(2015) & SCA & Q & EUR & 2 & 0 & 1 \\ 
10 & GFACON & Real Gross Fixed Capital Formation: Construction                        & CLV(2015) & SCA & Q & EUR & 2 & 0 & 1 \\ 
11 & GFAMG  & Real Gross Fixed Capital Formation: Machinery and Equipment             & CLV(2015) & SCA & Q & EUR & 2 & 0 & 1 \\
12 & AHRDI  & Adjusted Households Real Disposable Income                              & \%change  & SCA & Q & EUR & 0 & 0 & 0 \\ 
13 & AHFCE  & Actual Final Consumption Expenditure of Households                      & \%change  & SCA & Q & EUR & 0 & 0 & 0 \\ 
14 & GNFCPS & Gross Profit Share of Non-Financial Corporations                        & Percent   & SCA & Q & EUR & 0 & 0 & 0 \\ 
15 & GNFCIR & Gross Investment Share of Non-Financial Corporations                    & Percent   & SCA & Q & EUR & 0 & 0 & 0 \\ 
16 & GHIR   & Gross Investment Rate of Households                                     & Percent   & SCA & Q & EUR & 0 & 1 & 0 \\ 
17 & GHSR   & Gross Households Savings Rate                               			  & Percent   & SCA & Q & EUR & 0 & 0 & 0 \\ 
\hline
\multicolumn{9}{c}{(2) \textbf{Labor Market Indicators}}\\
\hline
18 & TEMP   & Total Employment (domestic concept)                                  & 1000-ppl & SCA & Q & EUR & 2 & 1 & 1 \\ 
19 & EMP    & Employees (domestic concept)                                         & 1000-ppl & SCA & Q & EUR & 2 & 1 & 1 \\ 
20 & SEMP   & Self Employment (domestic concept)                                   & 1000-ppl & SCA & Q & EUR & 2 & 0 & 1 \\ 
21 & THOURS & Hours Worked: Total                                                  & 2015=100 & SCA & Q & EUR & 2 & 0 & 1 \\
22 & EMPAG  & Quarterly Employment: Agriculture, Forestry, Fishing                 & 1000-ppl & SCA & Q & EUR & 2 & 1 & 0 \\ 
23 & EMPIN  & Quarterly Employment: Industry                                       & 1000-ppl & SCA & Q & EUR & 2 & 0 & 0 \\ 
24 & EMPMN  & Quarterly Employment: Manufacturing                                  & 1000-ppl & SCA & Q & EUR & 2 & 0 & 0 \\ 
25 & EMPCON & Quarterly Employment: Construction                                   & 1000-ppl & SCA & Q & EUR & 2 & 0 & 1 \\ 
26 & EMPRT  & Quarterly Employment: Wholesale/Retail trade, transport, food        & 1000-ppl & SCA & Q & EUR & 2 & 1 & 1 \\ 
27 & EMPIT  & Quarterly Employment: Information and Communication                  & 1000-ppl & SCA & Q & EUR & 2 & 1 & 1 \\ 
28 & EMPFC  & Quarterly Employment: Financial and Insurance activities             & 1000-ppl & SCA & Q & EUR & 2 & 0 & 1 \\ 
29 & EMPRE  & Quarterly Employment: Real Estate                                    & 1000-ppl & SCA & Q & EUR & 2 & 0 & 0 \\ 
30 & EMPPR  & Quarterly Employment: Professional, Scientific, Technical activities & 1000-ppl & SCA & Q & EUR & 2 & 1 & 1 \\ 
31 & EMPPA  & Quarterly Employment: PA, education, health ad social services       & 1000-ppl & SCA & Q & EUR & 2 & 1 & 1 \\ 
32 & EMPENT & Quarterly Employment: Arts and recreational activities               & 1000-ppl & SCA & Q & EUR & 2 & 1 & 0 \\ 
33 & UNETOT & Unemployment: Total                                                  & \%active & SA  & M & EUR & 0 & 3 & 0 \\ 
34 & UNEO25 & Unemployment: Over 25 years                                          & \%active & SA  & M & EUR & 0 & 1 & 0 \\ 
35 & UNEU25 & Unemployment: Under 25 years                                         & \%active & SA  & M & EUR & 0 & 1 & 0 \\
36 & RPRP   & Real Labour Productivity (person)                                    & 2015=100 & SCA & Q & EUR & 2 & 1 & 1 \\ 
37 & WS     & Wages and salaries                                                   & CP       & SCA & Q & EUR & 2 & 1 & 0 \\ 
38 & ESC    & Employers' Social Contributions                                      & CP       & SCA & Q & EUR & 2 & 1 & 0 \\ 
\hline
\multicolumn{9}{c}{(3) \textbf{Credit Aggregates}}\\
\hline
39 & TAS.SDB   & Total Economy - Assets: Short-Term Debt Securities           & MLN\euro & MSA & Q & EUR & 2 & 0 & 1 \\ 
40 & TAS.LDB   & Total Economy - Assets: Long-Term Debt Securities            & MLN\euro & MSA & Q & EUR & 2 & 1 & 0 \\ 
41 & TAS.SLN   & Total Economy - Assets: Short-Term Loans                     & MLN\euro & MSA & Q & EUR & 2 & 1 & 0 \\ 
42 & TAS.LLN   & Total Economy - Assets: Long-Term Loans                      & MLN\euro & MSA & Q & EUR & 2 & 1 & 1 \\ 
43 & TLB.SDB    & Total Economy - Liabilities: Short-Term Debt Securities     & MLN\euro & MSA & Q & EUR & 2 & 0 & 1 \\ 
44 & TLB.LDB    & Total Economy - Liabilities: Long-Term Debt Securities      & MLN\euro & MSA & Q & EUR & 2 & 1 & 1 \\ 
45 & TLB.SLN    & Total Economy - Liabilities: Short-Term Loans               & MLN\euro & MSA & Q & EUR & 2 & 1 & 1 \\ 
46 & TLB.LLN    & Total Economy - Liabilities: Long-Term Loans                & MLN\euro & MSA & Q & EUR & 2 & 1 & 1 \\ 
47 & NFCAS     & Non-Financial Corporations: Total Financial Assets           & MLN\euro & MSA & Q & EUR & 2 & 1 & 0 \\ 
48 & NFCAS.SLN & Non-Financial Corporations - Assets: Short-Term Loans        & MLN\euro & MSA & Q & EUR & 2 & 1 & 0 \\ 
49 & NFCAS.LLN & Non-Financial Corporations - Assets: Long-Term Loans         & MLN\euro & MSA & Q & EUR & 2 & 1 & 0 \\ 
50 & NFCLB      & Non-Financial Corporations: Total Financial Liabilities     & MLN\euro & MSA & Q & EUR & 2 & 1 & 0 \\ 
51 & NFCLB.SLN  & Non-Financial Corporations - Liabilities - Short-Term Loans & MLN\euro & MSA & Q & EUR & 2 & 1 & 1 \\ 
52 & NFCLB.LLN  & Non-Financial Corporations - Liabilities - Long-Term Loans  & MLN\euro & MSA & Q & EUR & 2 & 1 & 1 \\ 
53 & GGAS       & General Government: Total Financial Assets                  & MLN\euro & MSA & Q & EUR & 2 & 1 & 0 \\ 
54 & GGAS.SLN  & General Government - Assets: Short-Term Loans                & MLN\euro & MSA & Q & EUR & 2 & 1 & 0 \\ 
55 & GGAS.LLN  & General Government - Assets: Short-Term Loans                & MLN\euro & MSA & Q & EUR & 2 & 0 & 1 \\ 
56 & GGLB       & General Government: Total Financial Liabilities             & MLN\euro & MSA & Q & EUR & 2 & 1 & 1 \\ 
57 & GGLB.SLN   & General Government - Liabilities: Short-Term Loans          & MLN\euro & MSA & Q & EUR & 2 & 1 & 0 \\ 
58 & GGLB.LLN   & General Government - Liabilities: Long-Term Loans           & MLN\euro & MSA & Q & EUR & 2 & 0 & 1 \\ 
59 & HHAS       & Households: Total Financial Assets                          & MLN\euro & MSA & Q & EUR & 2 & 1 & 1 \\ 
60 & HHAS.SLN  & Households - Assets: Short-Term Loans                        & MLN\euro & MSA & Q & EUR & 2 & 1 & 0 \\ 
61 & HHAS.LLN  & Households - Assets: Long-Term Loans                         & MLN\euro & MSA & Q & EUR & 2 & 1 & 0 \\
62 & HHLB       & Households: Total Financial Liabilities                     & MLN\euro & MSA & Q & EUR & 2 & 2 & 0 \\
63 & HHLB.SLN   & Households - Liabilities: Short-Term Loans                  & MLN\euro & MSA & Q & EUR & 2 & 0 & 0 \\
64 & HHLB.LLN   & Households - Liabilities: Long-Term Loans                   & MLN\euro & MSA & Q & EUR & 2 & 2 & 0 \\
\hline
\hline
\end{tabular}%
}
\begin{tablenotes}
\scriptsize
\item \noindent In absence of available data on durable and non-durable goods for the Euro Area, we follow \cite{casalis2022cyclical} and build the aggregate\\ series of durable consumption (CONSD) aggregating the data for the 20 individual Euro Area countries. Since data for services and non-durable\\ 
goods are unavailable for many individual countries as well, we build an aggregate measure of non-durable goods (CONSND) which also includes\\ semi-durable goods and services. 
 \end{tablenotes}
\end{threeparttable}
\end{footnotesize}
\label{tab::dataEA}
\end{table}
\newpage
\newgeometry{left=1.5cm,right=1.5cm,top=2cm,bottom=2.5cm}
\begin{table}[h!]\ContinuedFloat
\caption{Data description: Euro Area}
\begin{footnotesize}
\resizebox{7.4in}{!}{%
\begin{tabular}{  c | c | l | c | c | c | c | c | c | c}
\hline
\hline
\textbf{ID} & \textbf{Ticker} & \textbf{Series} & \textbf{Unit} & \textbf{SA} & \textbf{F} & \textbf{Source} & \textbf{Trans} & \textbf{Trend}  & \textbf{Idio}\\ 
\hline
\hline
\multicolumn{9}{c}{(4) \textbf{Labor Costs}}\\
\hline
65 & ULCIN  & Nominal Unit Labor Costs: Industry                                       & 2016=100 & SCA & Q & EUR & 2 & 1 & 0 \\ 
66 & ULCMQ  & Nominal Unit Labor Costs: Mining and Quarrying                           & 2016=100 & SCA & Q & EUR & 2 & 1 & 0 \\ 
67 & ULCMN  & Nominal Unit Labor Costs: Manufacturing                                  & 2016=100 & SCA & Q & EUR & 2 & 1 & 0 \\ 
68 & ULCCON & Nominal Unit Labor Costs: Construction                                   & 2016=100 & SCA & Q & EUR & 2 & 1 & 0 \\ 
69 & ULCRT  & Nominal Unit Labor Costs: Wholesale/Retail Trade, Transport, Food, IT    & 2016=100 & SCA & Q & EUR & 2 & 1 & 0 \\ 
70 & ULCFC  & Nominal Unit Labor Costs: Financial Activities                           & 2016=100 & SCA & Q & EUR & 2 & 1 & 0 \\ 
71 & ULCRE  & Nominal Unit Labor Costs: Real Estate                                    & 2016=100 & SCA & Q & EUR & 2 & 1 & 0 \\ 
72 & ULCPR  & Nominal Unit Labor Costs: Professional, Scientific, Technical activities & 2016=100 & SCA & Q & EUR & 2 & 1 & 0 \\ 
\hline 
\multicolumn{9}{c}{(5) \textbf{Exchange Rates}}\\
\hline
73 & REER42 & Real Exchange Rate (42 main industrial countries) & 2010=100 & NSA & M & EUR & 2 & 0 & 0 \\
74 & ERUS   & Exchange Rate (US dollar)                         & 2010=100 & NSA & M & EUR & 2 & 0 & 0 \\
\hline 
\multicolumn{9}{c}{(6) \textbf{Interest Rates}}\\
\hline
75 & IRT3M & 3-Months Interest Rates                  & Percent & NSA & M & EUR & 0 & 0 & 1 \\ 
76 & IRT6M & 6-Months Interest Rates                  & Percent & NSA & M & EUR & 0 & 0 & 1 \\ 
77 & LTIRT & Long-Term Interest Rates (EMU Criterion) & Percent & NSA & M & EUR & 0 & 0 & 1 \\ 
\hline
\multicolumn{9}{c}{(7) \textbf{Industrial Production and Turnover}}\\
\hline
78 & IPMN     & Industrial Production Index: Manufacturing              & 2021=100 & SCA & M & EUR & 2 & 0 & 1 \\ 
79 & IPCAG    & Industrial Production Index: Capital Goods              & 2021=100 & SCA & M & EUR & 2 & 0 & 1 \\ 
80 & IPCOG    & Industrial Production Index: Consumer Goods             & 2021=100 & SCA & M & EUR & 2 & 1 & 0 \\ 
81 & IPDCOG   & Industrial Production Index: Durable Consumer Goods     & 2021=100 & SCA & M & EUR & 2 & 1 & 0 \\ 
82 & IPNDCOG  & Industrial Production Index: Non Durable Consumer Goods & 2021=100 & SCA & M & EUR & 2 & 1 & 0 \\ 
83 & IPING    & Industrial Production Index: Intermediate Goods         & 2021=100 & SCA & M & EUR & 2 & 0 & 1 \\ 
84 & IPNRG    & Industrial Production Index: Energy                     & 2021=100 & SCA & M & EUR & 2 & 1 & 0 \\ 
\hline
85 & TRNMN    & Turnover Index: Manufacturing              & 2021=100 & SCA & M & EUR & 2 & 1 & 1 \\ 
86 & TRNCAG   & Turnover Index: Capital Goods              & 2021=100 & SCA & M & EUR & 2 & 1 & 1 \\ 
87 & TRNDCOG  & Turnover Index: Durable Consumer Goods     & 2021=100 & SCA & M & EUR & 2 & 0 & 0 \\ 
88 & TRNNDCOG & Turnover Index: Non Durable Consumer Goods & 2021=100 & SCA & M & EUR & 2 & 1 & 0 \\ 
89 & TRNING   & Turnover Index: Intermediate Goods         & 2021=100 & SCA & M & EUR & 2 & 0 & 0 \\ 
90 & TRNNRG   & Turnover Index: Energy                     & 2021=100 & SCA & M & EUR & 2 & 0 & 0 \\ 
\hline
\multicolumn{9}{c}{(8) \textbf{Prices}}\\
\hline 
91  & PPICAG   & Producer Price Index: Capital Goods                               & 2021=100 & MSA & M & EUR  & 3 & 0 & 0 \\ 
92  & PPIDCOG  & Producer Price Index: Durable Consumer Goods                      & 2021=100 & MSA & M & EUR  & 3 & 0 & 0 \\ 
93  & PPINDCOG & Producer Price Index: Non Durable Consumer Goods                  & 2021=100 & MSA & M & EUR  & 3 & 0 & 0 \\ 
94  & PPIING   & Producer Price Index: Intermediate Goods                          & 2021=100 & MSA & M & EUR  & 3 & 0 & 0 \\ 
95  & PPIFD    & Producer Price Index: Food                                        & 2021=100 & MSA & M & EUR  & 3 & 0 & 1 \\ 
96  & HICPOV   & Harmonized Index of Consumer Prices: Overall Index                & 2010=100 & SCA & M & ECB  & 3 & 0 & 1 \\ 
97  & HICPNEF  & Harmonized Index of Consumer Prices: All Items: no Energy \& Food & 2010=100 & SCA & M & ECB  & 3 & 3 & 0 \\ 
98  & HICPG    & Harmonized Index of Consumer Prices: Goods                        & 2010=100 & SCA & M & ECB  & 3 & 3 & 0 \\ 
99  & HICPSV   & Harmonized Index of Consumer Prices: Services                     & 2010=100 & SCA & M & ECB  & 3 & 3 & 0 \\
100 & HICPNG   & Harmonized Index of Consumer Prices: Energy                       & 2010=100 & MSA & M & EUR  & 3 & 3 & 0 \\
101 & HICPFD   & Harmonized Index of Consumer Prices: Food                         & 2010=100 & MSA & M & EUR  & 3 & 3 & 0 \\
102 & DFGDP    & Real Gross Domestic Product Deflator                              & 2015=100 & SCA & Q & EUR  & 3 & 0 & 0 \\  
103 & HPRC     & Residential Property Prices (BIS)                                 & MLN\euro & SCA & Q & FRED & 3 & 0 & 0 \\ 
104 & POIL     & Crude Oil Prices: Brent - Europe                            & \euro/{barrel} & MSA & Q & FRED & 3 & 3 & 0 \\ 
105 & PNGAS    & Global price of Natural gas, EU                             & \euro/{MMbtu}  & MSA & Q & FRED & 3 & 3 & 0 \\ 
\hline
\multicolumn{9}{c}{(9) \textbf{Confidence Indicators}}\\
\hline
106 & ICONFIX  & Industrial Confidence Indicator  			  & Index    & SA & M & EUR  & 0 & 0 & 1 \\ 
107 & CCONFIX  & Consumer Confidence Indicator       			  & Index    & SA & M & EUR  & 0 & 0 & 0 \\ 
108 & ESENTIX  & Economic Sentiment Indicator      			  & Index    & SA & M & EUR  & 0 & 0 & 1 \\ 
109 & KCONFIX  & Construction Confidence Indicator 			  & Index    & SA & M & EUR  & 0 & 0 & 1 \\ 
110 & RTCONFIX & Retail Confidence Indicator       			  & Index    & SA & M & EUR  & 0 & 0 & 0 \\ 
111 & SCONFIX  & Services Confidence Indicator        		  & Index    & SA & M & EUR  & 0 & 1 & 1 \\ 
112 & BCI      & Cyclically-Adjusted Business Confidence Index & 2010=100 & SA & M & OECD & 0 & 0 & 1 \\ 
113 & CCI      & Cyclically-Adjusted Consumer Confidence Index & 2010=100 & SA & M & OECD & 0 & 0 & 0 \\ 
\hline
\multicolumn{9}{c}{(10) \textbf{Monetary Aggregates}}\\
\hline
114 & CURR & Money Stock: Currency & MLN\euro & SCA & M & ECB & 2 & 1 & 1 \\
115 & M1   & Money Stock: M1       & MLN\euro & SCA & M & ECB & 2 & 1 & 1 \\ 
116 & M2   & Money Stock: M2       & MLN\euro & SCA & M & ECB & 2 & 1 & 1 \\ 
\hline
\multicolumn{9}{c}{(11) \textbf{Others}}\\
\hline
117 & SHIX  & Share Prices                   & 2010=100 & SA  & M & OECD & 2 & 0 & 1  \\ 
118 & CAREG & Passenger's Cars Registrations & 1000U   & SCA & M & ECB  & 2 & 0 & 1  \\ 
\hline
\hline
\end{tabular}%
}
\end{footnotesize}
\end{table}

\newpage
\newgeometry{left=2.5cm,right=2.5cm,top=2.5cm,bottom=2.5cm}

%
%
\setcounter{table}{0}
\setcounter{figure}{0}
\setcounter{equation}{0}
\section{Assumptions}\label{app:ass}
In this section, we state all the formal assumptions underlying the model outlined in Section \ref{sec::mod}, and we provide both econometric and economic justifications for these assumptions. Throughout the text, we let $t_{\scaleto{19\text{Q4}}{4pt}}$,  $t_{\scaleto{20\text{Q1}}{4pt}}$ and $t_{\scaleto{21\text{Q4}}{4pt}}$ denote 2019:Q4, 2020:Q1 and 2021:Q4, respectively.

\subsubsection*{Identifying assumptions for the space spanned by the factors.}

\begin{enumerate}[(1)]
\item The number of factors $q$ is such that $q<n$ and is independent of $n$.

\item The $q$-dimensional vector  $\mathbf f_t$ is such that $\E[\Delta\mathbf f_t]=\mathbf 0$ and $\E[\Delta \mathbf f_t\Delta \mathbf f_t^\prime]=\mathbf I$.

\item The $n \times q$ matrix  
$\boldsymbol{\Lambda}= \left(\bm{\lambda}_{1} \cdots \bm{\lambda}_{n}\right)^{\prime}$, with $\bm\lambda_i=(\lambda_{i1}\cdots\lambda_{iq})^\prime$, $1\le i\le n$, is such that $\lim_{n \to \infty} n^{-1}{\boldsymbol{\Lambda}'\boldsymbol{\Lambda}}= \mathbf{H}$ positive definite.

\item The scalar  $g_t$ is such that $\E[\Delta g_t]=0$ and $\E[(\Delta g_t)^2]=1$.

\item The $n$-dimensional vector $\boldsymbol{\gamma}= \left({\gamma}_{1} \cdots \gamma_{n}\right)^{\prime}$, is such that $\lim_{n \to \infty} n^{-1}{\boldsymbol{\gamma}'\boldsymbol{\gamma}}>0$.

\item The $q$-dimensional vector $\mathbf{u}_t$ is such that $\E[\mathbf u_t]=\mathbf 0$, $\E[\mathbf u_t\mathbf u_{t}^\prime]=\bm\Sigma_u$ is positive definite, and $\E[\mathbf u_t\mathbf u_{t-k}^\prime]=\mathbf 0$ for $k\ne 0$. Moreover, $s_t$ is deterministic with $s_t>0$.

\item The idiosyncratic innovations $e_{it}$, $1\le i\le n$, are such that $\E[e_{it}]=0$, $\E[e_{it}^2]=\sigma_{e_i}^2>0$ for all $t$. Moreover, there exist
 finite constants $M_{ij}>0$ independent of $t$ and $0<\rho<1$ independent of $t$, $i$, and $j$ such that 
$\vert \E[e_{it}e_{j,t-k}] \vert\le M_{ij} \rho^{\vert k\vert}$ for all $k\in\mathbb Z$, with $\sum_{i=1, i\ne j}^n M_{ij}\le M$ and $\sum_{j=1, j\ne i}^n M_{ij}\le M$ for some finite constant $M>0$ independent of $i$, $j$, and $n$.

\item $\mathbb{E}[e_{it}\mathbf u_{s}]=\mathbf 0$, for all $i,t,s$.
\end{enumerate}

Assumptions (1)-(3) require the $q$ factors $\mathbf{f}_t$ to be pervasive so that they have a non-negligible effect on the variables of interest \citep{bai2004panic,barigozzi_large-dimensional_2021}. Following \citet{stockcomovement}, these assumptions are extended in parts (4) and (5) to the Covid factor, $g_t$, where pervasiveness stems from the common nature of the Covid shock affecting most of the series included in the dataset. Including both 2020 and 2021 in the Covid period is consistent with the evolution of the pandemic in Europe.

Assumption (6) assumes white noise innovations whose volatility changes over time after the Covid shock---time-varying volatility is not a prominent feature in the pre-2020 sample \citep{jarocinski2018inflation}. Accounting for the change in volatility due to Covid has proven to be fundamental both for estimation and forecasting, and here we adopt an approach similar to \cite{lenza_how_2022} by introducing a scaling term $s_t$ modeled independently for each period starting from 2020:Q1. \cite{lenza_how_2022} analyze monthly US data and impose an exponential decay for $s_t$ starting in June 2020. In contrast, we estimate one parameter for each period starting in 2020:Q1 because many series exhibit large variation even after the first half of 2020, which is not surprising given that (i) mobility restriction measures in the EA were much more restrictive than in the US, lasted for longer, and were also implemented in 2021, and (2) the Russia-Ukraine war had a much larger impact on Europe by pushing natural gas prices (and gasoline prices to a lesser extent) to the roof, and in creating a lot of macro-financial uncertainty. Moreover, as \citet{morley_estimating_2023} pointed out, quarterly data do not allow for a sharp identification of the decay parameter. 

\indent Assumption (7) allows the idiosyncratic innovations to be mildly cross-sectionally correlated and serially correlated with summable autocovariances, thus compatible with stationary ARMA dynamics \citep{bai2004panic,barigozzi_large-dimensional_2021}. Last, Assumption (8) requires the idiosyncratic and factor innovations to be uncorrelated at all leads and lags, a requirement consistent with the idea of global macroeconomic shocks being unrelated to local dynamics.

\subsubsection*{Assumptions on the dynamic specifications on the non-stationary idiosyncratic components and the secular components.}
\begin{enumerate}
  \item [(9)] Let $\mathcal I_1$ be the set of indexes such that $\xi_{it}\sim I(1)$ if $i\in\mathcal I_1$, then $n_I=\#\{i: i\in\mathcal I_1\}$ is such that $0<n_I<n$.
  \item [(10)] Let $\mathcal I_b$ be the set of indexes such that $b_{it}\ne 0$ if $i\in\mathcal I_b$, then $n_B=\#\{i:\ i\in\mathcal I_b\}$ is such that $0<n_B<n$. Moreover, $\mathrm{D}_{i0}= a_i\ne 0$, for all $i$.

  \item [(11)] Let $\widehat{\sigma}^2_{\Delta y_i}$ and $\widehat{\sigma}^2_{y_i}$ be the sample variances of $\Delta y_{it}$ and $y_{it}$ respectively, computed for $1\le t \le t_{\scaleto{19\text{Q4}}{4pt}}$ and $t_{\scaleto{21\text{Q4}}{4pt}}+1 \le t \le T$.
  \begin{enumerate}
    \item[(a)] Let $\mathcal{L}_1:=\{\text{\small GDP, HHLB, HHLB.LLN} \}$ be the set of indexes such that for $i\in \mathcal{L}_1$ we have ${\sigma}^2_{\eta_i}\ne 0$, then  $\E[\eta_{it}]=0$ and we set ${\sigma}^2_{\eta_i}=(1600\widehat{\sigma}^2_{\Delta y_i})^{-1}$. 
    \item[(b)] Let $\mathcal{L}_0:=\{\text{\small UNETOT, HICPOV, HICPNEF, HICPG, HICPSV, HICPFD, POIL, PNGAS} \}$ be the set of indexes such that for $i\in \mathcal{L}_0$ we have $\sigma^2_{\epsilon_i}\ne 0$, then $\E[\epsilon_{it}]=0$ and we set $\sigma^2_{\epsilon_i}=(800\widehat{\sigma}^2_{y_i})^{-1}$.
  \end{enumerate}
\end{enumerate}

Assumption (9) allows the idiosyncratic component to be $I(1)$ for some, but not all, of the series. This assumption is crucial when estimating the model on a large dataset. Imposing the assumption of all idiosyncratic components being $I(0)$ would be overly restrictive, as it implies cointegration for any $q$-dimensional vector of series \citep{barigozzi_large-dimensional_2021}. While cointegration may hold for certain series, it is highly unlikely to hold for many others. To accommodate potential cointegration, we allow only a limited number $n_I$ of variables to possess a non-stationary idiosyncratic component. Our dataset, where only $n_I=57$ out of $n=118$ series exhibit a non-stationary idiosyncratic component, supports this assumption.

Assumption (10) allows for a non-stationary secular component for some, but not all, of the variables in the dataset. This modeling choice is coherent with the properties of a standard macroeconomic dataset. Specifically, variables related to the real sector of the economy, such as consumption or investments, commonly display a distinct (upward) trend. Conversely, this may not hold for other variables, such as inflation rates or interest rates, for example. This intuition finds support in the empirical data, where only $n_B=58$ out of $n=118$ series exhibit a linear trend. 

Assumptions (10) and (11a) imply that GDP, households' financial liabilities and long-term loans, have a secular component given by the local linear trend model
\begin{align}
\mathrm D_{it} &= a_i + b_{it},\qquad b_{it} = b_{i,t-1}+\eta_{it}, \qquad i\in \mathcal{L}_1,
\end{align}
where $a_i=\mathrm D_{i0}$. 

We introduce a local-linear trend for GDP to capture the gradual drift in the secular decline in long-run output growth documented both for the US and the EA \citep{cette2016pre,antolin-diaz_tracking_2017, gordon2018declining}. The literature has identified several factors contributing to this slowdown, with particular emphasis on declining productivity growth. This decline has been more pronounced in the EA due to heterogeneity between core and peripheral countries, as peripheral countries are experiencing a larger misallocation of economic resources \citep{cette2016pre}. Therefore, it is crucial to accurately account for these features to assess GDP's long-run dynamics. This assessment is essential for estimating potential output, as it avoids spuriously inflating the output gap with unexplained predictable variation \citep{ng2018comments}.

We introduce a local-linear trend for household financial liabilities and long-term loans, which constitute about 85\% of total household liabilities, to capture the slowdown in their average growth rates that occurred since the GFC. 

Assumptions (10) and Assumption (11b) implies that the unemployment rate, all consumer price inflation indexes, oil and natural gas prices have a secular component given by the local level model
\begin{align}
\mathrm D_{it} &= a_{it},\qquad a_{it} = a_{i,t-1}+\epsilon_{it},\qquad i\in\mathcal{L}_0,
\end{align}
This specification captures relevant labor and demographic factors that may affect the unemployment rate secular trend, such as, for example, the aging of the population and the misallocation of resources in the labor market due to \enquote{soft budget constraints} or stringent labor market policies can lead to a mismatch between employers’ needs and the skill-set of the unemployed \citep{cette2016pre}. Similarly, this specification also allows us to account for the slowdown in inflation occurred after the GFC. 

All the other variables in the dataset have either a deterministic linear trend or a constant mean, i.e,
$\mathrm D_{it} = a_{i}+ b_i t$ if $ i\in\mathcal I_b$, or
$\mathrm D_{it} = a_{i}$ otherwise. Although it is technically possible to model a time-varying component for all the variables in the dataset, such an approach would introduce complexities in the estimation framework, with the number of latent states increasing linearly with the number of series. 

In Assumption (11) we fix the variances of the stochastic secular components following \cite{del2017safety} in order to effectively capture the gradual and persistent nature of the secular trends. This specification implies that when $i\in \mathcal{L}_1$, the standard deviation of the secular trend is approximately $1\%$ over 100 years, while when $i\in \mathcal{L}_0$, the standard deviation of the secular trend is approximately $1\%$ over 50 years. This choice consistent with the notion of a slow-moving secular component.

\subsubsection*{Assumptions on the dynamics of the factors, trend, and cycles.}

\begin{enumerate}
\item  [(12)] The polynomial {\normalfont $\textrm{det}( \mathbf{I} - \sum_{j=1}^{p}\mathbf A_j z^{j})=0$} has 1 root in $z=1$ and the remaining $q-1$ roots in $\abs{z}>1$.
\item [(13)] The $q$-dimensional vector $\boldsymbol{\psi}$ is such that $\boldsymbol{\beta}^\prime\boldsymbol{\psi}=\mathbf 0$, where $\boldsymbol{\beta}$ is the $q\times (q-1)$  matrix having as columns the cointegrating vectors of $\mathbf f_t$, i.e., such that $\boldsymbol{\beta}^\prime \mathbf f_t$ is weakly stationary.
\item [(14)] The $q$-dimensional vector $\boldsymbol{\omega}_t$ is weakly stationary and such that $\E[\bm\omega_t]=\mathbf 0$ and $\E[\bm\omega_t\bm\omega_t^\prime]=\bm\Sigma_\omega$ is positive definite.
\item [(15)] The scalar $\nu_t$ is such that $\E[\nu_t]=0$ and $\E[\nu_t^2]=\sigma^2_\nu>0$.
\item  [(16)] $\E[\nu_t \bm\omega_t]=\mathbf 0$ for all $t$.
\end{enumerate}

Assumption (12) imposes that 1 common trend drives the non-stationarity in the common factors, hence, that the factors are cointegrated with $q-1$ cointegrating relations---our data provide strong support for the presence of just one common trend. This is a standard assumption in the literature, which often assumes that common productivity trend is the sole driver of long-run economic growth \citep[see, e.g.,][]{del2007fit}. 

Assumptions (13) and (14) imply that $\bm\omega_t$, defined in \eqref{sbeq::obsTR}, belongs to the cointegration space of the common factors. This view is consistent with theoretical models assuming that the output gap represents deviations from long-run equilibria determined by a common productivity trend \citep{del2007fit}. 

Assumption (15) assumes that $\nu_t$ is a stochastic process---hence $\tau_t$ is a common stochastic trend---but it does not constraint $\nu_t$ to be a white noise---hence $\tau_t$ to be a random walk. Indeed, our estimates suggest that $\nu_t$ is autocorrelated, in line with the theoretical arguments by \citet{lippi_diffusion_1994}.

Finally, Assumption (16) implies contemporaneous orthogonality between potential output and the output gap, which is also assumed in the non-parametric approaches used by \citet{barigozzi_measuring_2021}.

Then, the extended state-space form of the model is given by:
\begingroup
\allowdisplaybreaks
\begin{subequations}
\begin{align}
& y_{it}\ =\ \mathrm{D}_{it} + \bm\lambda_{i}^\prime\mathbf {f}_{t} + \gamma_i\hspace{1pt} g_t \mathbb I_{\,\text{\tiny $ t_{\scaleto{20\text{Q1}}{4pt}}\!\le\! t\!\le\! t_{\scaleto{21\text{Q4}}{4pt}}$}} +  \zeta_{it} + z_{it},  \hspace{26pt} 1\le i\le n,\ 1\le t \le T, \label{eq::stsp1} \\
& \mathbf{f}_t\ =\sum_{j=1}^{p}\mathbf A_j\mathbf{f}_{t-j} +  \{s_t \mathbb I_{\,\text{\tiny $t\!\ge\! t_{\scaleto{20\text{Q1}}{4pt}}$}} + (1- \mathbb I_{\,\text{\tiny $t\!\ge\! t_{\scaleto{20\text{Q1}}{4pt}}$} })\}\mathbf{u}_t,\hspace{22pt} \mathbf{u}_t\stackrel{i.i.d.}{\sim} (\mathbf{0},\bs{\Sigma}_u), \\
& \mathrm{D}_{i t}=\left\{\begin{array}{lll}
a_i+b_i t & \text { if } i \in \mathcal{I}_b, & \\
\mathrm{D}_{i t-1}+b_{i t-1}, \hspace{3pt} b_{i t}=b_{i t-1}+\eta_{it} & \text { if } i\in\mathcal{L}_1, & \eta_{it}{\sim} (0, \sigma_{\eta_{i}}^2), \\
\mathrm{D}_{i t-1}+\epsilon_{i,t} & \text { if } i\in\mathcal{L}_0, & \epsilon_{it} {\sim} (0, \sigma_{\epsilon_i}^2),\\
a_i & \text { otherwise}, & 
\end{array}\right. \\[3pt]
& \zeta_{i t}=\left\{ \begin{array}{lll}
\xi_{i t},  \hspace{3pt} \xi_{it}=\xi_{i,t-1}+e_{it}& \hspace{50pt}\text { if } i \in \mathcal{I}_1, & \hspace{6pt} e_{it} {\sim} (0, \sigma_{e_i}^2),\\
0  & \hspace{50pt}\text { if } i \notin \mathcal{I}_1, & \\
\end{array}
\right. \\[3pt]
& z_{it}=\left\{ \begin{array}{lll}
z_{it}^*& \hspace{128pt}\text { if } i \in \mathcal{I}_1, & \hspace{5pt} z_{it}^*\stackrel{i.i.d.}{\sim}(0,R_i), \\
e_{it} \color{black}  & \hspace{128pt}\text { if } i \notin \mathcal{I}_1, & \hspace{5pt} e_{it}{\sim} (0, R_i), \\
\end{array}
\right. \\[3pt]
& R_{i}=\left\{ \begin{array}{lll}
\sigma^2_{z} & \hspace{126pt}\text { if } i \in \mathcal{I}_1, &\\
\sigma^2_{e_i} \color{black}  & \hspace{126pt}\text { if } i \notin \mathcal{I}_1. & \\
\end{array}
\right. 
\label{eq::stspF}
\end{align}
\end{subequations}
\endgroup

As defined in Assumption (9), $\mathcal{I}_1$ denotes the set of series with an $I(1)$ idiosyncratic component. As defined in Assumption (10), $\mathcal{I}_b$ denotes the set of series with a deterministic linear trend. Finally, as defined in Assumption (11), $\mathcal{L}_1$ denotes the set of series with a time-varying trend modeled as a local-linear trend, while $\mathcal{L}_0$ denotes the set of series with a time-varying mean modeled as a random walk.

%
%
\setcounter{table}{0}
\setcounter{figure}{0}
\setcounter{equation}{0}
\section{Estimation in detail} \label{sec::estdetail}

In this section, we provide details on the estimation procedure described in Section \ref{sec::fattoni}.

\subsection*{Estimating the dynamic factor model}

\subsubsection*{Initialization}
In order to apply the Kalman filter and smoother, we need initial estimates of all the quantities described in Equations \eqref{eq::stsp1}-\eqref{eq::stspF}, with the exception of $\sigma_{\eta_i}^2$, $\sigma_{\epsilon_i}^2$, both set as in Assumption 11, and $\sigma_z^2$ set to $10^{-2}$, as suggested by \citet{opschoor2023slow} who show that smaller values might be detrimental for the performance of the algorithm.
 
We denote with the superscript \enquote{19} all quantities computed with data up to 2019. Let $\breve{y}_{it}^{\scaleto{19}{4pt}} = (y_{t}^{\scaleto{19}{4pt}} - \breve{a}_i^{\scaleto{(0),19}{5pt}} - \breve{b}_i^{\scaleto{(0),19}{5pt}}\cdot t)/\widehat{\sigma}_{\Delta y_i^{\scaleto{19}{3pt}}}^2$, where $\widehat{\sigma}_{\Delta y_i^{\scaleto{19}{3pt}}}^2$ is the sample variance of $\Delta y_{it}^{\scaleto{19}{4pt}}$, and $\breve{a}_i^{\scaleto{(0),19}{5pt}}$ and $\breve{b}_i^{\scaleto{(0),19}{5pt}}$ are estimated by regressing $y_{it}^{\scaleto{19}{4pt}}$ on a constant and a time trend, whenever $i\in \mathcal{I}_b$ or $i\in\mathcal{L}_1$. If $i\in \mathcal{I}_a$ or $i \in\mathcal{L}_0$ we let $\breve{y}_{it} = y_{it}^{\scaleto{19}{4pt}}-\breve{a}_i^{\scaleto{(0),19}{5pt}}$, where $\breve{a}_i^{\scaleto{(0),19}{5pt}}$ is the sample average of $y_{it}$. The standardized slopes are denoted as $\widehat{b}_i^{\scaleto{(0),19}{5pt}} = \breve{b}_i^{\scaleto{(0),19}{5pt}}/\widehat{\sigma}^2_{\Delta y_{i}^{\scaleto{19}{3pt}}}$. We initialize the loadings using the estimator of \cite{barigozzi_large-dimensional_2021}:  the $n\times q$ matrix of estimated loadings $\widehat{\bs{\Lambda}}^{\scaleto{(0),19}{5pt}} = (\widehat{\bs{\lambda}}^{\scaleto{(0),19}{5pt}}_1,\ldots, \widehat{\bs{\lambda}}^{\scaleto{(0),19}{5pt}}_q)'$ is obtained by principal components on the standardized first differences of the data, i.e. $(\Delta y_{it}^{\scaleto{19}{4pt}}-\overline{\Delta y_i^{\scaleto{19}{4pt}}})/\widehat{\sigma}_{\Delta y_i^{\scaleto{19}{3pt}}}^2$, where $\overline{\Delta y_i^{\scaleto{19}{4pt}}}$ is the sample mean of $\Delta y_{it}^{\scaleto{19}{4pt}}$. Given the loadings, we also obtain a first estimate of the $q$ common factors, $\widehat{\mathbf{f}}_t^{\scaleto{(0),19}{5pt}} = n^{-1}\widehat{\bs{\Lambda}}^{\scaleto{(0),19}{5pt}'}\breve{\mathbf{y}}^{\scaleto{19}{4pt}}_t$
and of the idiosyncratic components, $\widehat{\xi}_{it}^{\scaleto{(0),19}{5pt}} = \breve{y}_{it}^{\scaleto{19}{4pt}} - \widehat{\bs{\lambda}}_i^{\scaleto{(0),19}{5pt}'}\widehat{\mathbf{f}}_t^{\scaleto{(0),19}{5pt}}$. Furthermore, we obtain $ \widehat{\mathbf{A}}_j^{\scaleto{(0),19}{5pt}}$, $j=1,\ldots, p$, by fitting a VAR($p$) on $\widehat{\mathbf{f}}_t^{\scaleto{(0),19}{5pt}}$. Given the residuals $\widehat{\mathbf{u}}_t^{\scaleto{(0),19}{5pt}} = \widehat{\mathbf{f}}_t^{\scaleto{(0),19}{5pt}} - \widehat{\mathbf{A}}^{\scaleto{(0),19}{5pt}}\widehat{\mathbf{f}}_{t-1}^{\scaleto{(0),19}{5pt}}$, where $\widehat{\mathbf{A}}^{\scaleto{(0),19}{5pt}}$ is the companion form representation of the autoregressive matrices $\widehat{\mathbf{A}}_1^{\scaleto{(0),19}{5pt}}, \ldots, \widehat{\mathbf{A}}_p^{\scaleto{(0),19}{5pt}}$, an estimate of the latent covariance is given by $\widehat{\bs{\Sigma}}_u^{\scaleto{(0),19}{5pt}} = \widehat{\Cov}(\widehat{\mathbf{u}}_t^{\scaleto{(0),19}{5pt}})$, where $\widehat{\Cov}$ is the sample covariance matrix.
Finally, when $i\notin \mathcal{I}_1$ we set $\widehat{R}_i^{\scaleto{(0),19}{5pt}} = (\widehat{\sigma}_{e_i}^{\scaleto{(0),19}{5pt}})^2 = \widehat{\Var}(\widehat{\xi}_{it}^{\scaleto{(0),19}{5pt}})$, where $\widehat{\Var}$ is the sample variance. And when $i\in\mathcal I_1$ we set $(\widehat{\sigma}_{e_i}^{\scaleto{(0),19}{5pt}})^2 = \widehat{\operatorname{Var}}(\Delta \widehat{\xi}_{i t}^{\scaleto{(0),19}{5pt}})$. 

\begin{table}[h!] \caption{Initialization of states for the Kalman filter}
\centering
\begin{tabular}{l c}\hline  \hline\\[-8pt]
$\mathbf{f}_{0 \mid 0}^{\scaleto{19}{4pt}}= \widehat{\mathbf{f}}_0^{\scaleto{(0),19}{5pt}}$\\[2pt] $\operatorname{vec}\left(\mathbf{P}_{0 \mid 0}^{\scaleto{19}{4pt}}\right)=\left(\mathbf{I}_{pq^2}-\widehat{\mathbf{A}}^{\scaleto{(0),19}{5pt}} \otimes \widehat{\mathbf{A}}^{\scaleto{(0),19}{5pt}}\right)^{-1} \operatorname{vec}\left(\widehat{\boldsymbol{\Sigma}}_u^{\scaleto{(0),19}{5pt}}\right)$ \\[-2pt] & \\  \hline\\[-8pt]
$\begin{array}{l}
\mathrm{D}_{i, 0 \mid 0}^{\scaleto{19}{4pt}}=\widehat{b}_i^{\scaleto{(0),19}{5pt}} \\[2pt]
\mathrm{D}_{i, 0 \mid 0}^{\scaleto{19}{4pt}}=0\\[2pt]
\mathrm{D}_{i, 0 \mid 0}^{\scaleto{19}{4pt}}=0 \\[2pt]
b_{i,0\giv 0}^{\scaleto{19}{4pt}} = \widehat{b}_i^{\scaleto{(0),19}{5pt}} \end{array}$ & $\begin{array}{l}
\text { if } i \in \{\mathcal{I}_b, \mathcal{L}_1\} \\[2pt]
\text { if } i\in \mathcal{L}_0 \\[2pt]
\text { if } i \notin \{\mathcal{I}_b, \mathcal{L}_0, \mathcal{L}_1 \}\\[2pt]
\text { if } i \in \{\mathcal{I}_b, \mathcal{L}_1\} \end{array}$\\ \\[-8pt] \hline\\[-8pt]
$\begin{array}{l}
P_{i, 0 \mid 0}^{\scaleto{D(19)}{5pt}}=\frac{1}{(1-0.99)^2} \sigma_{\eta_{i}}^{2} \\[2pt]
P_{i, 0 \mid 0}^{\scaleto{D(19)}{5pt}}={\sigma}_{\epsilon_i}^{2}\\[2pt]
P_{i, 0 \mid 0}^{\scaleto{D(19)}{5pt}}=0 \\[2pt]
P_{i, 0 \mid 0}^{\scaleto{b(19)}{5pt}}=\frac{1}{(1-0.99)^2} \sigma_{\eta_{i}}^{2}
\end{array}$ &\hspace{-18pt} $\begin{array}{l}
\text { if } i\in \mathcal{L}_1\\
\text { if } i\in \mathcal{L}_0\\[2pt]
\text { if } i \notin\left\{\mathcal{L}_0, \mathcal{L}_1\right\} \\[2pt]
\text { if } i\in \mathcal{L}_1\end{array}$\\ \\[-8pt] \hline\\[-8pt]
$\begin{array}{l}\zeta_{i, 0 \mid 0}^{\scaleto{19}{4pt}}=\widehat{\xi}_{i 1}^{\scaleto{(0),19}{5pt}} \\[2pt]
P_{i, 0 \mid 0}^{\scaleto{\zeta(19)}{5pt}}=\frac{1}{(1-0.99)^2} \widehat{\operatorname{Var}}\left(\Delta \widehat{\xi}_{i t}^{\scaleto{(0),19}{5pt}}\right)\end{array}$ & \hspace{-43pt}$\begin{array}{l}\text { if } i \in \mathcal{I}_1 \\[2pt]
\text { if } i \in \mathcal{I}_1\end{array}$\\\\[-8pt] \hline\hline
\end{tabular}
\label{tab::initS}

\end{table}

Table \ref{tab::initS} provides an overview of the initial values of the states for the Kalman filter. 
\subsubsection*{Step 1: Estimate the model up to 2019:Q4 (pre-Covid step)}
Given the initial values of the parameters and the states, we run the Kalman filter and smoother using a standardized version of the data in levels, up to 2019, that is:
\begin{equation*}
\tilde{y}_{it}^{19}\ =\ \left\{\begin{array}{lll}\frac{y_{i t}^{19}-\check{a}_i^{19}}{\widehat{\sigma}_{\Delta y_i^{\scaleto{19}{2pt}}}^{2}} & & \text { if } \hspace{2pt} i \in\left\{\mathcal{I}_b, \mathcal{L}_1\right\} \\ \frac{y_{i t}^{19}-\bar{y}_i^{19}}{\widehat{\sigma}_{\Delta y_i^{\scaleto{19}{2pt}}}^2} & & \text { otherwise }  \end{array}\right.
\end{equation*}
Given $\tilde{\mathbf{y}}_{t}^{\scaleto{19}{4pt}} = (\tilde{y}_{1,t},\ldots,\tilde{y}_{n,t})'$, we obtain a new estimate of the states, namely the factors $\mathbf{f}_{t\giv T}^{\scaleto{19}{4pt}}$, the time-varying secular components $\mathrm{\mathbf{D}}_{t\giv T}^{\scaleto{19}{4pt}}$  and slopes $\bs{\mathrm{b}}_{t\giv t}^{\scaleto{19}{4pt}}$,  and the non-stationary idiosyncratic components $\bs{\zeta}_{t\giv T}^{\scaleto{19}{4pt}}$ , along with the corresponding conditional covariances. 

Given the smoothed states, we estimate all the parameters as follows:
\begin{itemize}[-]
\item \textsc{Factor loadings}:
\begin{equation*}
\widehat{\boldsymbol{\lambda}}_i^{\scaleto{19}{4pt}'}\ =\ \left(\sum_{t=1}^T\left(\tilde{y}_{it}^{\scaleto{19}{4pt}}-\mathrm{D}_{i,t \giv T}^{\scaleto{19}{4pt}}-\zeta_{i,t \giv T}^{\scaleto{19}{4pt}}\right) \mathbf{f}_{t \giv T}^{\scaleto{19}{4pt}'}\right)\left(\sum_{t=1}^T \mathbf{f}_{t \giv T}^{\scaleto{19}{4pt}} \mathbf{f}_{t \giv T}^{\scaleto{19}{4pt}'} + \mathbf{P}_{t\giv T}^{\scaleto{19}{4pt}}\right)^{-1}
\end{equation*}
\item \textsc{Parameters of the law of motion of the common factors}:
    \begin{align*}
        \widehat{\mathbf{A}}^{\scaleto{19}{4pt}}\ &=\ \parT{\sum_{t=2}^T \mathbf{f}_{t\mid T}^{\scaleto{19}{4pt}}\mathbf{f}_{t-1\mid T}^{\scaleto{19}{4pt}'} + \mathbf{P}_{t,t-1\giv T}^{\scaleto{19}{4pt}}}\parT{\sum_{t=2}^T \mathbf{f}_{t-1\mid T}^{\scaleto{19}{4pt}}\mathbf{f}_{t-1\mid T}^{\scaleto{19}{4pt}'}+ \mathbf{P}_{t-1\giv T}^{\scaleto{19}{4pt}}}^{-1} \\[8pt]
\widehat{\bs{\Sigma}}_u^{\scaleto{19}{4pt}}\ &=\ \frac{1}{T}\parT{\sum_{t=2}^T \parT{\mathbf{f}_{t\mid T}^{\scaleto{19}{4pt}}\mathbf{f}_{t\mid T}^{\scaleto{19}{4pt}'} + \mathbf{P}_{t\giv T}^{\scaleto{19}{4pt}}} - \widehat{\mathbf{A}}^{\scaleto{19}{4pt}}\sum_{t=2}^T\parT{\mathbf{f}_{t\mid T}^{\scaleto{19}{4pt}}\mathbf{f}_{t-1\mid T}^{\scaleto{19}{4pt}'}+ \mathbf{P}_{t,t-1\giv T}^{\scaleto{19}{4pt}}}}
    \end{align*}
\item \textsc{Slopes of secular trend}:
\begin{equation*}
        \widehat{b}_i^{\scaleto{19}{4pt}}\ =\ \left(\sum_{t=1}^T\left(\tilde{y}_{i t}^{\scaleto{19}{4pt}}-\widehat{\boldsymbol{\lambda}}_i^{\scaleto{19}{4pt}'} \mathbf{f}_{t \mid T}^{\scaleto{19}{4pt}}-\zeta_{i, t \mid T}^{\scaleto{19}{4pt}}\right) t\right)\left(\sum_{t=1}^T t^2\right)^{-1}
    \end{equation*}
\item \textsc{Variance of $I(1)$ idiosyncratic components}:
\begin{align*}
\widehat{\sigma}_{e_i}^{2, \scaleto{19}{4pt}}\ =\ &\frac{1}{T} \sum_{t=2}^T\left(\zeta_{i, t \mid T}^{\scaleto{19}{4pt}} \zeta_{i, t \mid T}^{\scaleto{19}{4pt}'} + P_{i,t\giv T}^{\scaleto{\zeta(19)}{6pt}}\right)+\frac{1}{T} \sum_{t=2}^T\left(\zeta_{i t-1 \mid T}^{\scaleto{19}{4pt}} \zeta_{i t-1 \mid T}^{\scaleto{19}{4pt}'} + P_{i,t-1\giv T}^{\scaleto{\zeta(19)}{6pt}}\right) -\\[4pt]
& - \frac{2}{T} \sum_{t=2}^T\left(\zeta_{i t \mid T}^{\scaleto{19}{4pt}} \zeta_{i t-1 \mid T}^{\scaleto{19}{4pt}'} + P_{i,t,t-1\giv T}^{\scaleto{\zeta(19)}{6pt}}\right)
\end{align*}
\item \textsc{Covariance prediction error}:
\begin{align*}
 \widehat{R}_i^{\scaleto{19}{4pt}}\ =\ &\frac{1}{T} \sum_{t=1}^T\left\{\left(\tilde{y}_{i,t}^{\scaleto{19}{4pt}}-\widehat{\boldsymbol{\lambda}}_i^{\scaleto{19}{4pt}'} \mathbf{f}_{t \mid T}^{\scaleto{19}{4pt}}- \mathbb{I}_{i \in \mathcal{I}_b} \mathrm{D}_{i, t \mid T}^{\scaleto{19}{4pt}}- \mathbb{I}_{i \in \mathcal{I}_1}\zeta_{i, t \mid T}^{\scaleto{19}{4pt}}\right)^2 + \widehat{\boldsymbol{\lambda}}_i^{\scaleto{19}{4pt}'} \mathbf{P}_{t,T\giv T}^{\scaleto{19}{4pt}}\widehat{\boldsymbol{\lambda}}_i^{\scaleto{19}{4pt}} + \right.\\[4pt]
& + \left. \mathbb{I}_{i\in \mathcal{I}_b} P_{i,t\giv T}^{\scaleto{\mathrm{D}(19)}{6pt}} + \mathbb{I}_{i\in \mathcal{I}_1} P_{i,t\giv T}^{\scaleto{\zeta(19)}{6pt}} \right\}
\end{align*}
\end{itemize}
\subsubsection*{Step 2: Estimate the Covid factor and volatility (Covid step)}\label{sapp::step2}
Given the estimated parameters up to 2019:Q4, we run the Kalman filter and smoother using standardized data in levels for the entire sample, that is:
\begin{equation*}
\tilde{y}_{it}\ =\ \left\{\begin{array}{lll}\frac{y_{i t}-\check{a}_i}{\widehat{\sigma}_{\Delta y_i}^2} & & \text { if } \hspace{2pt} i \in\left\{\mathcal{I}_b, \mathcal{L}_1\right\} \\ \frac{y_{i t}-\bar{y}_i}{\widehat{\sigma}_{\Delta y_i}^2} & & \text { otherwise }  \end{array}\right.
\end{equation*}\\
In computing $\bar{y}_i$, $\widehat{\sigma}^2_{\Delta y_i}$ and $\breve{a}_i$, we treat Covid outliers as missing values. Given $\tilde{\mathbf{y}}_t = (\tilde{y}_{1,t},\ldots,\tilde{y}_{n,t})'$, we obtain the estimated states given the pre-Covid parameters. In doing so, we truncate the Kalman smoother in correspondence of 2020:Q1, to avoid spurious backward effects from the presence of Covid outliers. The estimated states are denoted as $\mathbf{f}_{t\giv T}^{\scaleto{(0)}{6pt}}$, $\mathrm{\mathbf{D}}_{t\giv T}^{\scaleto{(0)}{6pt}}$, $\mathbf{b}_{t\giv T}^{\scaleto{(0)}{6pt}}$ and $\bs{\zeta}_{t\giv T}^{\scaleto{(0)}{6pt}}$.\\
\indent Given the smoothed states, let:
\begin{equation*}
\widehat{\bs{\xi}}_t\ =\ \tilde{\mathbf{y}}_t - \widehat{\boldsymbol{\Lambda}}^{\scaleto{19}{4pt}}\mathbf{f}_{t\mid T}^{\scaleto{(0)}{6pt}} - \mathbf{D}_{t\mid T}^{\scaleto{(0)}{6pt}}
\end{equation*}
and denote as $\widehat{\bs{\Xi}} = (\bs{\xi}_1,\ldots, \bs{\xi}_t)'$ the $T\times n$ matrix of idiosyncratic components. Then we estimate the Covid factor by estimating the first principal component using the $n \times n$ variance-covariance matrix of the estimated idiosyncratic components from 2020:Q1 to 2021:Q4, denoted as $\widehat{\bs{\Sigma}}_{\bs{\Xi}^{\scaleto{C}{4pt}}}$. This is the procedure proposed by \cite{stockcomovement} that we modify to account for non-stationarity in the idiosyncratic component. This done by partitioning the matrix of idiosyncratic components during the Covid period as $\widehat{\bs{\Xi}}^{\scaleto{C}{4pt}} = (\widehat{\bs{\Xi}}^{\scaleto{C,1}{4pt}} \vert \widehat{\bs{\Xi}}^{\scaleto{C,0}{4pt}}) $, where $\widehat{\bs{\Xi}}^{\scaleto{C,1}{4pt}}$ and $\widehat{\bs{\Xi}}^{\scaleto{C,1}{4pt}}$ are the matrices of estimated idiosyncratic components in the period 2020:Q1 to 2021:Q4 for $i\in \mathcal{I}_1$ and $i\in \mathcal{I}_0$, respectively. Then, we estimate $\widehat{\bs{\Sigma}}_{\bs{\Xi}^C}$ (\citealp[Chapter 17]{hamilton2020time}; \citealp{bai2004estimating}):
\begin{equation*}
\widehat{\bs{\Sigma}}_{\bs{\Xi}^{\scaleto{C}{4pt}}}\ =\ \left[
\begin{array}{cc}
\frac{1}{(T^{\scaleto{C}{3pt}})^{2}} \widehat{\bs{\Xi}}^{\scaleto{C,1}{4pt}'}\widehat{\bs{\Xi}}^{\scaleto{C,1}{4pt}} & \frac{1}{(T^{\scaleto{C}{3pt}})^{3/2}} \widehat{\bs{\Xi}}^{\scaleto{C,1}{4pt}'}\widehat{\bs{\Xi}}^{\scaleto{C,0}{4pt}} \\[8pt]
\frac{1}{(T^{\scaleto{C}{3pt}})^{3/2}} \widehat{\bs{\Xi}}^{\scaleto{C,0}{4pt}'}\widehat{\bs{\Xi}}^{\scaleto{C,1}{4pt}} & \frac{1}{T^C} \widehat{\bs{\Xi}}^{\scaleto{C,0}{4pt}'}\widehat{\bs{\Xi}}^{\scaleto{C,0}{4pt}}
\end{array}
\right]
\end{equation*}
where $T^c = 8$ denotes the time-periods between 2020:Q1 and 2021:Q4.

Given $\widehat{\bs{\Sigma}}_{\bs{\Xi}^C} $, we obtain the Covid factor and the corresponding loadings as:
\begin{align*}
    \widehat{\boldsymbol{\gamma}}\ &=\ \sqrt{n} \cdot \widehat{\mathbf{V}}_{\bar{\boldsymbol{\Xi}}^C}
    \\[5pt]
    \widehat{\mathbf{g}}\ &=\ \frac{1}{\sqrt{n}}\cdot (\widehat{\boldsymbol{\Xi}}^{C} \widehat{\mathbf{V}}_{\bar{\boldsymbol{\Xi}}^C})  
\end{align*}
where $\widehat{\mathbf{g}}$ is the $T^C \times 1$ vector with entries $\widehat{g}_t$ and $\mathbf{\widehat{V}}_{\boldsymbol{\Xi}^C}$ is the $n \times 1$ eigenvector corresponding to the largest eigenvalue of $\widehat{\boldsymbol{\Sigma}}_{\widehat{\boldsymbol{\Xi}}^C}$. Given $\widehat{\mathbf{g}}$, the associated loadings are $\widehat{\boldsymbol{\gamma}} = (\widehat{\gamma}_1,\ldots,\widehat{\gamma}_n)'$.

Next, give the estimate of states and the Covid factor, we account for the presence of changes in the volatility after the Covid shock by modifying the \cite{lenza_how_2022} procedure to accommodate for quarterly data.
Let $s^*_t = s_t \mathbb I_{\,\text{\tiny $t\!\ge\! t_{\scaleto{20\text{Q1}}{4pt}}$}} + (1- \mathbb I_{\,\text{\tiny $t\!\ge\! t_{\scaleto{20\text{Q1}}{4pt}}$} })$, the likelihood writes as:
\begin{align*}
    {
        \mathcal{L}(\mathbf{f}^{\scaleto{(0)}{6pt}} \giv \mathbf{A}, \bs{\Sigma}_u, s^{*}_{\scaleto{1}{4pt}},\ldots, s^{*}_{\scaleto{T}{4pt}})}\ &{\propto\ \prod_{t=2}^T\left|(s_t^{*})^2 \bs{\Sigma}_u\right|^{-\frac{1}{2}} \cdot \exp \left\{-\frac{1}{2} \sum_{t=2}^T\left(\mathbf{f}_{t\mid T}^{\scaleto{(0)}{6pt}}-\mathbf{A} \mathbf{f}_{t-1\mid T}^{\scaleto{(0)}{6pt}}\right)^{\prime}\left(s_t^2 \bs{\Sigma}_u\right)^{-1}\left(\mathbf{f}_{t\mid T}^{\scaleto{(0)}{6pt}}-\mathbf{A} \mathbf{f}_{t-1\mid T}^{\scaleto{(0)}{6pt}}\right)\right\}} \\[7pt]
        &{\propto\ \parT{\prod_{t=2}^T (s_t^{*})^{-n}}\left| \bs{\Sigma}_u\right|^{-\frac{T-1}{2}} \cdot \exp \left\{-\frac{1}{2} \sum_{t=2}^T\left(\mathbf{f}_{t\mid T}^{*\scaleto{(0)}{6pt}}-\mathbf{A} \mathbf{f}_{t-1\mid T}^{*\scaleto{(0)}{6pt}}\right)^{\prime}\left(\bs{\Sigma}_u\right)^{-1}\left(\mathbf{f}_{t\mid T}^{*\scaleto{(0)}{6pt}}-\mathbf{A} \mathbf{f}_{t-1\mid T}^{*\scaleto{(0)}{6pt}}\right)\right\}}
    \end{align*}
where $\mathbf{f}_{t\giv T}^{*\scaleto{(0)}{6pt}}= \mathbf{f}_{t\giv T}^{\scaleto{(0)}{6pt}}/s^*_t$, with corresponding variance-covariance matrix $\mathbf{P}_{t\giv T}^{*\scaleto{(0)}{6pt}}$.\\
\indent The maximum-likelihood estimators of $\mathbf{A}$ and $\bs{\Sigma}_u$ given the factors are:
\begingroup
\allowdisplaybreaks
    \begin{align*}
        \check{\mathbf{A}}\ &=\ \left(\sum_{t=2}^T \mathbf{f}_{t \mid T}^{*\scaleto{(0)}{6pt}} \mathbf{f}_{t-1 \mid T}^{*\scaleto{(0)}{6pt}'} + \mathbf{P}_{t,t-1\giv T}^{*\scaleto{(0)}{6pt}}
        \right)\left(\sum_{t=1}^T \mathbf{f}_{t-1 \mid T}^{*\scaleto{(0)}{6pt}} \mathbf{f}_{t-1 \mid T}^{*\scaleto{(0)}{6pt}'} + \mathbf{P}_{t-1\giv T}^{*\scaleto{(0)}{6pt}}
        \right)^{-1} \\[8pt] 
        \check{\bs{\Sigma}}_u\ &=\ \frac{1}{T}\parT{\sum_{t=2}^T \parT{\mathbf{f}_{t\mid T}^{*\scaleto{(0)}{6pt}}\mathbf{f}_{t\mid T}^{*\scaleto{(0)}{6pt}'} + \mathbf{P}_{t\giv T}^{*\scaleto{(0)}{6pt}}} - \check{\mathbf{A}}\sum_{t=2}^T\parT{\mathbf{f}_{t\mid T}^{*\scaleto{(0)}{6pt}}\mathbf{f}_{t-1\mid T}^{*\scaleto{(0)}{6pt}'}+ \mathbf{P}_{t,t-1\giv T}^{*\scaleto{(0)}{6pt}}}}
    \end{align*}
    \endgroup\\ 
    Substituting $\check{\mathbf{A}}$ and $\check{\bs{\Sigma}}_u$ in the likelihood, we obtain the concentrated likelihood:\\
  \begin{equation*}
        \mathcal{L}(\mathbf{f}^{\scaleto{(0)}{6pt}}\giv \check{\mathbf{A}},\check{\bs{\Sigma}}_u, s_{\scaleto{1}{4pt}}^{*},\ldots, s_{\scaleto{T}{4pt}}^{*})\ =\ \prod_{t=2}^T (s_t^{*})^{-n} \cdot\left| \check{\bs{\Sigma}}_u \right|^{-\frac{T}{2}}
    \end{equation*}\\    
    By numerically maximizing the concentrated likelihood we obtain the volatility parameters $\widehat{s}_{\scaleto{t}{4pt}}$.

\subsubsection*{Step 3: Full sample estimation}   
Given the Covid factor estimated in Step 2, we obtain all the other parameters in the model:
\begin{itemize}[-]
\item \textsc{Factor loadings}:\\
\begin{equation*}
\widehat{\boldsymbol{\lambda}}_i'\ =\ \left(\sum_{t=1}^T\left(\tilde{y}_{it}-\mathrm{D}_{i,t \giv T}^{\scaleto{(0)}{6pt}}-\zeta_{i,t \giv T}^{\scaleto{(0)}{6pt}}\right) \mathbf{f}_{t \giv T}^{\scaleto{(0)}{6pt}'} - \widehat{\gamma}_i\widehat{g}_t\right)\left(\sum_{t=1}^T \mathbf{f}_{t \giv T}^{\scaleto{(0)}{6pt}} \mathbf{f}_{t \giv T}^{\scaleto{(0)}{6pt}'} + \mathbf{P}_{t\giv T}^{\scaleto{(0)}{6pt}}\right)^{-1}
\end{equation*}
\item \textsc{Slopes of secular trend}:
\begin{equation*}
        \widehat{b}_i\ =\ \left(\sum_{t=1}^T\left(\tilde{y}_{i t}-\widehat{\boldsymbol{\lambda}}_i' \mathbf{f}_{t \mid T}^{\scaleto{(0)}{6pt}}-\zeta_{i, t \mid T}^{\scaleto{(0)}{6pt}}- \widehat{\gamma}_i\widehat{g}_t\right) t\right)\left(\sum_{t=1}^T t^2\right)^{-1}
    \end{equation*}
\item \textsc{Variance of $I(1)$ idiosyncratic components}:
\begin{align*}
\widehat{\sigma}_{e_i}^{2}\ =\ &\frac{1}{T} \sum_{t=2}^T\left(\zeta_{i, t \mid T}^{\scaleto{(0)}{6pt}} \zeta_{i, t \mid T}^{\scaleto{(0)}{6pt}'} + P_{i,t\giv T}^{\scaleto{\zeta(0)}{6pt}}\right)+\frac{1}{T} \sum_{t=2}^T\left(\zeta_{i t-1 \mid T}^{\scaleto{(0)}{6pt}} \zeta_{i t-1 \mid T}^{\scaleto{(0)}{6pt}'} + P_{i,t-1\giv T}^{\scaleto{\zeta(0)}{6pt}}\right) -\\[4pt]
& - \frac{2}{T} \sum_{t=2}^T\left(\zeta_{i t \mid T}^{\scaleto{(0)}{6pt}} \zeta_{i t-1 \mid T}^{\scaleto{(0)}{6pt}'} + P_{i,t,t-1\giv T}^{\scaleto{\zeta(0)}{6pt}}\right)
\end{align*}
\item \textsc{Covariance prediction error}:
\begin{align*}
 \widehat{R}_i\ =\ &\frac{1}{T} \left\{\sum_{t=1}^T\left(\tilde{y}_{i,t}^{\scaleto{(0)}{6pt}}-\widehat{\boldsymbol{\lambda}}_i^{\scaleto{(0)}{6pt}'} \mathbf{f}_{t \mid T}^{\scaleto{(0)}{6pt}}- \mathbb{I}_{i \in \mathcal{I}_b} \mathrm{D}_{i, t \mid T}^{\scaleto{(0)}{6pt}}- \mathbb{I}_{i \in \mathcal{I}_1}\zeta_{i, t \mid T}^{\scaleto{(0)}{6pt}}- \widehat{\gamma}_i\widehat{g}_t\right)^2 + \widehat{\boldsymbol{\lambda}}_i^{\prime} \mathbf{P}_{t,T\giv T}^{\scaleto{(0)}{6pt}}\widehat{\boldsymbol{\lambda}}_i + \right.\\[4pt]
 & + \left. \mathbb{I}_{i\in \mathcal{I}_b} P_{i,t\giv T}^{\scaleto{\mathrm{D}(0)}{6pt}} + \mathbb{I}_{i\in \mathcal{I}_1} P_{i,t\giv T}^{\scaleto{\zeta(0)}{6pt}} \right\}
\end{align*}
\end{itemize}

Given the estimated parameters, using data net of the Covid component, i.e. $\tilde{\mathbf{y}}_{t} - \widehat{\bs{\gamma}}\widehat{g}_t$, we obtain a final estimates of all the states, $\mathbf{f}_{t\giv T}$, $\mathrm{\mathbf{D}}_{t\giv T}$, $\mathbf{b}_{t\giv T}$, $\bs{\zeta}_{t\giv T}$ and their conditional covariances with the Kalman filter and smoother. Note that, for this final run, the Kalman filter and smoother do not need to be truncated in 2019:Q4 because we have already controlled for the Covid pandemic.

\subsection*{Estimating the common trend}   
With the estimated factors, $\mathbf{f}_{t\giv T}$, we obtain an estimate of the trend and cyclical component by means of the EM algorithm. To run the algorithm, we need an initial estimate of the parameters $\boldsymbol{\psi}$, $\bm\Sigma_\omega$, and $\sigma_\nu^2$. 

\begin{compactitem}

\item [(a)] Let $\widehat\tau_t^{(0)}$ be the initial estimate of the common trend. We set $\widehat\tau_t^{(0)} = f_{j,{t\giv T}}$, where $f_{j,{t\giv T}}$ is the factor explaining the largest share of long-run dynamics. Specifically, let $\widehat{S}_j(\omega)$ denote the estimated spectral density of $\Delta f_{j,t\giv T}$ for frequency $\omega$, and $\widehat{s}_j(\omega)$ be the corresponding standardized spectral density. Let $\bar{s}_j$ be the integral of $\widehat{s}_j(\omega)$ over frequencies $\geq 8$ years. The common trend is initialized as the factor $f_{j,t\giv T}$ with the highest value of $\bar{s}_j$. This corresponds to an initial value of $\bs{\psi}$, denoted as $\widehat{\boldsymbol{\psi}}^{(0)}$, such that $\widehat{\psi}_j^{(0)} = 1$ and $\widehat{\psi}_{-j}^{(0)} = 0$. 

\item[(b)] Given $\widehat{\boldsymbol{\psi}}^{(0)}$ and $\widehat\tau_t^{(0)}$, we obtain an initial estimate of the cyclical component $\widehat{\bm\omega}_t^{(0)} = \mathbf{f}_{t\giv T}^{(0)}-\widehat{\bm\psi}^{(0)}\widehat\tau_t^{(0)}$

\item [(c)] We initialize $\sigma_\nu^2$ following \cite{del2017safety}, so that $\widehat{\sigma}_\nu^{2,(0)} = (400\widehat{\sigma}^{2}_{\Delta\tau})^{-1}$, where $\widehat{\sigma}^{2}_{\Delta\tau}$ is the sample variance of $\Delta\widehat\tau_t^{(0)}$. We chose a very small value for the variance of $\widehat{\sigma}_\nu^{2,(0)}$ to incorporate our prior assumption of a slow-moving trend.

\item [(d)] Lastly, since by construction $\widehat{\bm\omega}_t^{(0)}$ has a sample covariance matrix of reduced rank $(q-1)$, in order to run the EM algorithm we initialize this covariance as $\widehat {\bm\Sigma}_\omega^{(0)}= T^{-1}\sum_{t=1}^T\widehat{\bm\omega}_t^{(0)}\widehat{\bm\omega}_t^{(0)\prime} + \kappa\mathbf{I}_q$, where we set $\kappa= 10^{-2}$, in agreement with the recommendations by \citet{opschoor2023slow}.
\end{compactitem}
\vspace{5pt} 

\noindent Once the initial estimates for the algorithm have been computed, in the E-step we run the Kalman filter and smoother to obtain a new estimate of the trend, namely $\tau_{t\giv T}^{(1)}$, along with an estimate of its conditional variance and covariance, $P_{t\giv T}^{(1)}$ and $P_{t,t-1\giv T}^{(1)}$, respectively. The smoothed trend is then used to estimate the parameters in the M-step. This procedure is repeated iteratively until convergence.
 
For a generic iteration $k$ of the algorithm, we estimate the parameters as follow:
\begin{itemize}[-]
\item \textsc{Trend loadings}:
\begin{equation*}
\widehat{\bs{\psi}}^{\scaleto{(k)}{6pt}}\ =\ \parT{\sum_{t=1}^T \mathbf{f}_{t\giv T}\tau_{t\giv T}^{\scaleto{(k)}{6pt}}}\parT{\sum_{t=1}^T\tau_{t\giv T}^{2 \scaleto{(k)}{6pt}} + P_{t\giv T}^{\tau\scaleto{(k)}{6pt}}}^{-1}
\end{equation*}

\item \textsc{Variance of common trend}:
\begin{align*}
\widehat{\sigma}_{\nu}^{2\scaleto{(k)}{6pt}}\ =\ &\frac{1}{T} \sum_{t=2}^T\left(\tau_{t \mid T}^{2\scaleto{(k)}{6pt}} + P_{t\giv T}^{\tau\scaleto{(k)}{6pt}}\right)+\frac{1}{T} \sum_{t=2}^T\left(\tau_{t-1 \mid T}^{2\scaleto{(k)}{6pt}} + P_{t-1\giv T}^{\tau\scaleto{(k)}{6pt}}\right) -\\[4pt]
& - \frac{2}{T} \sum_{t=2}^T\left(\tau_{t \mid T}^{\scaleto{(k)}{6pt}} \tau_{t-1 \mid T}^{\scaleto{(k)}{6pt}} + P_{t,t-1\giv T}^{\tau\scaleto{(k)}{6pt}}\right)
\end{align*}

\item \textsc{Covariance of transitory component}
\begin{equation*}
\widehat{\bs{\Sigma}}_{\omega}^{\scaleto{(k)}{6pt}}\ =\ \frac{1}{T}\sum_{t=1}^T\parG{\parT{\mathbf{f}_{t\giv T} - \widehat{\bs{\psi}}^{\scaleto{(k)}{6pt}}\tau_{t\giv T}^{\scaleto{(k)}{6pt}}}\parT{\mathbf{f}_{t\giv T} - \widehat{\bs{\psi}}^{\scaleto{(k)}{6pt}}\tau_{t\giv T}^{\scaleto{(k)}{6pt}}}' + \widehat{\bs{\psi}}^{\scaleto{(k)}{6pt}} P_{t\giv T}^{\tau\scaleto{(k)}{6pt}} \widehat{\bs{\psi}}^{\scaleto{(k)}{6pt}'}}
\end{equation*}

\end{itemize}
The algorithm is stopped using the likelihood-based criterion of \cite{doz2012quasi}, with a threshold of $10^{-3}$. At convergence, we obtain an estimate of the trend and transitory components, $\tau_{t\giv T}$ and $\bs{\omega}_{t\giv T}$, respectively, along with the estimated parameters $\widehat{\bs{\psi}}$, $\widehat{\sigma}_{\nu}^{2}$ and $\widehat{\bs{\Sigma}}_{\omega}$.

Ultimately, given the estimated trend and transitory components, the estimated output gap and potential output are defined as:
\begin{align*}
\widehat{\text{PO}}_t\ &=\ \mathrm{D}_{\text{\tiny GDP},t\giv T} + \widehat{\bs{\lambda}}'_{\text{\tiny GDP} }\widehat{\bs{\psi}}{\tau}_{t\giv T},\label{POT}\\
\widehat{\text{OG}}_t\ &=\ \widehat{\bs{\lambda}}'_{\text{\tiny GDP}}\bs{\omega}_{t\giv T}
\end{align*}

%
%
\setcounter{table}{0}
\setcounter{figure}{0}
\setcounter{equation}{0}
\section{Confidence bands}
\label{app::confbands}
To obtain confidence bands for our quantities of interest, we follow the procedure outlined in \cite{barigozzi_measuring_2021}.  In particular, we simulate all the states in the model using the simulation smoother of \cite{durbin2002simple}, and we generate all the stationary residuals of the model using a stationary block bootstrap procedure \citep{politis1994stationary}.
In practice, we have an estimate of all the states, namely $\mathbf{f}_{t\giv T}$, $\mathrm{\mathbf{D}}_{t\giv T}$, $\mathbf{b}_{t\giv T}$ and $\bs{\zeta}_{t\giv T}$ , an estimate of the Covid factor $\widehat{g}_t$ and the estimated volatility parameters $\widehat{s}_t$, $t\!\ge\! t_{\scaleto{20\text{Q1}}{4pt}}$. Then, the algorithm is structured as follows:
\begin{enumerate}
\item Simulate the states by the simulation smoother \citep{durbin2002simple}
\begin{enumerate}
\item Common factors:
\begin{enumerate}
\item simulate \hspace{2.8pt}$\tilde{\hspace{-2.8pt}\mathbf{f}}_1^{(b)} \sim N\left(\mathbf{f}_{1\giv T}, \mathbf{P}_{1 \mid T}\right)$;
\item simulate $\tilde{\mathbf{u}}_t^{(b)} \sim N\left(\mathbf{0}_q, \check{\boldsymbol{\Sigma}}_u\right)$;
\item for $t=2, \ldots, T$ generate \hspace{2.8pt}$\tilde{\hspace{-2.8pt}\mathbf{f}}_t^{(b)}=\sum_{k=1}^p \check{\mathbf{A}}_k \hspace{2.8pt}\tilde{\hspace{-2.8pt}\mathbf{f}}_{t-k}^{(b)}+\{s_t \mathbb I_{\,\text{\tiny $t\!\ge\! t_{\scaleto{20\text{Q1}}{4pt}}$}} + (1- \mathbb I_{\,\text{\tiny $t\!\ge\! t_{\scaleto{20\text{Q1}}{4pt}}$} })\}\tilde{\mathbf{u}}_t^{(b)}$.
\end{enumerate}
\item $I(1)$ idiosyncratic components. For each $i \in \mathcal{I}_1$:
\begin{enumerate}
\item simulate $\tilde{\xi}_{i 1}^{(b)} \sim N\left(\xi_{i, 1\giv T}, P_{i, 1 \mid T}^\zeta\right)$;
\item simulate $\tilde{e}_{i t}^{(b)} \sim N\left(0, \widehat{\sigma}_{e_i}^2\right)$;
\item for $t=2, \ldots, T$ generate $\tilde{\zeta}_{i t}^{(b)}=\tilde{\xi}_{i t}^{(b)}$; $\tilde{\xi}_{i t}^{(b)} = \tilde{\xi}_{i t-1}^{(b)}+\tilde{e}_{i t}^{(b)}$.
\end{enumerate}
\item Time-varying secular components:
\begin{itemize}
\item $i\in\mathcal{L}_1$:
\begin{enumerate}
\item simulate $\tilde{b}_{i, 1}^{(b)} \sim N\left(b_{i, 1\giv T}, P_{i,1 \mid T}^{b}\right)$, and set \hspace{-1.5pt}$\tilde{\hspace{1.5pt}\mathrm{D}}_{i, 1}^{(b)}=\tilde{b}_{i, 1}^{(b)}$;
\item simulate $\tilde{\eta}_t^{(b)} \sim N\left(0, \sigma_{\eta_{i}}^2\right)$;
\item for $t=2, \ldots, T$ generate $\tilde{b}_{i, t}^{(b)}=\tilde{b}_{i, t-1}^{(b)}+\tilde{\eta}_t^{(b)}$;
\item for $t=2, \ldots, T$ generate $\hspace{-1.5pt}\tilde{\hspace{1.5pt}\mathrm{D}}_{i, t}^{(b)}=\hspace{-1.5pt}\tilde{\hspace{1.5pt}\mathrm{D}}_{i, t-1}^{(b)}+\tilde{b}_{i,t}^{(b)}$.
\end{enumerate}
\item $i\in\mathcal{L}_0$:
\begin{enumerate}
\item simulate $\hspace{-1.5pt}\tilde{\hspace{1.5pt}\mathrm{D}}_{i, 1}^{(b)} \sim N\left(b_{i, 1\giv T}, P_{i,1 \mid T}^{\mathrm{D}}\right)$
\item simulate $\tilde{\epsilon}_t^{(b)} \sim N\left(0, \sigma_{\epsilon_{i}}^2\right)$;
\item for $t=2, \ldots, T$ generate $\hspace{-1.5pt}\tilde{\hspace{1.5pt}\mathrm{D}}_{i, t}^{(b)}=\hspace{-1.5pt}\tilde{\hspace{1.5pt}\mathrm{D}}_{i, t-1}+\tilde{\epsilon}_t^{(b)}$.
\end{enumerate}
\end{itemize}
\end{enumerate}
\item Simulate the stationary residuals of the model, $\mathbf{z}_t = (z_{1,t},\ldots,z_{n,t})'$, using a stationary block-bootstrap \citep{politis1994stationary} with an average block length of four quarters. Denote the resulting simulated residuals as $\tilde{\mathbf{z}}_t^{(b)} = (\tilde{z}_{1,t}^{(b)},\ldots,\tilde{z}_{n,t}^{(b)})'$. 
\item Generate the data. For $t=1,\ldots, T$, generate:
\begin{enumerate}
\item $\tilde{y}_{i t}^{(b)}=\hspace{-1.5pt}\tilde{\hspace{1.5pt}\mathrm{D}}_{i t}^{(b)}+ \widehat{\boldsymbol{\lambda}}_{i}' \hspace{2.8pt}\tilde{\hspace{-2.8pt}\mathbf{f}}_{t}^{(b)}+\widehat{\gamma}_i\widehat{g}_t +\tilde{\zeta}_{i t}^{(b)} + \tilde{z}_{i t}^{(b)}$, for $i \in\{\mathcal{L}_0, \mathcal{L}_1\}$.
\item $\tilde{y}_{i t}^{(b)}=\mathrm{D}_{i, t\giv T}+ \widehat{\boldsymbol{\lambda}}_{i}' \hspace{2.8pt}\tilde{\hspace{-2.8pt}\mathbf{f}}_{t}^{(b)}+ \widehat{\gamma}_i\widehat{g}_t +\tilde{\zeta}_{i t}^{(b)} + \tilde{z}_{i t}^{(b)}$, for all other variables.
\end{enumerate}
where $\tilde{\zeta}_{i t}^{(b)} = 0$ $\forall t = 1,\ldots,T$ if $i\notin \mathcal{I}_1$
\item Using $\tilde{\mathbf{y}}_t^{(b)} = (\tilde{y}_{1,t}^{(b)},\ldots, \tilde{y}_{n,t}^{(b)})'$, estimate the model as described in Appendix \ref{sec::estdetail} to get a new estimate of the loadings, $\widehat{\bs{\Lambda}}^{(b)} = (\widehat{\bs{\lambda}}_{1}^{(b)'},\ldots,\widehat{\bs{\lambda}}_{N}^{(b)'})'$, and all the other parameters in the model, as well as a new estimate of the states $\mathbf{f}_{t\giv T}^{(b)}, \mathrm{\mathbf{D}}_{t\giv T}^{(b)}, \mathbf{b}_{t\giv T}^{(b)}$ and $\boldsymbol{\zeta}_{t\giv T}^{(b)}$.
\item Center the estimated states: $\hspace{2.8pt}\bar{\hspace{-2.8pt}\mathbf{f}}_{t\giv T}^{(b)}=\mathbf{f}_{t\giv T}-\hspace{2.8pt}\tilde{\hspace{-2.8pt}\mathbf{f}}_t^{(b)}$, $\hspace{-1.5pt}\bar{\hspace{1.5pt}\mathbf{D}}_{t\giv T}^{(b)}=$ $\mathrm{\mathbf{D}}_{t\giv T}-\hspace{-1.5pt}\tilde{\hspace{1.5pt}\mathrm{\mathbf{D}}}_{t}^{(b)}+\mathrm{\mathbf{D}}_{t\giv T}^{(b)}$, $ \hspace{-2.8pt}\bar{\hspace{2.8pt}\mathbf{b}}_{t\giv T}^{(b)} = \mathbf{b}_{t\giv T} -  \hspace{-2.8pt}\tilde{\hspace{2.8pt}\mathbf{b}}_{t}^{(b)} + \mathbf{b}_{t\giv T}^{(b)}$ and $  \hspace{1.5pt}\bar{\hspace{-1.5pt}\boldsymbol{\zeta}}_t^{(b)}=\boldsymbol{\zeta}_{t\giv T}-\hspace{1.5pt}\tilde{\hspace{-1.5pt}\boldsymbol{\zeta}}_t^{(b)}+\boldsymbol{\zeta}_{t\giv T}^{(b)}$.
\item Run the trend cycle decomposition on the estimated factors $\hspace{2.8pt}\bar{\hspace{-2.8pt}\mathbf{f}}_t^{(b)}$ to get a new estimate of the common trend $\tau_{t\giv T}^{(b)}$, the transitory component $\boldsymbol{\omega}_{t\giv T}^{(b)}$, and the parameter $\widehat{\boldsymbol{\psi}}^{(b)}$.
\item Estimate potential output as $\widehat{\text{PO}}_t^{(b)}=\hspace{-1.5pt}\bar{\hspace{1.5pt}\mathrm{D}}_{\mathrm{GDP}, t\giv T}^{(b)}+ \widehat{\boldsymbol{\lambda}}_{\mathrm{GDP}}^{(b)^{\prime}} \widehat{\boldsymbol{\psi}}^{(b)} \tau_{t\giv T}^{(b)}$, and the output gap as $\widehat{\text{OG}}_t^{(b)}= \widehat{\boldsymbol{\lambda}}_{\mathrm{GDP}}^{(b)^{\prime}}  \boldsymbol{\omega}_{t\giv T}^{(b)}$.
\end{enumerate}
Repeating this procedure $B$ times, we obtain a distribution of  the output gap: $\{\widehat{\text{OG}}_t^{(b)}, b=1,\ldots, B\}$. Then, we construct the $(1-\alpha)$ confidence interval as $[\widehat{\text{OG}}_t+z_{\alpha / 2} \widehat{\sigma}_t^{\scaleto{\text{OG}}{4pt}}, \widehat{\text{OG}}_t+z_{1-\alpha / 2} \widehat{\sigma}_t^{\scaleto{\text{OG}}{4pt}}]$, where $\widehat{\sigma}_t^{\scaleto{\text{OG}}{4pt}}$ is the sample standard deviation of $\{\widehat{\text{OG}}_t^{(b)}-\widehat{\text{OG}}_t\}$ and $z_{\alpha/2}= -z_{1-\alpha/2}$ is the $\alpha/2$-th quantile of a standard normal distribution.

\setcounter{table}{0}
\setcounter{figure}{0}
\setcounter{equation}{0}
%
%
\setcounter{table}{0}
\setcounter{figure}{0}
\setcounter{equation}{0}
\section{Accounting for Covid} \label{sec::covid}

\subsection{Why adjusting for the Covid shock matters} \label{sbsec::noCOVpre}
\begin{figure}[ht!]\caption{Output gap when using pre-Covid parameters or without Covid adjustment} \label{fig::covidgaps}
\centering \footnotesize \sc \smallskip
\setlength{\tabcolsep}{.01\textwidth}        
\begin{tabular}{lcc}
& Potential output growth & Output gap \\
\rotatebox{90}{No Covid adjustment} & 
\raisebox{-.15\height}{\includegraphics[trim={2cm 9.1cm 2.2cm 9.5cm},clip,width = 0.475\textwidth]{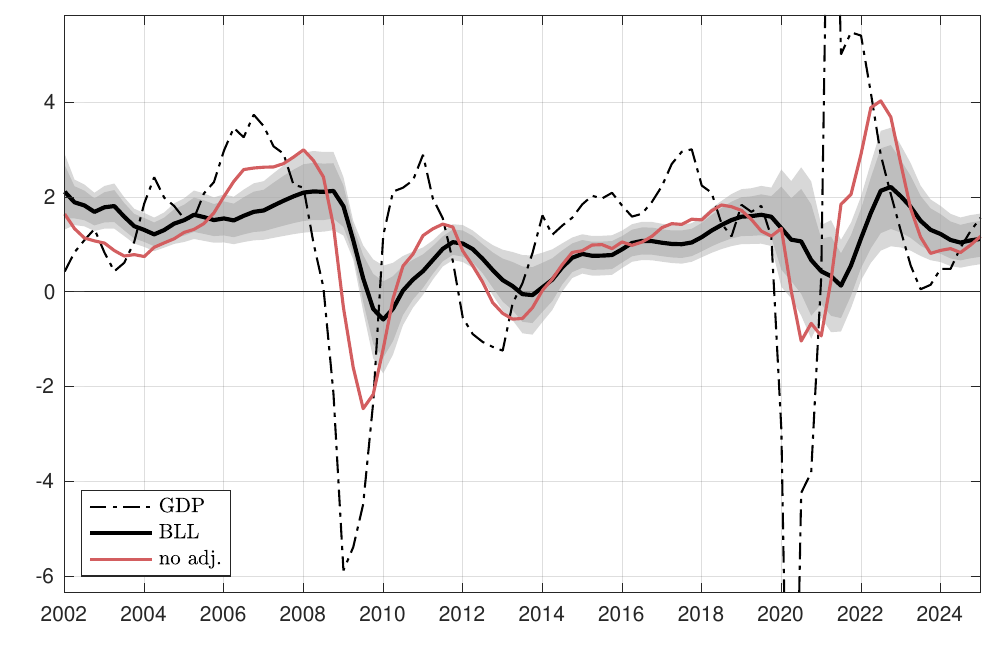}}&
\raisebox{-.15\height}{\includegraphics[trim={2cm 9.1cm 2.2cm 9.5cm},clip,width = 0.475\textwidth]{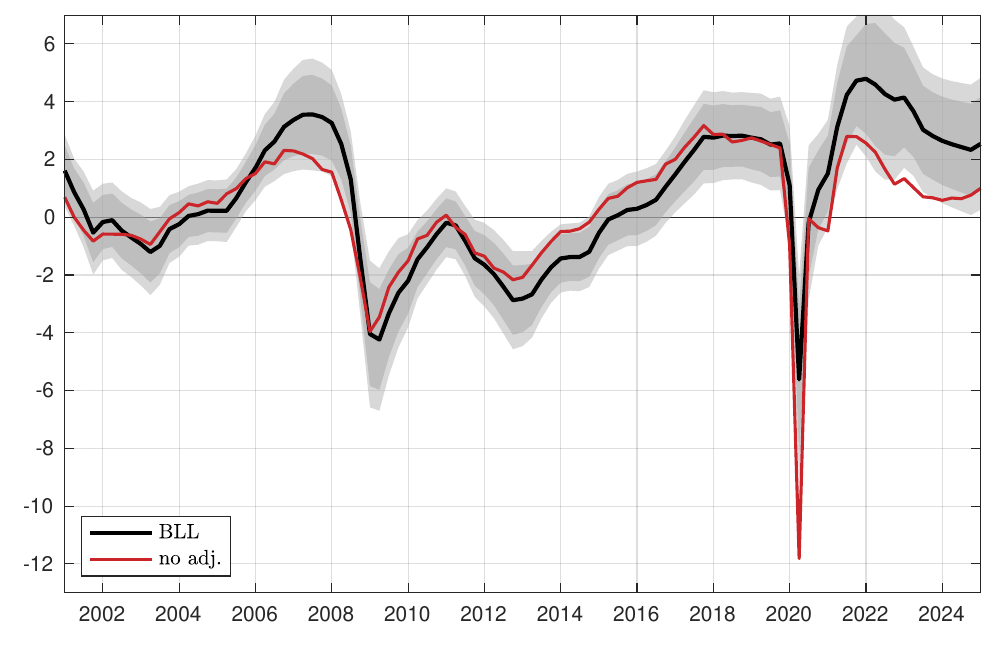}}\\[-6pt]
&&\\
\rotatebox{90}{Pre-Covid parameters} & 
\raisebox{-.15\height}{\includegraphics[trim={2cm 9.1cm 2.2cm 9.5cm},clip,width = 0.475\textwidth]{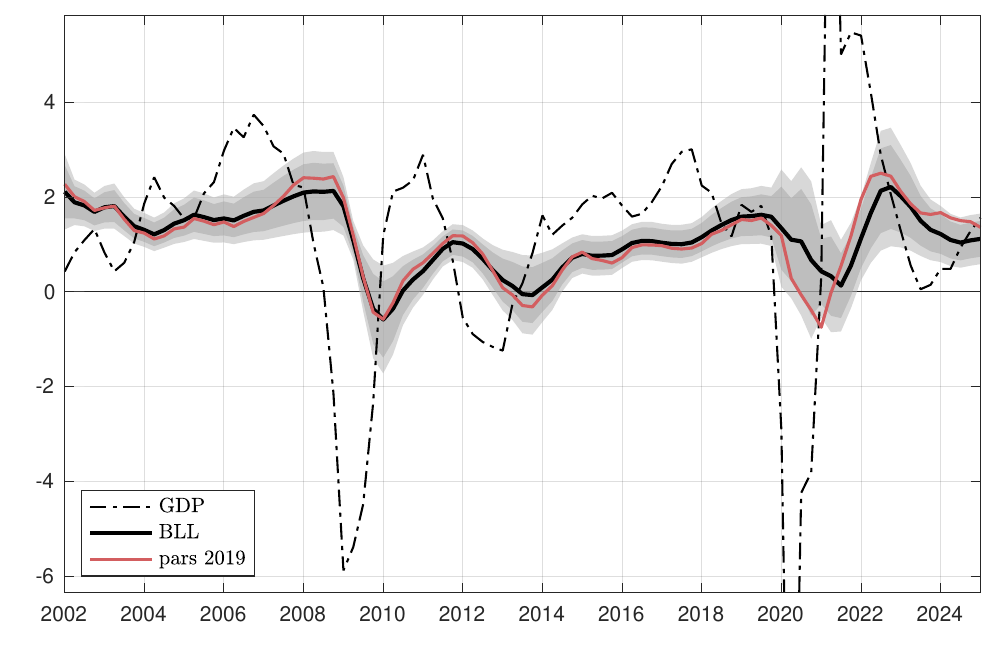}}&
\raisebox{-.15\height}{\includegraphics[trim={2cm 9.1cm 2.2cm 9.5cm},clip,width = 0.475\textwidth]{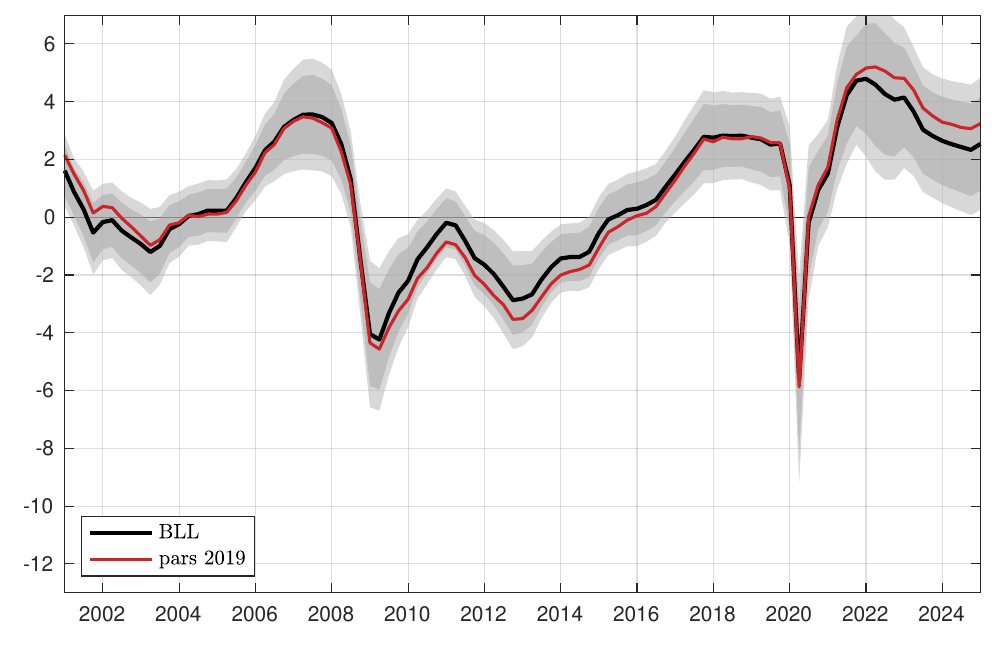}}\\[-6pt]
&&\\
\end{tabular}
   
\begin{tabular}{p{\textwidth}}\scriptsize 
Notes: \rm The black solid line is our benchmark estimate and the grey shaded areas are the 68$\%$ and 84$\%$ confidence bands, the black dashed line is GDP YoY growth rate, the red lines are the estimates obtained with two alternative estimation strategies: 1. (left) fixing parameters estimated up to 2019:Q4 (pars 2019); 2. (right) estimating the model with no adjustment for Covid (no adj.). 
\end{tabular}
\end{figure}

Section \ref{sec::fattoni} explained how we accounted for the Covid shock when estimating the model. The upper plots in Figure \ref{fig::covidgaps} compare our benchmark estimates with the one we would have obtained if we had estimated the model over the full sample without applying any adjustment for the Covid shock (\enquote{no adj.}).  As can be seen, ignoring the Covid shock affects the estimates of potential output and the output gap throughout the sample, which is undesirable; moreover, ignoring the Covid shock distorts the estimate of common dynamics in 2020 and 2021.

Having shown that accounting for the Covid shock is necessary, the question is whether a strategy different from the one we adopted would have been desirable. For example, what if we had estimated all parameters up to 2019:Q4 and then extracted the states by simply truncating the Kalman smoother? Despite being effective, this strategy is sub-optimal because estimating the parameters up to 2019 becomes less and less justifiable as new data come in. Moreover, if the increase in volatility induced by the Covid shock turns out to be very persistent, confidence intervals would be underestimated because they only account for the pre-Covid volatility regime. By accounting for the Covid shock, we avoid both these issues.

The lower plots in Figure \ref{fig::covidgaps} compare our benchmark estimate with the one obtained by estimating the parameters up to the last quarter of 2019 (\enquote{pars 2019}). The two estimates are very similar up to the pandemic, after which the estimate using the up-to-2019 parameters points towards much larger fluctuations in potential output growth and a much larger output gap, which, if taken at face value, signals a very tight economy.

\subsection{Covid volatiliy}\label{sbsec::covV}
Section \ref{sec::fattoni} explained how we accounted for the effect of the Covid shock on the volatility of the common factors. Specifically, we follow \cite{lenza_how_2022} and introduce a factor $s_t$ that scales the volatility of the common factors from the beginning of the pandemic onward. \cite{lenza_how_2022} analyzed monthly US data and impose an exponential decay for $s_t$ starting in June 2020. In contrast, we estimate one parameter for each period starting in 2020:Q1. This choice is motivated by both the different impact and policy response of the Covid pandemic in Europe and by the fact that, as \cite{morley_estimating_2023} pointed out, quarterly data do not allow for a sharp identification of the decay parameter.

\begin{figure}[h!]\caption{Output gap estimate with alternative Covid volatility}\label{fig::stf}
\centering \footnotesize \sc \smallskip

\setlength{\tabcolsep}{.01\textwidth}        
\begin{tabular}{cc}
Estimated Covid volatility & Output gap \\
\includegraphics[trim={2cm 9.1cm 2.2cm 9.5cm},clip,width = 0.475\textwidth]{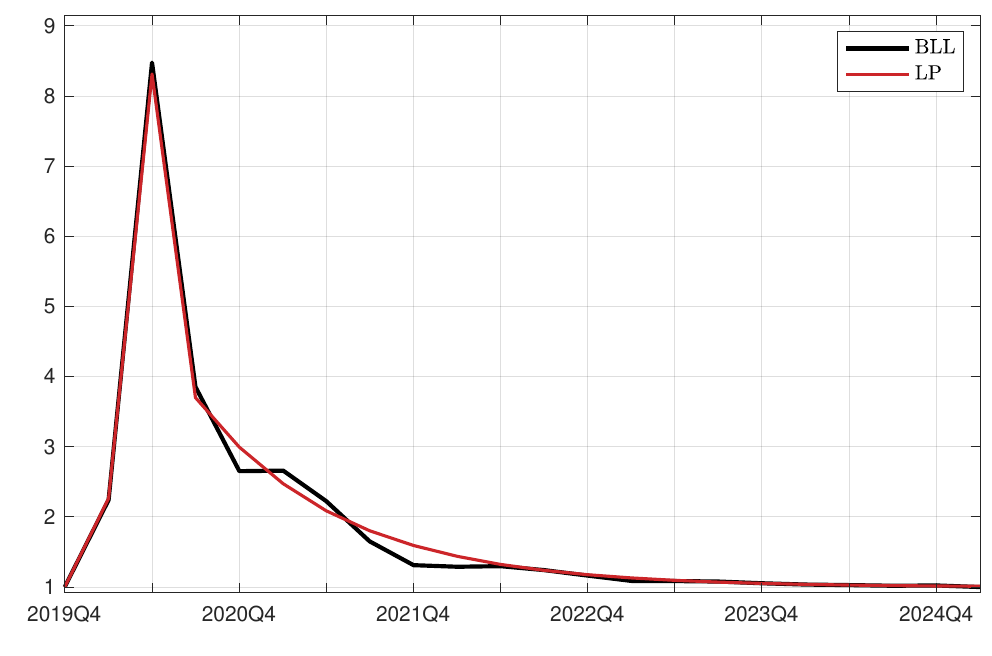} &
\includegraphics[trim={2cm 9.1cm 2.2cm 9.5cm},clip,width = 0.475\textwidth]{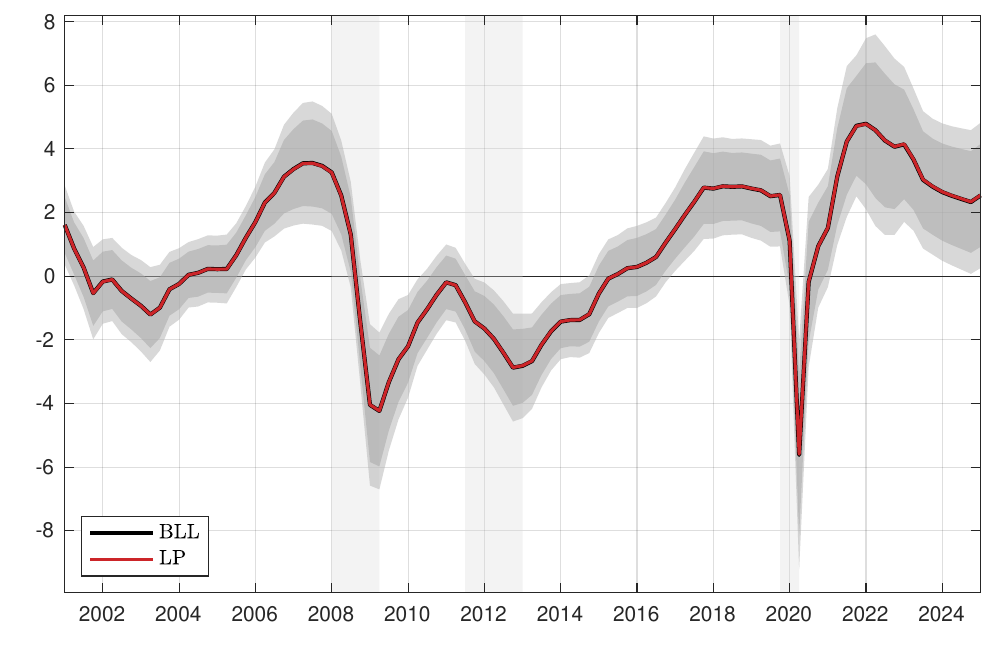} 
\end{tabular}

\begin{tabular}{p{.98\textwidth}} \scriptsize
Notes: \rm In the left plot, the black line is our estimate of Covid volatility ($s_t$), the red  line is the estimate obtained by assuming the exponential decay parametrization of the Covid volatility as in \cite{lenza_how_2022}. In the right plot, the black  line is our benchmark estimate and the grey shaded areas are the 68$\%$ and 84$\%$ confidence bands, the red  line is the estimate obtained with the exponential decay parametrization of the Covid volatility.
\end{tabular}
\end{figure}

The left plot in Figure \ref{fig::stf} shows the estimated scaling factor $s_t$  obtained under both our parametrization (black line) and under the exponential decay parametrization proposed by \cite{lenza_how_2022} (red line). Our estimate of the volatility closely tracks the evolution of the pandemic, as it spikes in the first two quarters of 2020 when mobility restrictions were most stringent in Europe, and pick-up again in 2021:Q1 when the spread of the Delta variant reached its peak. Moreover, we estimate that the volatility is very persistent---the decay we get is very close to 1 ($\approx 0.98$)---much more persistent than estimated by \cite{lenza_how_2022}, who estimated their model on monthly US data. This difference is likely due to the evolution of the pandemic in Europe, where mobility restriction measures were much more restrictive than in the US, lasted for longer, and were also implemented in 2021. Moreover, the Russia-Ukraine war had a much larger impact on Europe by pushing natural gas (and gasoline prices to a lesser extent) to the roof and creating a lot of macro-financial uncertainty. This result motivates the need to allow for time variation in the factor volatility until the end of the sample. 

As shown in the right plot in Figure \ref{fig::stf}, the two parametrizations lead to virtually identical results.

\subsection{Alternative estimator for the Covid factor} \label{sbsec::covF}
The approach we described in Appendix \ref{sec::estdetail} to estimate the Covid factor has the benefit of retaining all the information in $\widehat{\bs{\xi}}_t$, but it has the problem of relying only on eight data points. That said, since we are only interested in the first eigenvector of $\widehat{\bs{\Sigma}}_{\widehat{\boldsymbol{\Xi}}^C}$, and given the extent to which the series co-moved during the Covid period, even a few data points should be informative. However, the estimates could be imprecise, thereby motivating an alternative estimation strategy.

As an alternative approach, we estimate the Covid factor by estimating the first principal component using the $T^C \times T^C$ variance-covariance matrix of the estimated idiosyncratic components from 2020:Q1 to 2021:Q4, denoted as $\tilde{\bs{\Sigma}}_{\bs{\Xi}^C}$. In order to estimate $\tilde{\bs{\Sigma}}_{\bs{\Xi}^C}$, we consider $\widehat{\xi}_{it}$ if $i\in\mathcal{I}_0$ and $\Delta \widehat{\xi}_{it}$ if $i\in \mathcal{I}_1$, i.e. we take first-differences of all non-stationary idiosyncratic components.\footnote{In this case $\widehat{\bs{\xi}}_{i}$ will be a $T-1$ vector denoting at time $t$ the level of the idiosyncratic component for $i\in \mathcal{I}_0$ and the growth rate of the idiosyncratic component for $i\in \mathcal{I}_1$.} We obtain the Covid factor and the corresponding loadings as:
\begin{align*}
    \tilde{\mathbf{g}}\ &=\ \sqrt{T^C}\cdot \tilde{\mathbf{V}}_{\bar{\boldsymbol{\Xi}}^C}\\[5pt]
     \tilde{\boldsymbol{\gamma}}\ &=\ \frac{1}{\sqrt{T^C}} \cdot  (\widehat{\boldsymbol{\Xi}}^{C} \tilde{\mathbf{V}}_{\bar{\boldsymbol{\Xi}}^C}) 
\end{align*}\\
where $\tilde{\mathbf{g}}$ is the $T^C \times 1$ vector with entries $\tilde{g}_t$ and $\mathbf{\tilde{V}}_{\boldsymbol{\Xi}^C}$ is the $T^C \times 1$ eigenvector corresponding to the largest eigenvalue of $\tilde{\boldsymbol{\Sigma}}_{\widehat{\boldsymbol{\Xi}}^C}$. Given $\tilde{\mathbf{g}}$, the associated loadings are $\tilde{\boldsymbol{\gamma}} = (\tilde{\gamma}_1,\ldots,\tilde{\gamma}_n)'$.

This second strategy allows us to estimate $\tilde{\boldsymbol{\Sigma}}_{\widehat{\boldsymbol{\Xi}}^C}$ with $N$ data points, thereby yielding a more precise estimate. However, this comes at the cost of missing important information due to differencing of the non-stationary idiosyncratic component. Which one of the two approaches is better?

\begin{figure}[h!]\caption{Output gap and potential output growth using the alternative Covid factor}\label{fig::ogF1}

\centering \footnotesize \sc \smallskip
\begin{tabular}{cc}
\setlength{\tabcolsep}{.01\textwidth}       
Potential Output: YoY Growth Rates & Output Gap: Levels \\
\includegraphics[trim={2cm 9.1cm 2cm 9.5cm},clip,width = 0.475\textwidth]{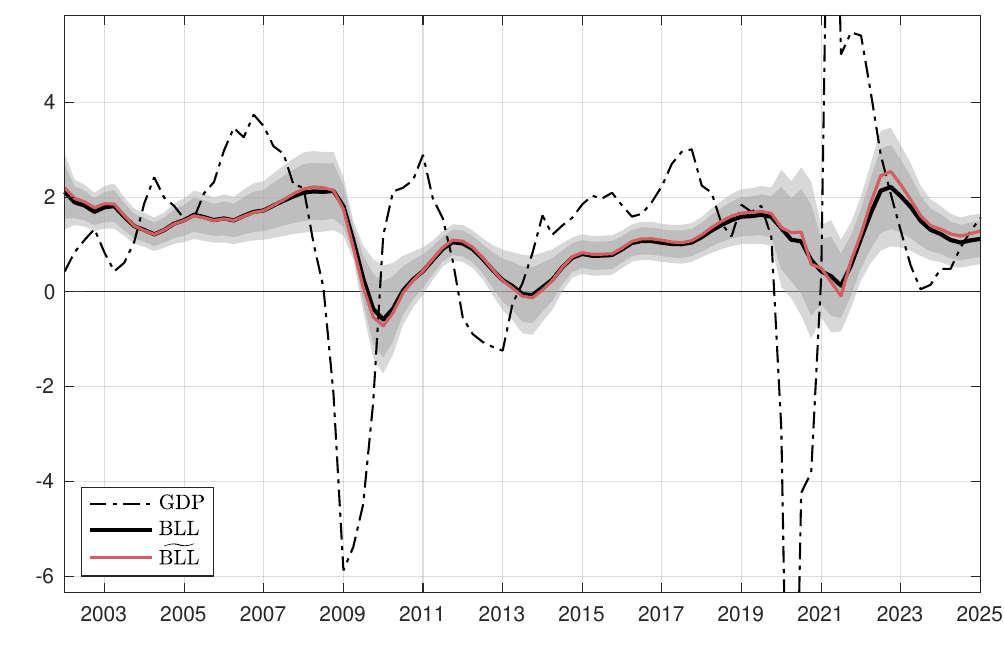} &
\includegraphics[trim={2cm 9.1cm 2cm 9.5cm},clip,width = 0.475\textwidth]{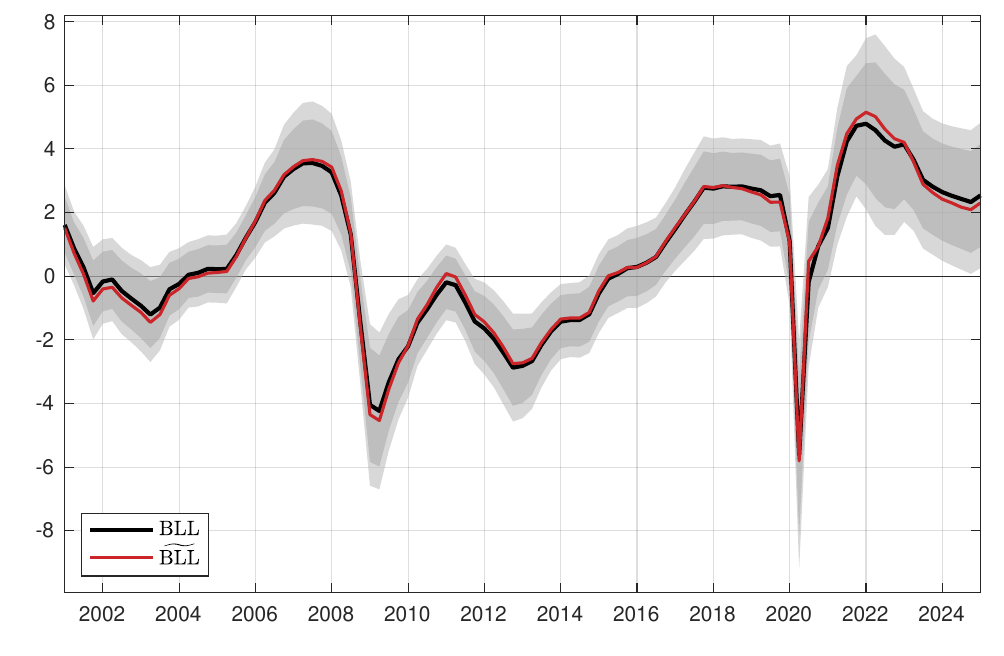} \\
\end{tabular} 

\begin{tabular}{p{.98\textwidth}} \scriptsize
Notes: \rm The black solid line is our benchmark estimate and the grey shaded areas are the 68$\%$ and 84$\%$ confidence bands. The red solid lines are the estimates obtained with the alternative Covid factor ($\widetilde{\text{BLL}}$). The level of the output gap is the percentage deviation from potential.
\end{tabular}      
\end{figure}

As a robustness exercise, in this Appendix we look at what would have been the output gap estimate, had we adopted the second strategy to estimate the Covid factor that we just laid out. As shown in Figure \ref{fig::ogF1}, the results obtained with the alternative Covid factor are almost identical to those obtained in the benchmark specification. This is not surprising, since the co-movoments observed in most of the series during the Covid period are so large to be easily identified even with a limited range of observations. 

%
%
\setcounter{table}{0}
\setcounter{figure}{0}
\setcounter{equation}{0}
\section{Modeling time-varying parameters} \label{app::notvpars}
To address the robustness of our results to the calibration of time-varying parameters variances in Equation \eqref{eq::obseq}, we compare our output gap and potential output estimates with those obtained by: (i) adopting a different calibration for the time-varying parameters' variances, and (ii) removing time-variation from \textit{all} series.

\begin{figure}[ht!]\caption{Output gap and potential output growth using TV parameters calibrations}
\label{fig::othTVpars}
\centering \footnotesize \sc \smallskip
\begin{tabular}{cc}
\setlength{\tabcolsep}{.01\textwidth}       
Potential Output: YoY Growth Rates & Output Gap: Levels \\
\includegraphics[trim={2cm 9.1cm 2.2cm 9.5cm},clip,width = 0.475\textwidth]{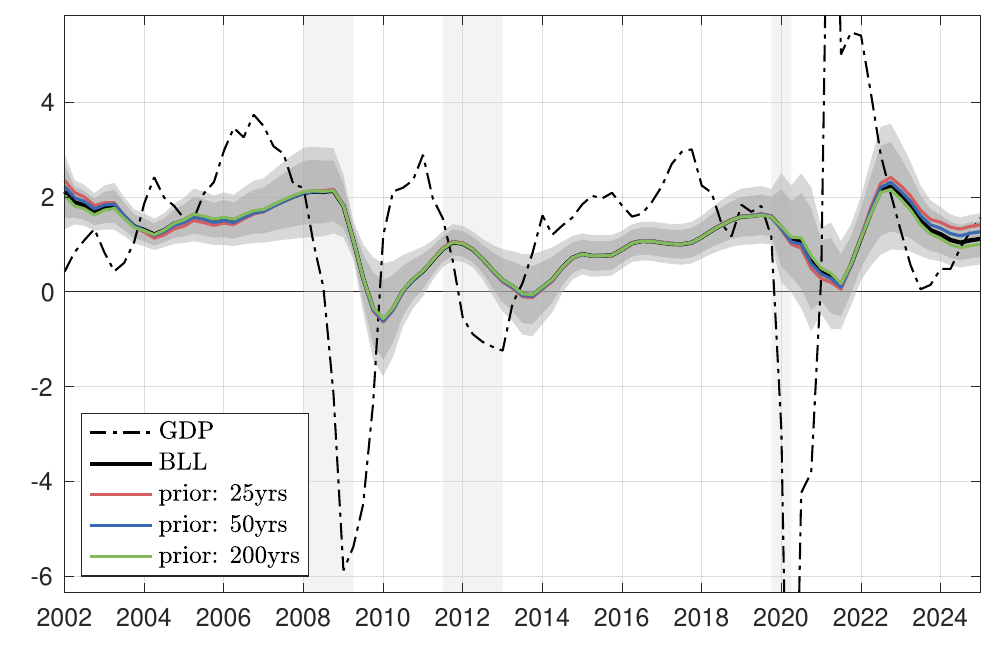} &
\includegraphics[trim={2cm 9.1cm 2.2cm 9.5cm},clip,width = 0.475\textwidth]{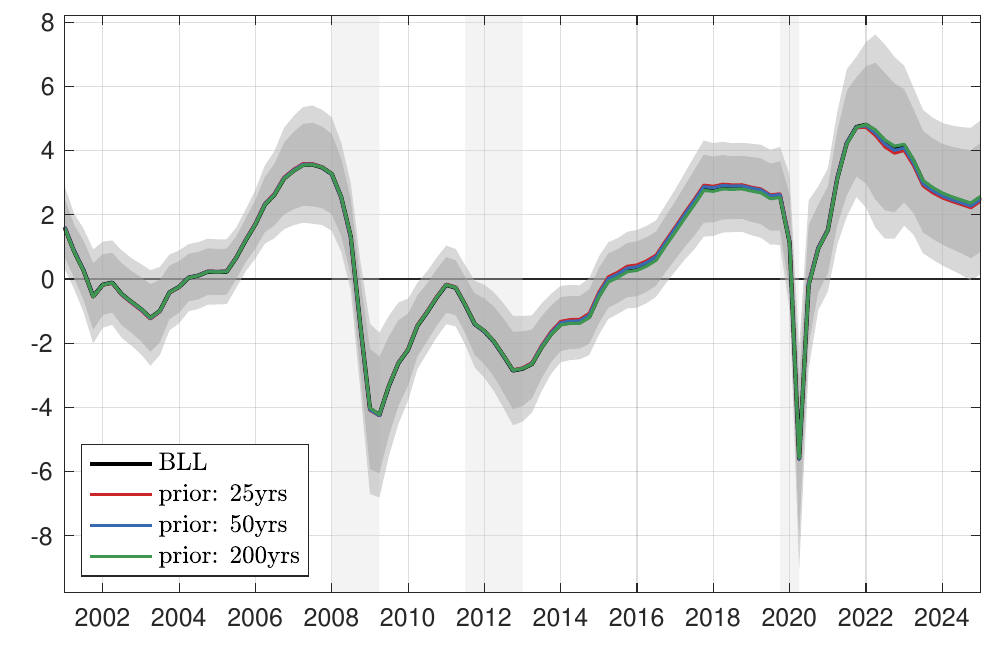} \\
\end{tabular} 
\begin{tabular}{p{.98\textwidth}} \scriptsize
Notes: \rm The black solid line is our benchmark estimate, the grey shaded areas are the 68$\%$ and 84$\%$ confidence bands, the black dashed line is GDP YoY growth rate, the red line is the estimate obtained without time-varying parameters (no TV pars). The level of the output gap is the percentage deviation from potential.
\end{tabular}   
\end{figure}

Figure \ref{fig::othTVpars} compares the benchmark estimates of the output gap (right plot) and YoY potential output growth (left plot) with those obtained under alternative calibrations of the parameter variances. Specifically, in our benchmark specification we calibrate the variances of these parameters so that the standard deviation of the local-linear trends is approximately $1\%$ over 100 years, while for time-varying means, it is approximately $1\%$ over 50 years. We consider three alternative calibration, scaling these variances by a factor of 0.5, 2, and 4. These scalings correspond to standard-deviation changes of about 1\% over roughly 200, 50, and 25 years for local-linear trends, and over 100, 25, and 12.5 years for time-varying means. The results from Figure \ref{fig::othTVpars} show that our estimates are robust to alternative calibration.
 
\begin{figure}[ht!]
\caption{Output gap and potential output growth imposing no TV parameters}
\label{fig::noTVpars}
\centering \footnotesize \sc \smallskip
\begin{tabular}{cc}
\setlength{\tabcolsep}{.01\textwidth}       
Potential Output: YoY Growth Rates & Output Gap: Levels \\
\includegraphics[trim={2cm 9.1cm 2.2cm 9.5cm},clip,width = 0.475\textwidth]{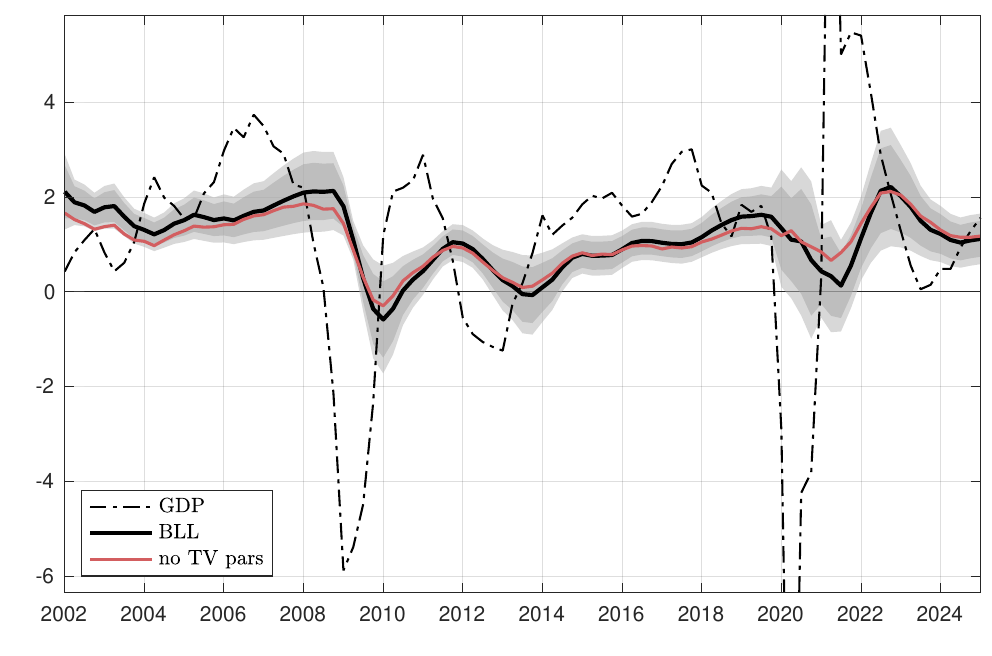} &
\includegraphics[trim={2cm 9.1cm 2.2cm 9.5cm},clip,width = 0.475\textwidth]{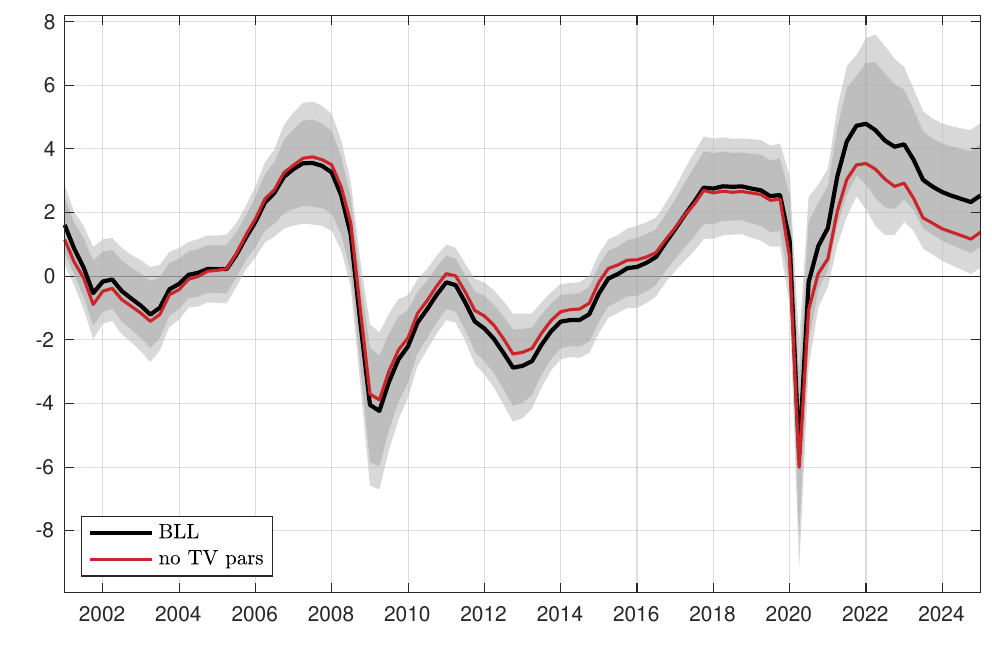} \\
\end{tabular} 
\begin{tabular}{p{.98\textwidth}} \scriptsize
Notes: \rm The black solid line is our benchmark estimate, the grey shaded areas are the 68$\%$ and 84$\%$ confidence bands, the black dashed line is GDP YoY growth rate, the red line is the estimate obtained without time-varying parameters (no TV pars). The level of the output gap is the percentage deviation from potential.
\end{tabular}      
\end{figure}

Furthermore, Figure \ref{fig::noTVpars} compares the benchmark estimates of the output gap (right plot) and YoY potential output growth (left plot) with those obtained  without allowing for a local linear trend for GDP, household liabilities, and long-term loans and no time-varying mean for the unemployment rate and inflation indicators. Eliminating time variation in the secular trends produces a flatter path for potential output growth---and, consequently, a lower output gap---in the post-pandemic period. This result shows that allowing for a time-varying trend for GDP is crucial to properly capture the slowdown in potential output in the latter part of the sample.

%
%
\setcounter{table}{0}
\setcounter{figure}{0}
\setcounter{equation}{0}
\section{Comparison with alternative measures} \label{sec::altTC}
\begin{figure}[h!]\caption{Output gap estimates with alternative methods} \label{fig::othergaps}
\centering \footnotesize \sc \smallskip
\setlength{\tabcolsep}{.01\textwidth}    
\begin{tabular}{p{.31\textwidth}p{.31\textwidth}p{.31\textwidth}}
\centering \textcolor{white}{0} HP filter  &
\centering \textcolor{white}{0} \citeauthor{hamilton2018you} filter &
\textcolor{white}{000000} Boosted HP filter \\
\includegraphics[trim={2cm 9.1cm 2.2cm 9.5cm},clip,width = 0.31\textwidth]{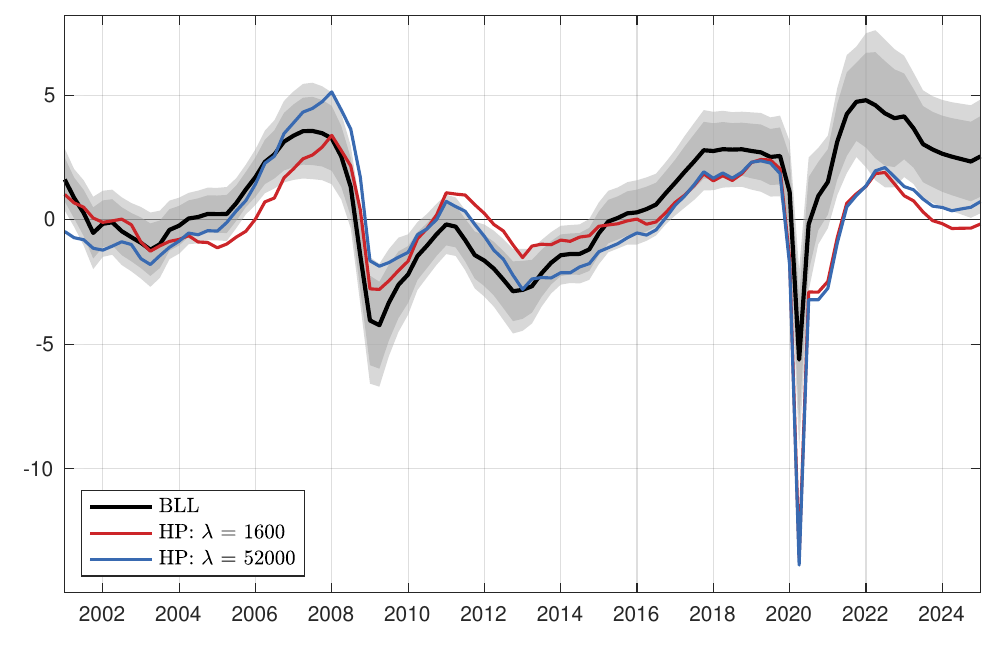}& 
\includegraphics[trim={2cm 9.1cm 2.2cm 9.5cm},clip,width = 0.31\textwidth]{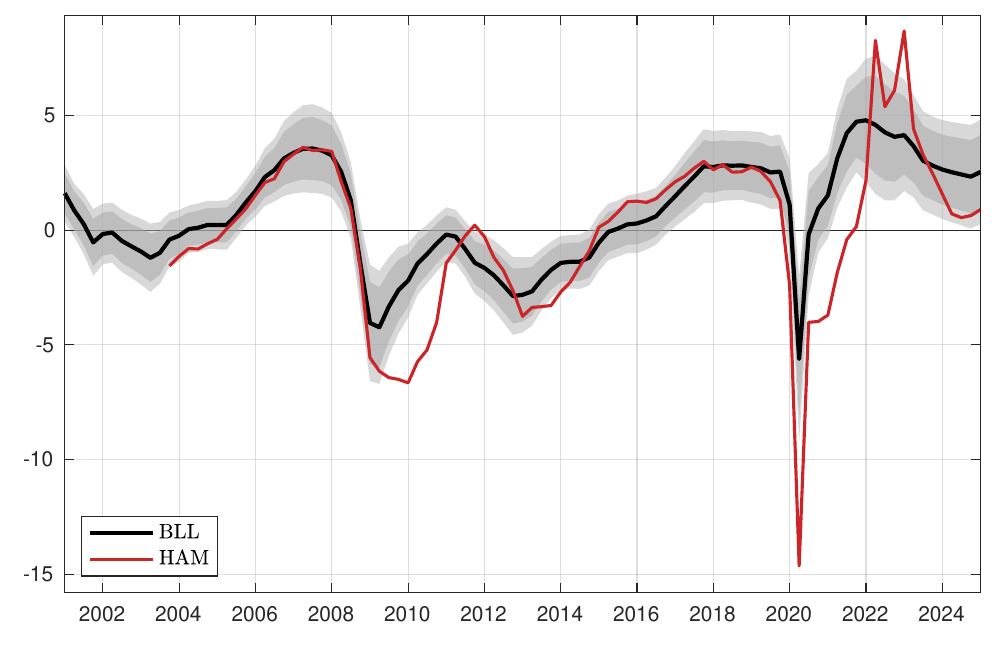}& 
\includegraphics[trim={2cm 9.1cm 2.2cm 9.5cm},clip,width = 0.31\textwidth]{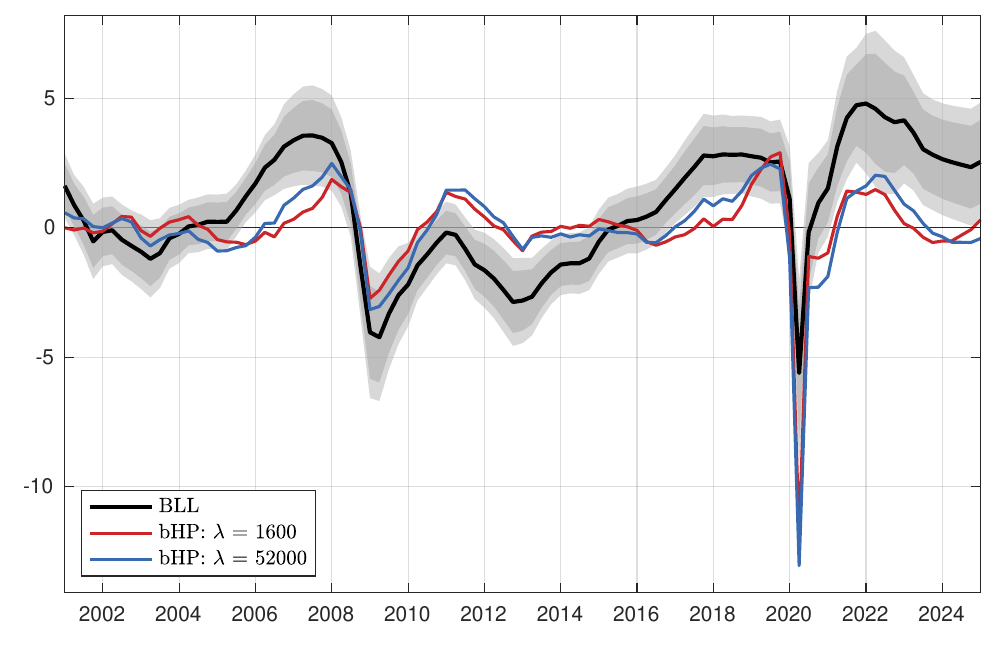}\\[-6pt]
&&\\
\centering \textcolor{white}{0} Christiano-Fitzgerald filter &\centering \textcolor{white}{0} Butterworth filter &
\textcolor{white}{0000000} Multivariate BN  \\
\raisebox{-.55\height}{\includegraphics[trim={2cm 9.1cm 2.2cm 9.5cm},clip,width = 0.31\textwidth]{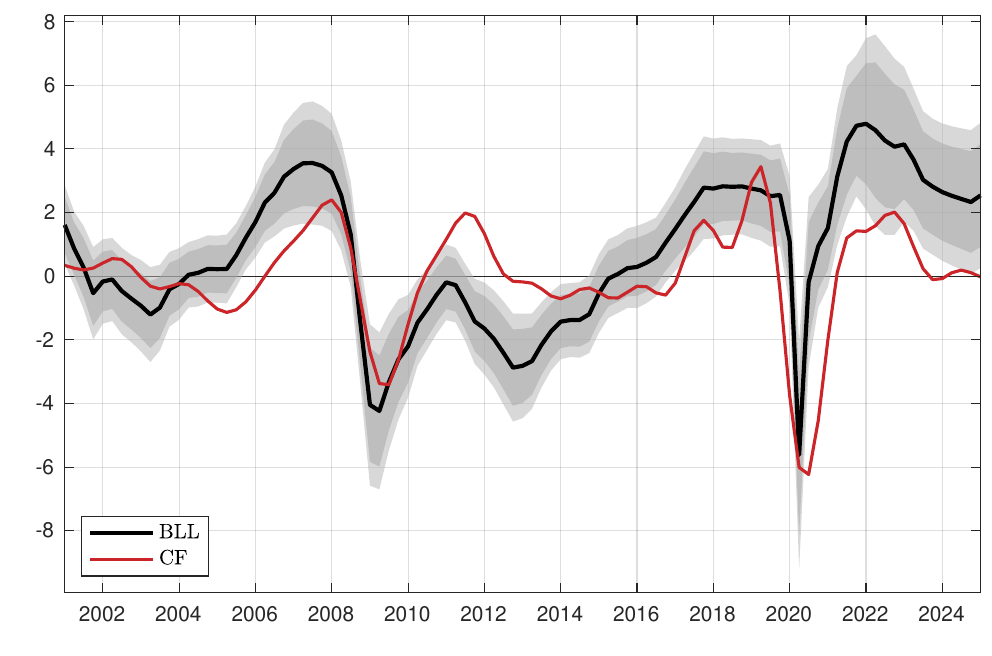}} &
\raisebox{-.55\height}{\includegraphics[trim={2cm 9.1cm 2.2cm 9.5cm},clip,width = 0.31\textwidth]{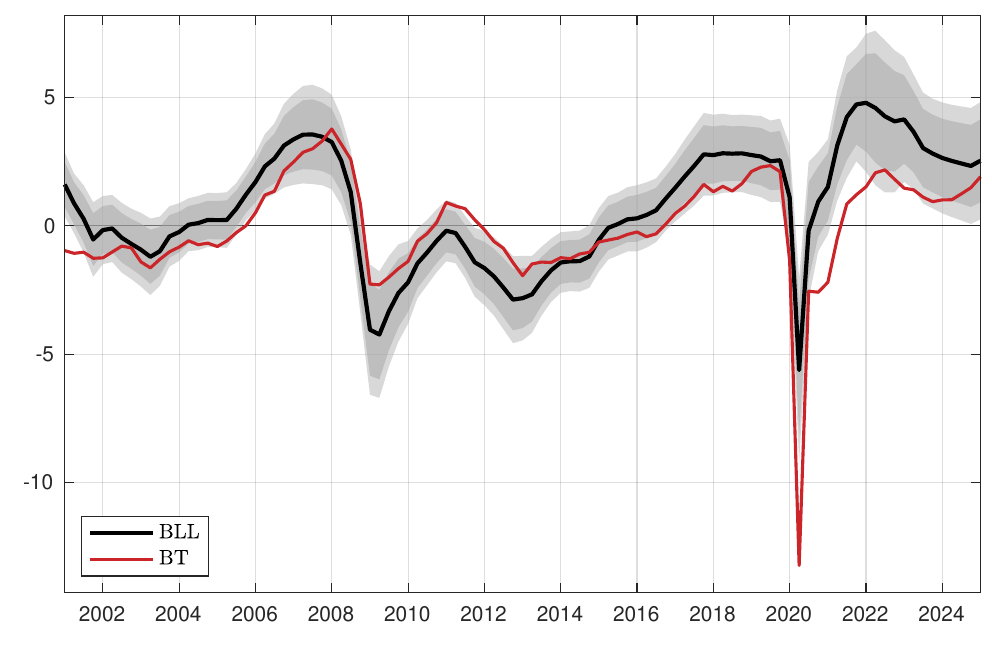}}&
\raisebox{-.55\height}{\includegraphics[trim={2cm 9.1cm 2.2cm 9.5cm},clip,width = 0.31\textwidth]{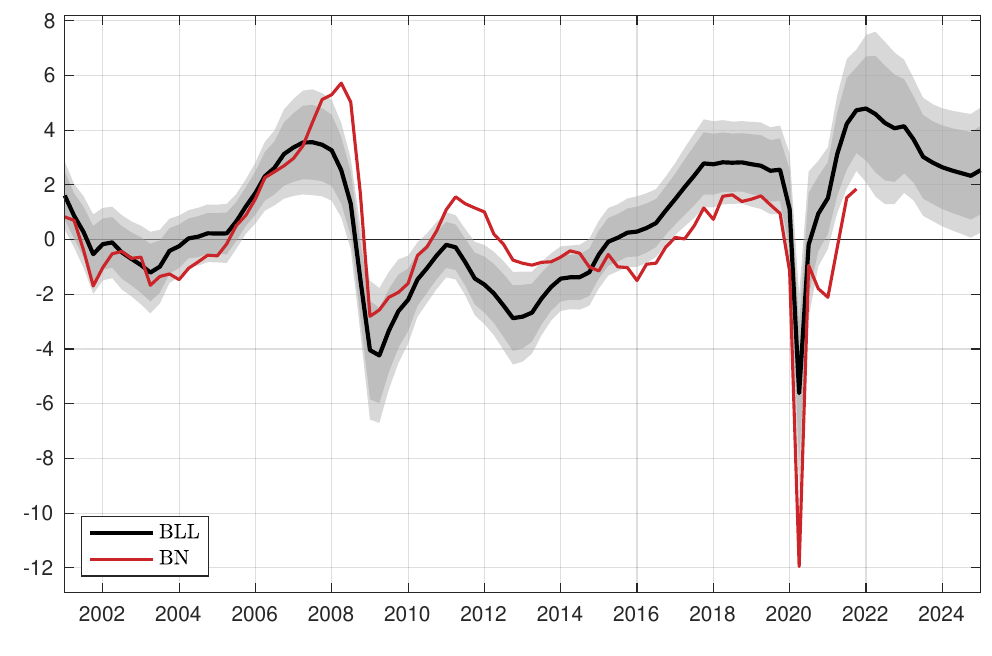}} 
\end{tabular}
\begin{tabular}{p{.98\textwidth}} \scriptsize 
Notes: \rm The black line is our benchmark estimate and the grey shaded areas are the 68$\%$ and 84$\%$ confidence bands, the red and blue lines are alternative estimates.\\
\end{tabular}
\end{figure}

In this section, we compare our estimates of potential output and the output gap with those obtained using four different univariate filters and one multivariate approach:
\begin{compactenum}[1)]
  \item The \cite{hodrick1997postwar} filter (HP), with two different values for the smoothing parameter $\lambda$: (i) $\lambda=1600$, commonly used for quarterly data, (ii) $\lambda=51200$, as proposed by \cite{borio2014financial} to capture variability at lower frequencies.
  
  \item The \cite{hamilton2018you} filter (Ham), where the trend is the $8-$step ahead forecast of quarterly GDP growth, obtained using the $4$ most recent values of quarterly GDP for each time $t$, and the cycle is the residual obtained from this regression.

  \item The boosted HP filter (bHP) of \cite{phillips2021boosting}, which improves on the standard HP filter by applying the filter recursively on the residuals extracted from previous iterations. The number of iterations $m$ is a tuning parameter that controls the intensity of the updating, and it is chosen to minimize the information criterion proposed by the authors.
   
  \item The \cite{christianofitzgerald} filter (CF), with cutoff frequencies for the transitory component between 8 and 32 quarters.   
   
  \item A Butterworth filter (BT) for the transitory component, as proposed by \cite{canova2022faq}. This filter can be cast in state-space form, and its squared-gain function defines the frequencies attributed to the cycles. Here, we employ a first-order polynomial $n=1$, with a cutoff point for the frequency set at $\omega=0.04$ and scale $G_0=1$. 
  
  \item The multivariate Beveridge-Nelson (BN) decomposition based on a large Bayesian VAR, as proposed by \cite{morley_estimating_2023}. The authors estimate the Euro Area output gap from 1999:Q1 to 2021:Q3. Here, due to the lack of availability of their data, we keep their original estimates, truncating the figure in correspondence with our starting point, i.e. 2001:Q1.
\end{compactenum}
Figure \ref{fig::othergaps} presents the results of this exercise. Overall, the output gap obtained with our methodology aligns with those estimated with univariate models in terms of peaks and troughs. However, there are several differences in terms of shape and amplitude.

%
%
\setcounter{table}{0}
\setcounter{figure}{0}
\setcounter{equation}{0}
\section{Alternative estimation strategies for the common trend}
\label{sec::altTCest}

\subsection{A Vector-Error Correction model for the common factors}\label{sbsec::VECM}
An alternative to our multi-step strategy to estimate common trends and common cycles is to estimate a Vector Error Correction Model (VECM) on the common factors, thereby estimating the factors, the trend, and the cycle jointly in a single step. In this section, we compare our estimates of potential output growth and the output gap with those obtained from this alternative strategy.

The VECM with one common trend ($q-1$ cointegrating relations) corresponding to the VAR(2) in \eqref{eq::faceq} reads as:
\begin{equation}\label{eq:VECMnov25}
\Delta\mathbf{f}_t\ =\ \bs{\alpha}(\bs{\beta}'\mathbf{f}_{t-1} + \bs{\delta}) + \boldsymbol{\Gamma}\Delta\mathbf{y}_{t-1} + \bs{\varepsilon}_t
\end{equation}
where $\bs{\beta}$ and $\bs{\alpha}$ are $q\times (q-1)$ matrices---the restricted constant, $\bs{\delta}$, in the VECM corresponds to the constant term we included in the VAR. Let $\bs{\beta}_{\perp}$ denote the $(q-1)\times 1-$dimensional orthogonal complement of $\bs{\beta}$, such that $\bs{\beta}'\bs{\beta}_{\perp} = \mathbf{0}_{(q-1)\times 1}$. Then, following \cite{kasa1992common}, we can decompose the factors as:
\begin{equation}
\mathbf{f}_t\ =\ \bs{\beta}_{\perp}(\bs{\beta}_{\perp}'\bs{\beta}_{\perp})^{-1}\bs{\beta}_{\perp}'\mathbf{f}_t + \bs{\beta}(\bs{\beta}'\bs{\beta})^{-1}\bs{\beta}'\mathbf{f}_t\ =\ \bs{\beta}_{\perp}(\bs{\beta}_{\perp}'\bs{\beta}_{\perp})^{-1}\tau_t + \bs{\beta}(\bs{\beta}'\bs{\beta})^{-1}\mathbf{c}_t,
\end{equation}
where $\tau_t$ is the common trend, and $\mathbf{c}_t$ the common cycles.
Using the notation of equation \eqref{sbeq::obsTR}, we have that $\bm\psi=\bs{\beta}_{\perp}(\bs{\beta}_{\perp}'\bs{\beta}_{\perp})^{-1}$, and $\bm\omega=\bs{\beta}(\bs{\beta}'\bs{\beta})^{-1}\mathbf{c}_t.$

We estimate the parameters of the VECM in \eqref{eq:VECMnov25}, following the standard methodology by \citet{johansen1988statistical} applied on the initial estimate of the factors obtained as in \citet{barigozzi_large-dimensional_2021}. To account for the Covid pandemic, we apply the same adjustment used in the benchmark model. The estimated parameters are then used to recover the coefficients of the implied VAR in levels, which serve as inputs for the Kalman smoother.

\begin{figure}[ht!]\caption{Output gap and potential output year-on-year growth estimates with a VECM} \label{fig::VECMgap}
\centering \footnotesize \sc \smallskip
\setlength{\tabcolsep}{.01\textwidth}        
\begin{tabular}{cc}
Potential output growth & Output gap \\
\raisebox{-.15\height}{\includegraphics[trim={2cm 9.1cm 1.8cm 9.5cm},clip,width = 0.475\textwidth]{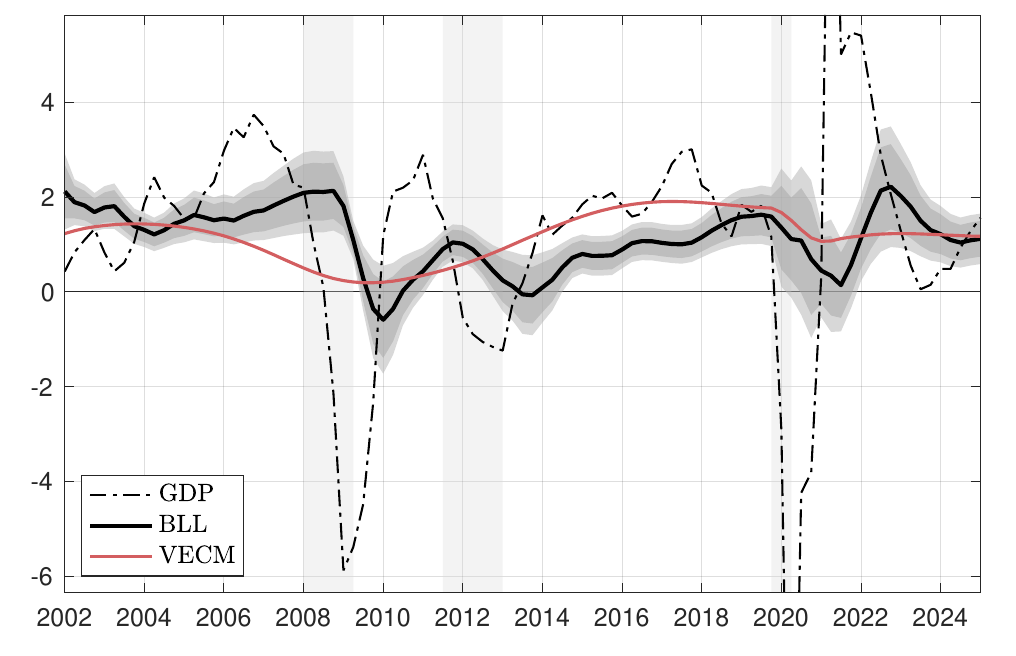}}&
\raisebox{-.15\height}{\includegraphics[trim={2cm 9.1cm 1.8cm 9.5cm},clip,width = 0.475\textwidth]{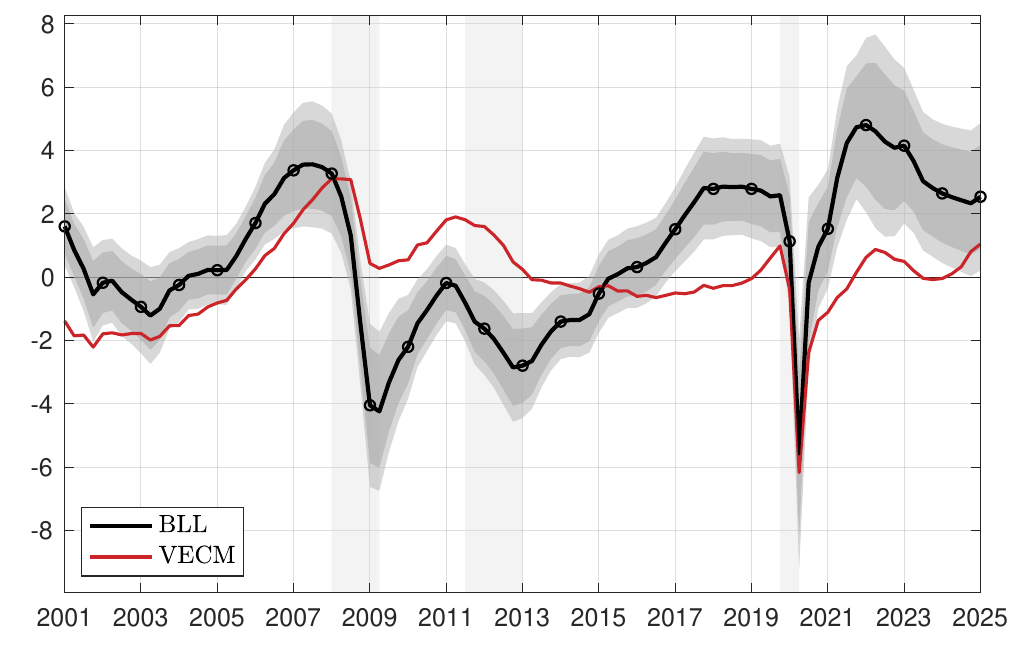}}
\end{tabular}
\begin{tabular}{p{.98\textwidth}} \scriptsize
Notes: \rm The black solid line is our benchmark estimate, the grey shaded areas are the 68$\%$ and 84$\%$ confidence bands, the black dashed line is GDP YoY growth rate, the red line is the estimate obtained with the VECM. The level of the output gap is the percentage deviation from potential.
\end{tabular}
\end{figure}

Figure \ref{fig::VECMgap} shows the results obtained by fitting the VECM on the common factors. At face value, the output gap estimated with this approach (right panel) suggests that the EA economy operated above potential between the GFC and the SDR, and mostly below potential thereafter until the Covid pandemic. Moreover, the VECM estimates that potential output growth was close to zero even before the GFC. Finally, this approach leaves a sizeable share of GDP fluctuations unexplained---that is, captured by the idiosyncratic component $\xi_{i,t}$. We therefore interpret these findings as evidence that the VECM-based strategy is unstable and unreliable, which is unsurprising given that consistent estimation of cointegration relationships typically requires long time series.

\subsection{Identification à la Morley et al. (2023)}\label{app:morley}
{\cite{morley2023simple}, henceforth MTW, propose an alternative trend smoothing approach to correct the estimated trend whenever it displays some serial correlation in first differences despite being assumed to be a random walk. We can apply the MTW approach in our setting by estimating an ARMA(1,1) model on the first difference of the estimated common trend $\Delta\widehat{\tau}_t^{(0)}$. The trend estimate corrected as in MTW is given by
\begin{equation}
\Delta \widetilde{\tau}_t\ =\ \left(\frac{1+\widehat\theta}{1-\widehat{\phi}}\right)\widehat{\varepsilon}_t,\label{morley}
\end{equation}
where $\widehat\phi$ and $\widehat\theta$ are the estimated ARMA parameters, and $\widehat{\varepsilon}_t$ are the ARMA residuals. By cumulating $\Delta \tilde{\tau}_t$, we obtain the MTW estimate of the common trend, $\tilde{\tau}_t$. 

In practice, smoothing as in \eqref{morley} appears less efficient than our Kalman smoother approach, largely due to residual Covid-related volatility in the estimated factors, which affects the ARMA estimation. Excluding observations from 2020 and 2021, the sample variance of the differenced trend in our estimate is $\widehat{\text{Var}}(\Delta\widehat\tau_t)=\widehat{\sigma}\nu^2=0.092$, compared with $\widehat{\text{Var}}(\Delta\widetilde\tau_t)=(1+\widehat\theta)(1-\widehat{\phi})^{-1}\widehat{\sigma}{\varepsilon}^2=0.17$ for the MTW estimate in \eqref{morley}, where $\widehat{\sigma}_{\varepsilon}^2$ denotes the sample variance of $\widehat{\varepsilon}_t$. Although these values are relatively close, they diverge sharply when computed over the full sample, increasing to 0.095 and 0.52, respectively. This widening gap reflects the anomalous fluctuations in the ARMA residuals $\widehat{\varepsilon}_t$ during 2020–2021.

Figure \ref{fig::ogapMTW} compares our output gap estimate with the one obtained by applying the correction proposed by \cite{morley2023simple}. As expected, the two approaches yield very similar estimates, but in 2021 and 2022, when the MTW corrections reduce the estimate of the output gap from $+4\%$ to $+3\%$.

\begin{figure}[h!]\caption{Output gap estimate with identification \`a la \cite{morley2023simple}}\label{fig::ogapMTW}
\centering \footnotesize \sc \smallskip
\begin{tabular}{p{.5\textwidth}p{.5\textwidth}}
\includegraphics[trim={2cm 9.1cm 2.2cm 9.5cm},clip,width = 0.5\textwidth]{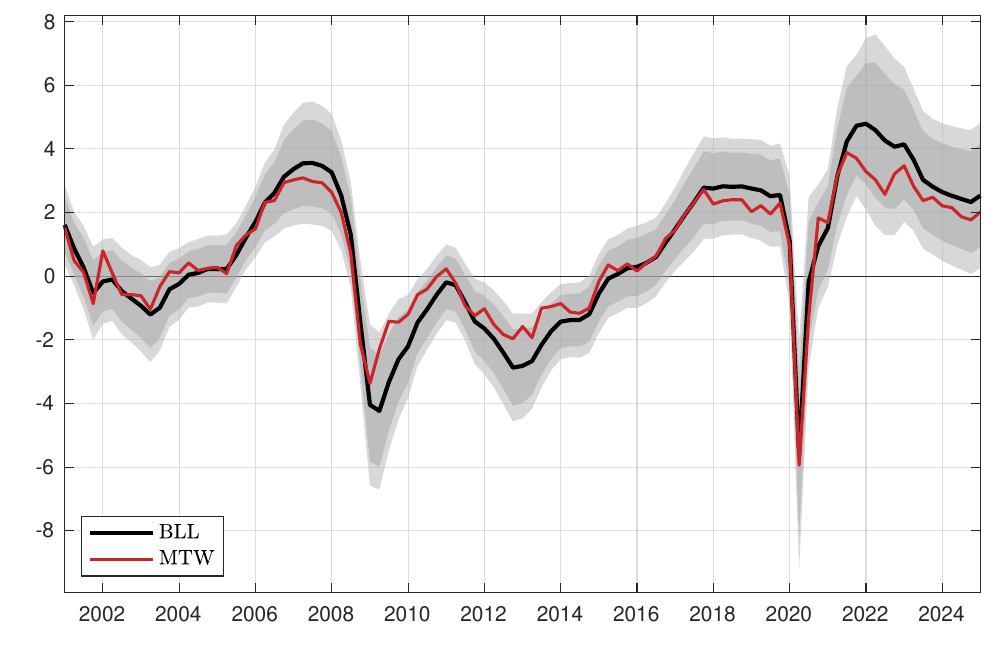}\\
\scriptsize Notes: \rm The black bold line is our estimate of the output gap. The grey shaded areas are the 68$\%$ and 84$\%$ confidence bands. The light red line is the estimated output gap obtained applying the correction of the preliminary estimated trend proposed by \cite{morley2023simple} (MTW).
\end{tabular}
\end{figure}

%
%
\setcounter{table}{0}
\setcounter{figure}{0}
\setcounter{equation}{0}
\section{Real-time reliability}\label{sec::realtime}
There is skepticism in the literature on the reliability of model-based output gap estimates in real time because of the size of end-of-sample revisions \citep{orphanides2002unreliability}. Model-based estimates of the output gap are subject to revisions in real time because new information leads to changes in both the estimates of the model parameters and the latent states. In this section, we assess the reliability of our output gap estimate through a \textit{quasi}-real-time exercise on expanding windows, where the first window begins in 2000:Q1 ends in 2015:Q1 ($T=57$).

Looking at the upper charts in Figure \ref{fig::QRT_ogap}, it is clear that the model is slow in recognizing how deep the 2012 Sovereign debt crisis is, only doing so once data for 2016 becomes available. As much as this result is disappointing, we cannot help but notice that 60 observations are probably too few to pin down the output gap accurately. As more information becomes available, the model's estimate of the output gap stabilizes, so much so that from 2017 onward, the \textit{quasi}-real-time estimate is very close to the final estimate.

\begin{figure}[ht!] \caption{Output Gap: quasi-real time, expanding window}\label{fig::QRT_ogap}
\centering \sc \smallskip
\setlength{\tabcolsep}{.01\textwidth}
\begin{tabular}{ccc}
\footnotesize 2015:Q4 & \footnotesize 2016:Q4 & \footnotesize 2017:Q4 \\
\includegraphics[trim={2cm 9.1cm 2.2cm 9.5cm},clip,width = 0.325\textwidth]{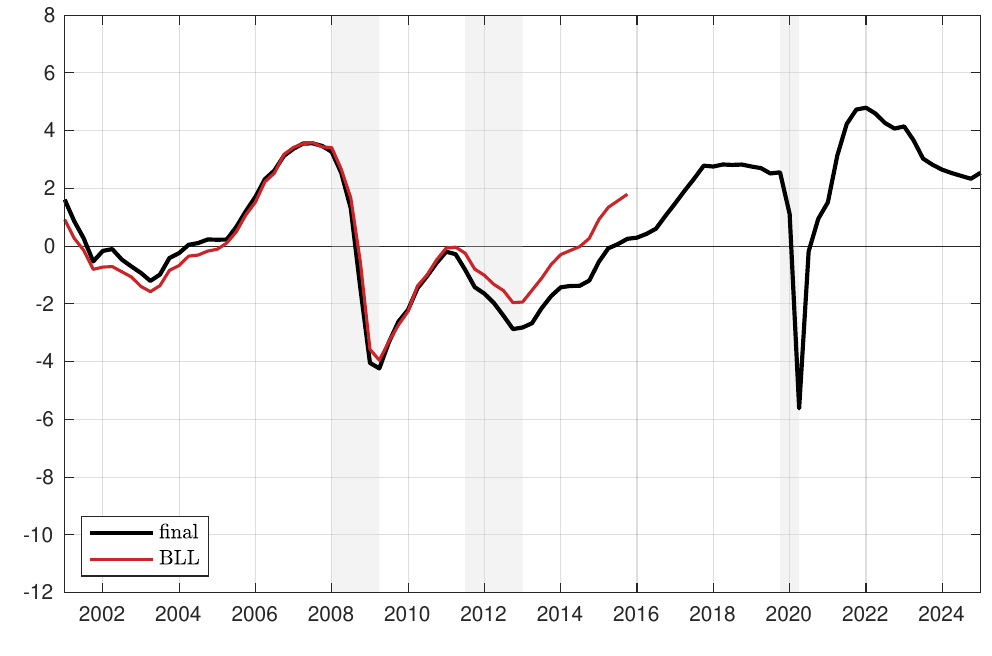} & 
\includegraphics[trim={2cm 9.1cm 2.2cm 9.5cm},clip,width = 0.325\textwidth]{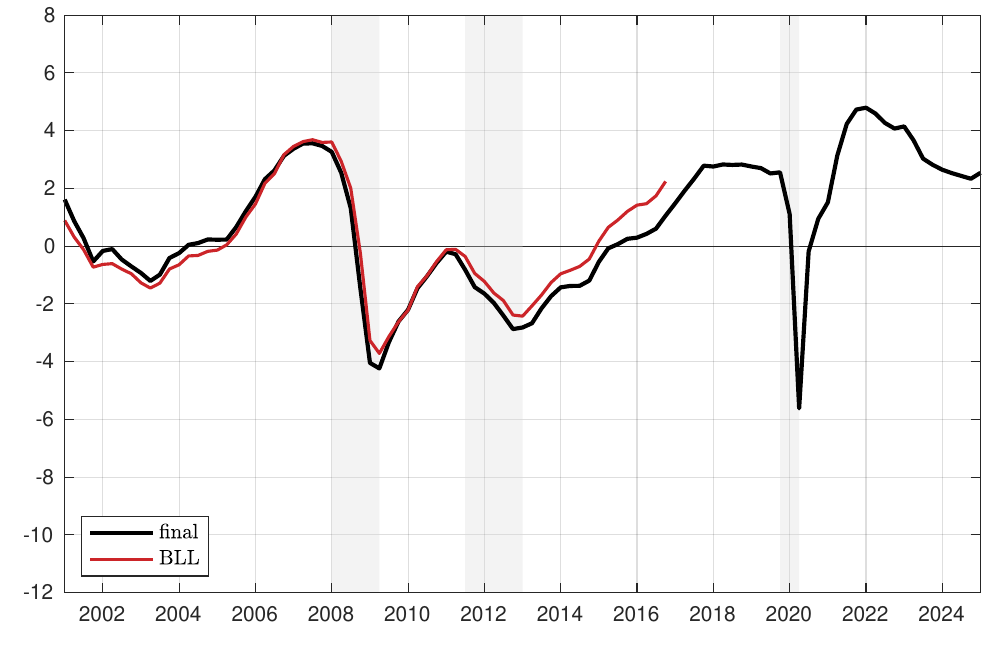} &
\includegraphics[trim={2cm 9.1cm 2.2cm 9.5cm},clip,width = 0.325\textwidth]{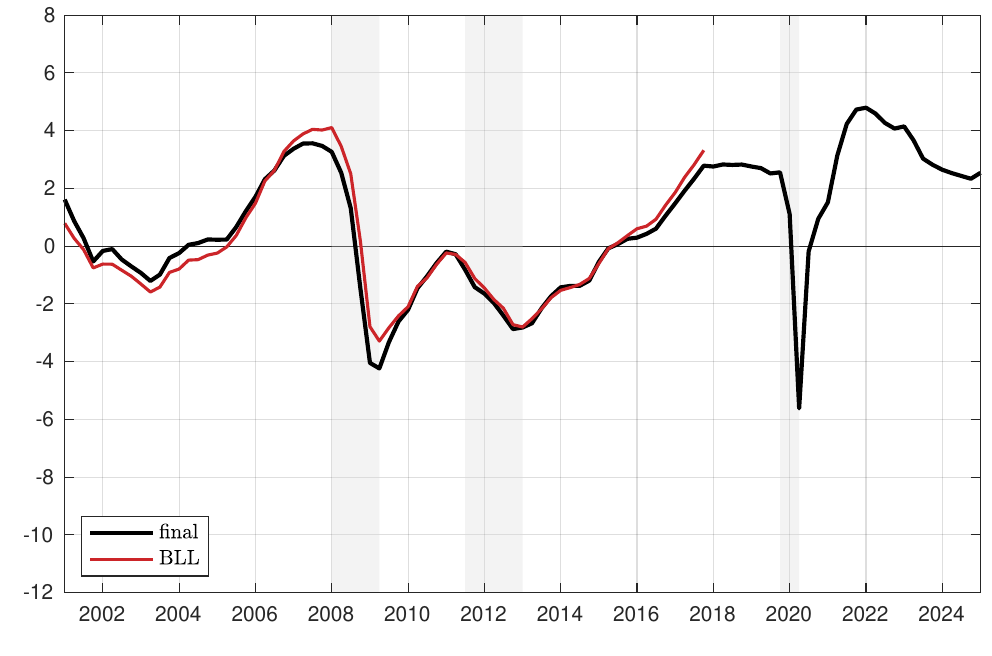} \\
\footnotesize 2018:Q4 & \footnotesize 2019:Q4 & \footnotesize 2020:Q4 \\
\includegraphics[trim={2cm 9.1cm 2.2cm 9.5cm},clip,width = 0.325\textwidth]{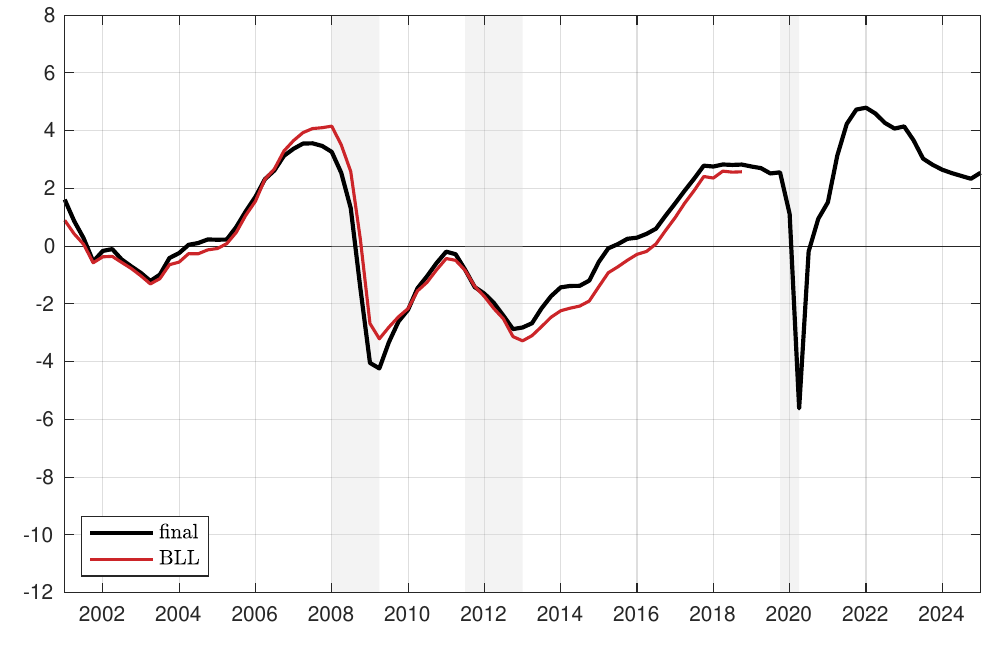} &
\includegraphics[trim={2cm 9.1cm 2.2cm 9.5cm},clip,width = 0.325\textwidth]{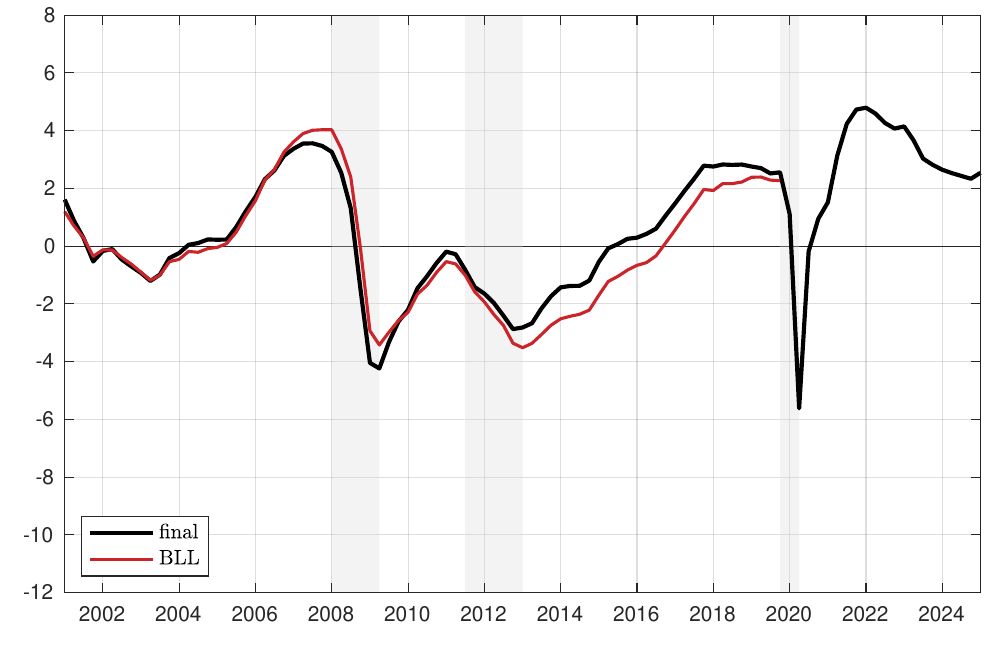} &
\includegraphics[trim={2cm 9.1cm 2.2cm 9.5cm},clip,width = 0.325\textwidth]{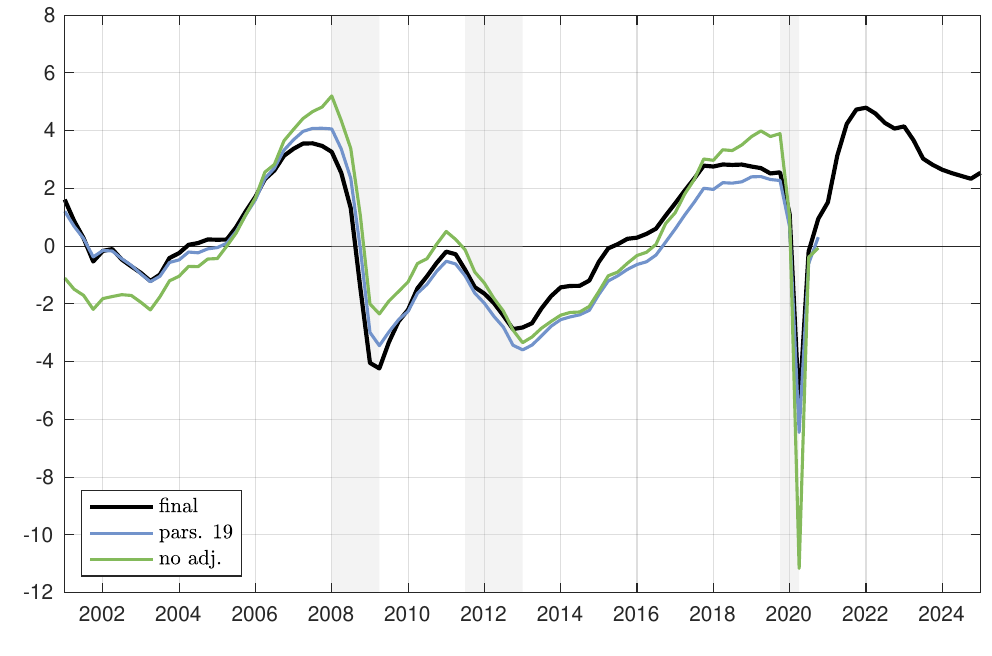} \\
\footnotesize 2021:Q4 & \footnotesize 2022:Q4 & \footnotesize 2023:Q4 \\
\includegraphics[trim={2cm 9.1cm 2.2cm 9.5cm},clip,width = 0.325\textwidth]{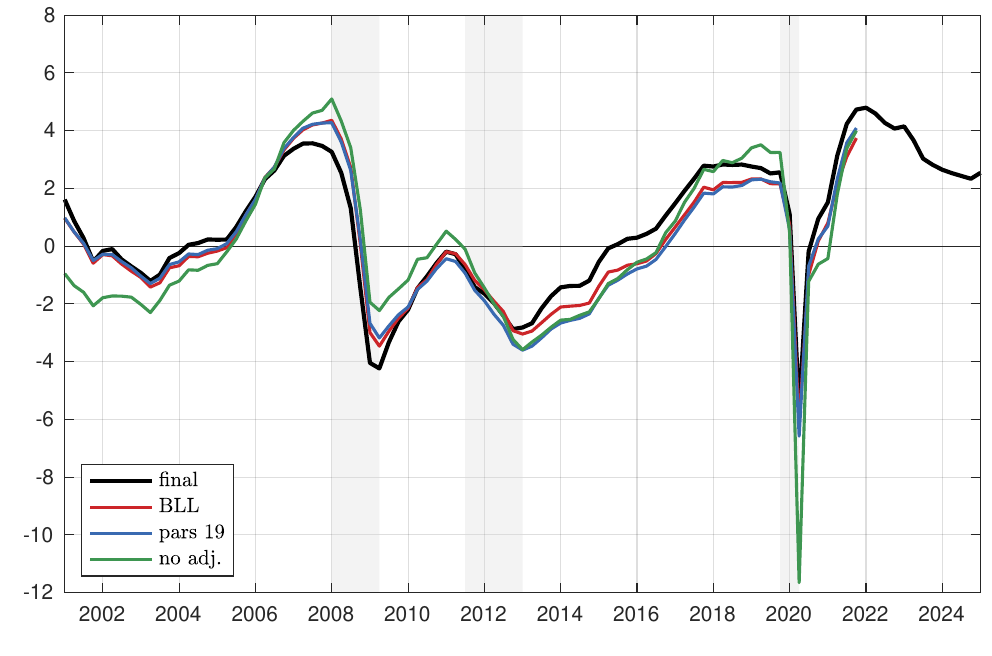} & 
\includegraphics[trim={2cm 9.1cm 2.2cm 9.5cm},clip,width = 0.325\textwidth]{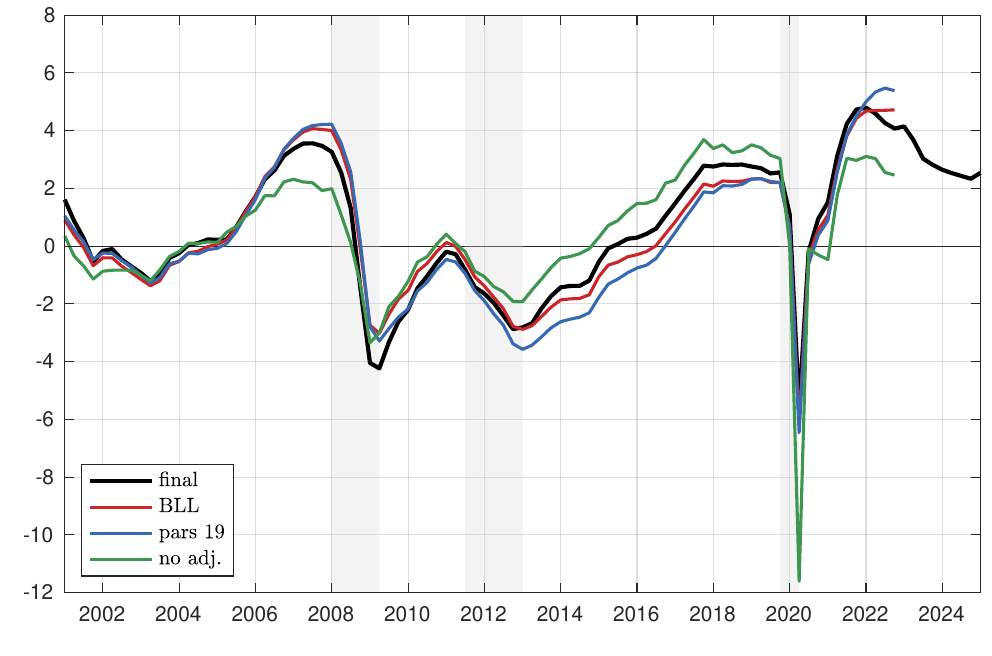} &
\includegraphics[trim={2cm 9.1cm 2.2cm 9.5cm},clip,width = 0.325\textwidth]{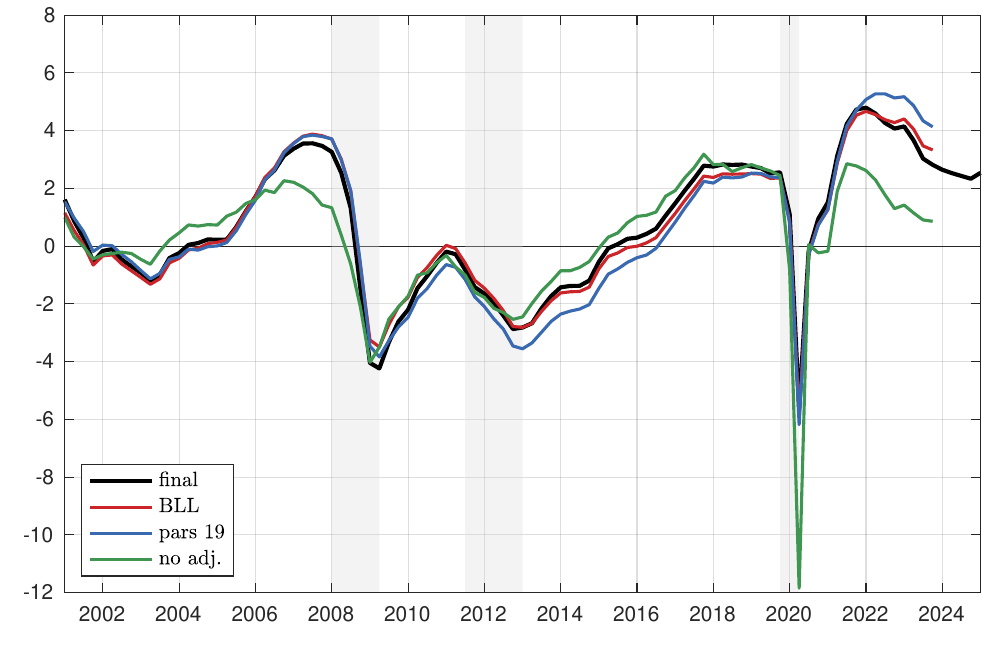}\\
\footnotesize 2024:Q4 &  & \\
\raisebox{-.65\height}{\includegraphics[trim={2cm 9.1cm 2.2cm 9.5cm},clip,width = 0.325\textwidth]{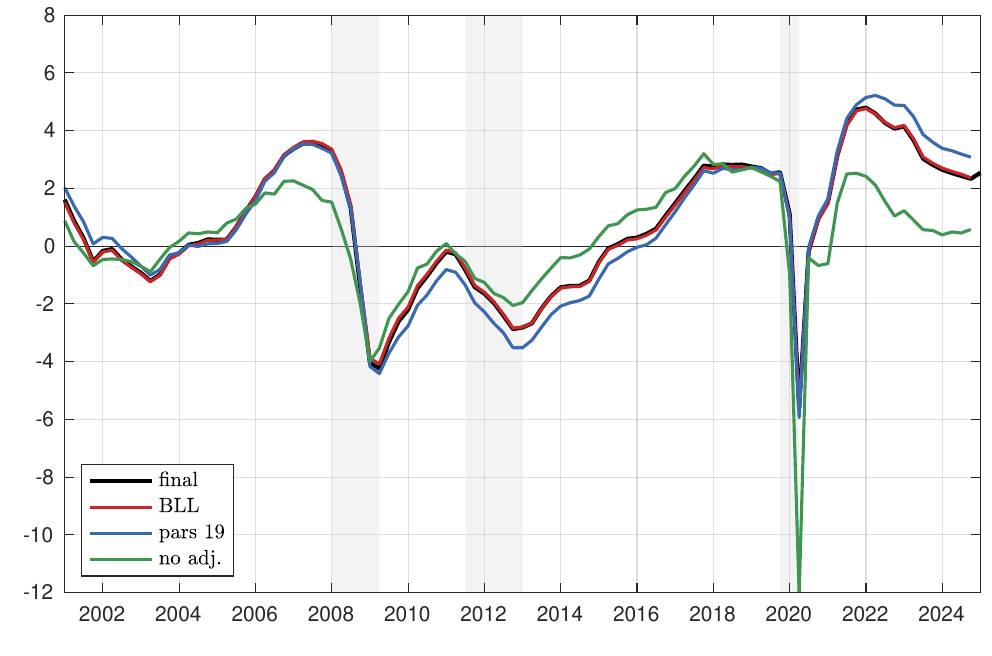}} & 
\multicolumn{2}{p{.65\textwidth}}{\scriptsize \textsc{Notes}: \rm The black line is the estimate of the output gap obtained using the final sample of data, up to 2024:Q4. The red line in Figure \ref{fig::QRT_ogap} displays the evolution of our \textit{quasi}-real-time estimate of the output gap obtained using the procedure outlined in Section \ref{sec::method}. The blue line shows the estimate we would have obtained without any adjustment for the Covid shock. Lastly, the green line is the estimate we would have obtained if we had frozen the parameter estimate at the value in 2019:Q4.} \\
\end{tabular}
\end{figure}

Moving to the Covid pandemic and its aftermath, it is evident that our strategy for adapting to the Covid shock was viable only starting from the second half of 2020 or even 2021. Moreover, it is reasonable to assume that anybody would have understood that doing nothing and allowing the Covid shock to affect the estimates was a huge mistake. Thus, we consider the \textit{quasi}-real-time performance of the simple strategy of freezing the parameters to pre-Covid data. As shown by the green line, if we had followed this approach, we would have had very reliable output gap estimates, as opposed to extreme estimates, had we chosen to do nothing. Finally, by comparing the red and the green lines, we can appreciate the impact of the adjustment we implemented to account for the Covid shock.

\begin{figure}[ht!]\caption{Output gap: \textit{quasi}-real time results} \label{fig::QRT_compare}
\centering \sc \smallskip
\setlength{\tabcolsep}{.005\textwidth}
\begin{tabular}{ccc}
\small Adjusted for Covid & \small HP & \small CF \\
\includegraphics[trim={2cm 9.1cm 2.2cm 9.5cm},clip,width = 0.325\textwidth]{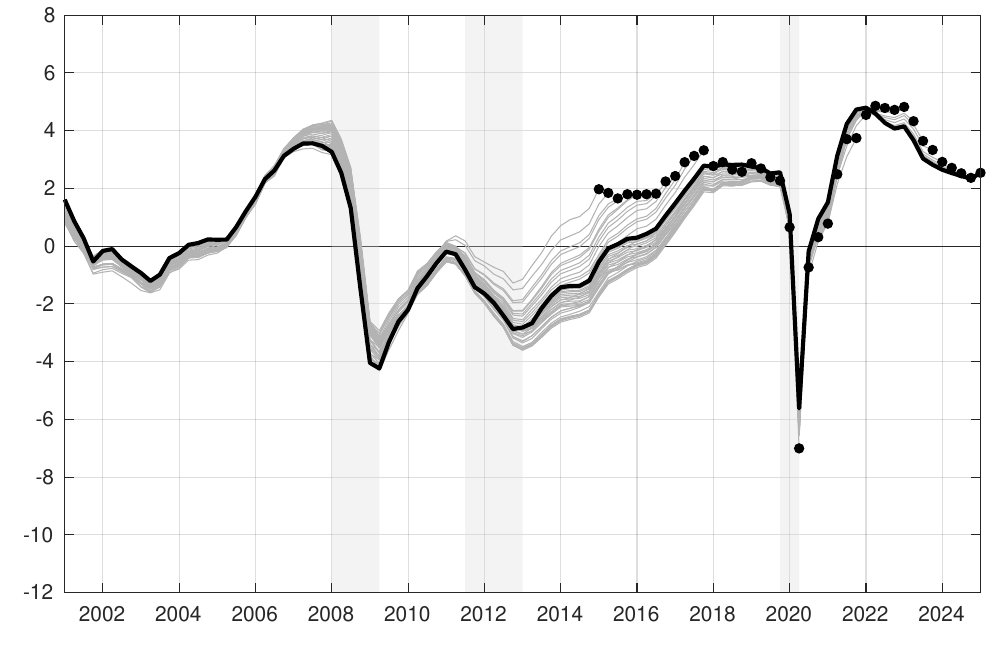} & \includegraphics[trim={2cm 9.1cm 2.2cm 9.5cm},clip,width = 0.325\textwidth]{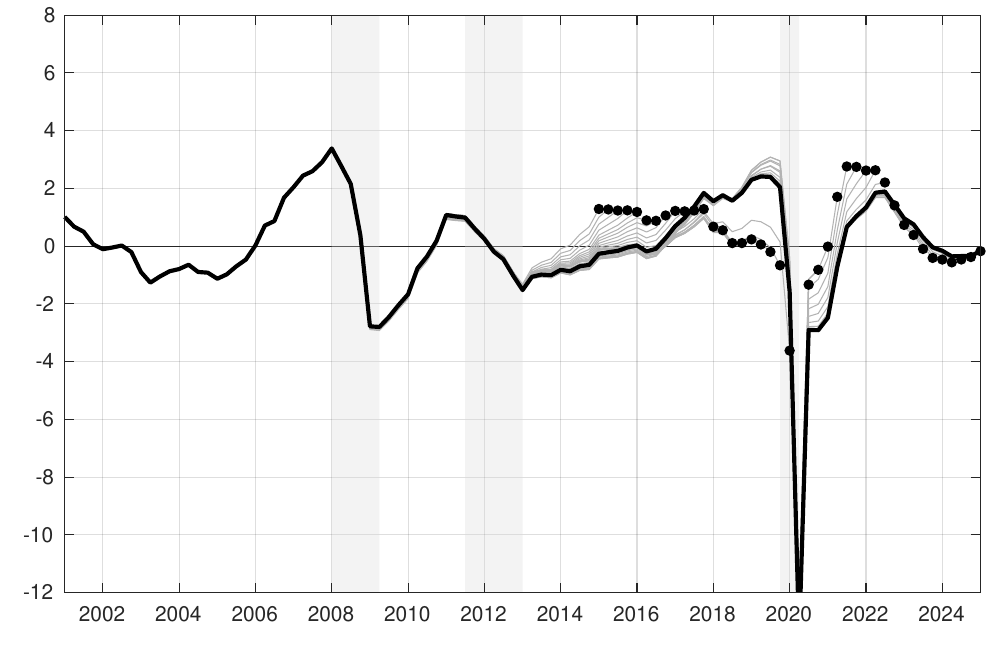}
& \includegraphics[trim={2cm 9.1cm 2.2cm 9.5cm},clip,width = 0.325\textwidth]{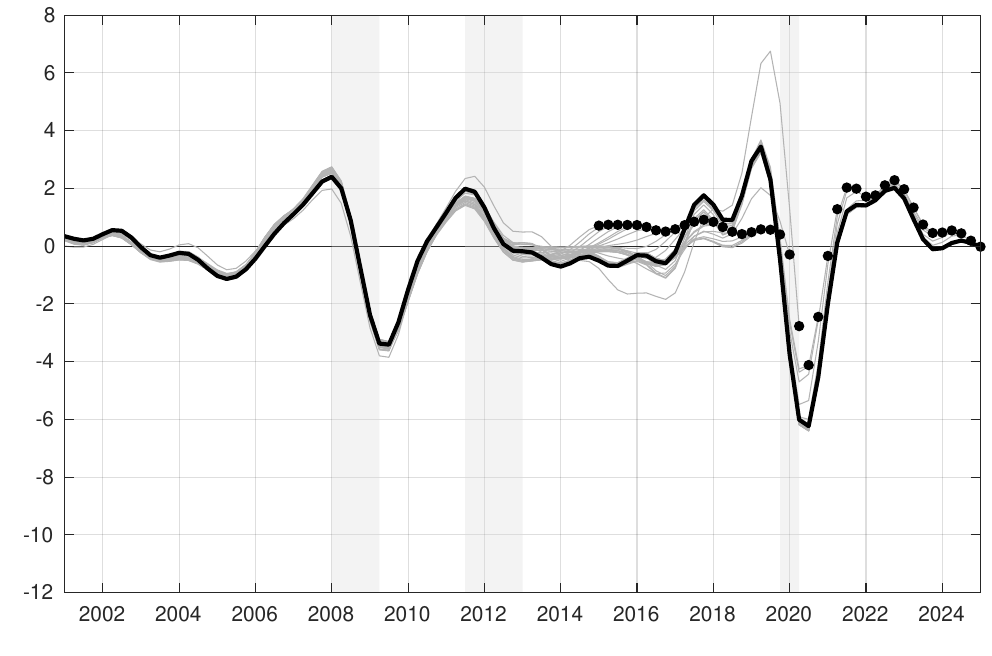}
\end{tabular}
\begin{tabular}{p{\textwidth}} \scriptsize
\textsc{Notes}: \rm the black line is the estimate of the output gap obtained over the full sample, and the thin grey lines are the estimate obtained on all the other subsamples. Each black dots represent the estimate of the output gap for quarter \textit{Q} and year \textit{Y} obtained on the sample ending at quarter \textit{Q} and year \textit{Y}.\\
\end{tabular}
\end{figure}

Figure \ref{fig::QRT_compare} compares the \textit{quasi}-real-time estimate of the output gap from our model with those obtained from an HP filter and a Christiano-Fitzgerald band-pass filter. As mentioned earlier, our estimate's weakness is that the one obtained on the sample ending in 2015 is quite different from the final estimate. However, its strength is that from 2017 onward, the \textit{quasi}-real-time estimates converge to the final and are very robust. In contrast, the HP filter and the Christiano-Fitzgerald filter yield \textit{quasi}-real-time estimates that are never to far from the final estimate. However, they converge to the final estimate very late, making them unreliable for real-time analysis.

\end{document}